%% file: main.tex
\documentclass[a4paper,11pt]{article}
\pdfoutput=1 

\usepackage{jheppub} 
\usepackage[T1]{fontenc} 
\usepackage{natbib}
\usepackage{here}
\usepackage{float}
\usepackage{multirow}
\usepackage{graphicx}
\usepackage{subcaption}
\usepackage{multirow}
\include{preamble}

\title{
    Two-Loop Integrals for Planar Five-Point One-Mass Processes
}

\preprint{CP3-20-18, FR-PHENO-2020-008}

\author[1]{Samuel Abreu,}
\affiliation[1]{Center for Cosmology, Particle Physics and Phenomenology (CP3), \\
Universit\'e Catholique de Louvain, 1348 Louvain-La-Neuve, Belgium}

\author[2]{Harald Ita,}
\affiliation[2]{Physikalisches Institut, Albert-Ludwigs-Universit\"at Freiburg, \\
D-79104 Freiburg, Germany}

\author[3]{Francesco Moriello,}
\affiliation[3]{Institut f\"ur Theoretische Physik, Eidgen\"ossische Technische Hochschule Z\"urich, \\
Wolfgang-Pauli-Strasse 27, 8093 Z\"urich, Switzerland}

\author[4]{Ben Page,}
\affiliation[4]{Institut de Physique Th\'eorique, CEA, CNRS, Universit\'e Paris-Saclay, \\
F-91191 Gif-sur-Yvette cedex, France}

\author[2]{Wladimir Tschernow,}

\author[3]{Mao Zeng}

\abstract{
We present the computation of a full set of planar five-point two-loop
master integrals with one
external mass. These integrals are an important ingredient for two-loop
scattering amplitudes for two-jet-associated W-boson production at leading color
in QCD.
We provide a set of pure integrals together with differential equations in
canonical form.
We obtain analytic differential equations efficiently from numerical samples 
over finite fields, fitting an ansatz built from symbol letters. The symbol
alphabet itself is constructed from cut differential equations and we find that
it can be written in a remarkably compact form.
We comment on the analytic properties of the integrals and confirm the 
extended Steinmann relations, which govern the double discontinuities 
of Feynman integrals, to all orders in $\epsilon$.
We solve the differential equations in terms of generalized power series on single-parameter contours in the space of Mandelstam invariants. This form of the solution trivializes the analytic continuation and the integrals can be evaluated in all kinematic regions with arbitrary numerical precision.  
}
\begin{document}

\maketitle

\input{introduction}

\input{kinematics}

\input{masters}

\input{differentialEquation}

\input{analyticStructure}

\input{seriesSolution}

\input{numerics}

\input{conclusions}

\section*{Acknowledgments}

We would like to thank V.~Del~Duca and L.~Dixon for inspiring discussions.
The work of S.A.~is supported by the Fonds de la
Recherche Scientifique--FNRS, Belgium.
The work of B.P.~is supported by the French Agence Nationale
pour la Recherche, under grant ANR--17--CE31--0001--01.
H.I.~thanks the Pauli Center of ETH Z\"urich and the University of Z\"urich for hospitality.          
W.T.'s work is funded by the German Research Foundation (DFG) within the Research Training Group GRK 2044.
The authors acknowledge support by the state of
Baden-W\"urttemberg through bwHPC.

\appendix

\input{app_kin}

\input{app_pure}

\bibliography{main.bib}
\end{document}

%% file: preamble.tex
\usepackage[utf8]{inputenc}

\usepackage{hyperref}
\usepackage{bbold}

\usepackage{graphicx,epsf}
\usepackage{amsmath,amsfonts,amssymb,amsbsy}

\usepackage{hyperref}
\usepackage{xcolor}
\hypersetup{
    colorlinks=true,
    allcolors={blue!60!black}
}

\usepackage{multirow}
\usepackage{here}
\usepackage{stackrel}
\usepackage{tikz}
\usetikzlibrary{calc}

\usepackage{ifdraft}
\usepackage[obeyFinal]{easy-todo}

\usepackage{placeins}
\usepackage{slashed} 

\usepackage{booktabs}
\usepackage{adjustbox}

\newcommand{\be}{\begin{equation}}
\newcommand{\ee}{\end{equation}}

\newcommand{\bea}{\begin{eqnarray}}
\newcommand{\eea}{\end{eqnarray}}

\newcommand{\eps}{\epsilon}
\newcommand{\eqnDiag}[1]{ \vcenter{\hbox{#1}} }

\def\eqn#1{eq.~(\ref{#1})}

\def\fig#1{fig.~{\ref{#1}}}

\def\tab#1{table~{\ref{#1}}}

\def\spa#1.#2{\left\langle#1\,#2\right\rangle}
\def\spb#1.#2{\left[#1\,#2\right]}
\def\spash#1.#2{\spa{\smash{#1}}.{\smash{#2}}}
\def\spbsh#1.#2{\spb{\smash{#1}}.{\smash{#2}}}
\def\sand#1.#2.#3{%
\left\langle\smash{#1}{\vphantom1}^{-}\right|{#2}%
\left|\smash{#3}{\vphantom1}^{-}\right\rangle}
\def\sandpp#1.#2.#3{%
\left\langle\smash{#1}{\vphantom1}^{+}\right|{#2}%
\left|\smash{#3}{\vphantom1}^{+}\right\rangle}
\def\sandpm#1.#2.#3{%
\left\langle\smash{#1}{\vphantom1}^{+}\right|{#2}%
\left|\smash{#3}{\vphantom1}^{-}\right\rangle}
\def\sandmp#1.#2.#3{%
\left\langle\smash{#1}{\vphantom1}^{-}\right|{#2}%
\left|\smash{#3}{\vphantom1}^{+}\right\rangle}

\def\trFive{{\rm tr}_5}
\def\trp{{\rm tr}_{+}}
\def\trm{{\rm tr}_{-}}
\def\offShellScale{p_1^2}
\def\offShellScaleSquared{p_1^4}

\def\dlog{$\mathrm{d}\log$}

\def\Sherpa{{\sc Sherpa}}
\newcommand*{\pySecDec}{py{\textsc{SecDec}}}

\def\trFive{{\rm tr}_5}
\newcommand*{\ncg}{\textrm{nc}}
\newcommand*{\zmz}{\textrm{zmz}}
\newcommand*{\mzz}{\textrm{mzz}}
\newcommand*{\zzz}{\textrm{zzz}}
\newcommand*{\sqrd}{\textrm{1-loop}^2}

%% file: introduction.tex

\section{Introduction}

Since the early history of quantum field theory, perturbative scattering
amplitudes have been a crucial tool in high-energy physics. As gauge-invariant
consequences of the underlying field theory, their analytic structure
unambiguously captures features of the theory which are not manifest in the action. 
It is then no surprise that they find many uses both in formal studies of field theory,
as well as in more traditional applications 
such as in making predictions for collider physics.
The computation of
scattering amplitudes has been a topic of intense study in recent years.
Nevertheless, and despite great recent
advances at the five-point frontier~\cite{Badger:2017jhb,Abreu:2017hqn,Badger:2018gip,
Abreu:2018zmy,Badger:2018enw,Chicherin:2018yne,Abreu:2018jgq,Abreu:2018aqd,
Chicherin:2019xeg,Abreu:2019rpt,Abreu:2019odu,Badger:2019djh,Hartanto:2019uvl},
it still represents a formidable challenge at the two-loop level.
Loop scattering amplitudes are multi-valued functions, whose branch-cut
structure depends on the kinematics and loop order.
This largely theory-independent analytic structure can be packaged and understood in various ways.
One practical presentation is through a collection of so-called `master integrals', in
terms of which all scattering processes with the same kinematics and loop order
can be linearly expanded. When these master integrals evaluate to
polylogarithmic functions, it is
also fruitful to understand the analytic structure in terms of the so-called
`symbol'~\cite{Goncharov:2010jf,Duhr:2011zq,Duhr:2012fh}. 
For instance, in maximally supersymmetric Yang-Mills theory, this
organization has led to the growth of a `bootstrap' program for the
amplitudes~\cite{Dixon:2011pw,Dixon:2011nj,Dixon:2013eka,Dixon:2014voa,Dixon:2014xca,
Dixon:2014iba,Dixon:2015iva,Caron-Huot:2016owq,Caron-Huot:2019bsq}, most
recently culminating in a computation at the seven-loop order~\cite{Caron-Huot:2019vjl}.

In this work, we contribute to the understanding of the analytic structure of multi-leg two-loop
scattering amplitudes, by computing the planar two-loop master integrals with one
massive and four massless external legs. These integrals are highly relevant
for QCD collider processes including four massless partons and a heavy particle
such as a massive vector boson. While these amplitudes have been computed 
numerically~\cite{Hartanto:2019uvl}, here we take the first steps towards their analytic
calculation.
In particular, these are strongly desirable for phenomenological studies
at the Large Hadron Collider (LHC)~\cite{Amoroso:2020lgh}.
The computation of multi-scale Feynman integrals relevant for massless QCD has received much attention
in the literature.  
At two-loops all relevant four-point integrals are known
\cite{Henn:2014lfa,Gehrmann:2015ora}, and recently the planar five-point two-loop
integrals have been evaluated analytically both in terms of multiple
polylogarithms \cite{Gehrmann:2015bfy,Papadopoulos:2015jft} and a more tailored set of pentagon
functions~\cite{Gehrmann:2018yef}. Important progress has also been made beyond
the planar limit \cite{Abreu:2018aqd,Chicherin:2018old}. Much
less is known about five-point two-loop integrals with an external mass, where
only partial results are known~\cite{Papadopoulos:2015jft,Papadopoulos:2019iam}.
The planar five-point one-mass integrals at two-loops are the main topic of the
present paper.

In the past few decades a great deal of progress has been made in novel
computational techniques for Feynman integrals. One of the most effective
is the differential equations method \cite{Kotikov:1990kg,
Kotikov:1991pm, Bern:1993kr, Remiddi:1997ny, Gehrmann:1999as}.
The method is particularly useful whenever a basis of integrals is available
such that the differential equation assumes a `canonical' form \cite{Henn:2013pwa}, where the
$\epsilon$ dependence factorizes and the matrix can be expressed in terms of
so-called `\dlog-forms'.
A canonical differential equation also naturally encodes the analytic structure of the integrals,
directly manifesting the `symbol alphabet'. Simultaneously, it provides an
important way of evaluating the master integrals, and so is the perfect
workhorse for our investigations. Nevertheless, the construction of a canonical 
differential equation is challenging because it requires a basis of `pure'
master integrals~\cite{ArkaniHamed:2010gh, Henn:2014qga}.
Devising an effective $D$-dimensional algorithm 
to find such a basis  is an active field of research
(see e.g.~\cite{Chicherin:2018old,Henn:2014qga,Lee:2014ioa,Prausa:2017ltv,
Gituliar:2017vzm,Meyer:2017joq,Meyer_2018,
Wasser:2018qvj, Abreu:2018rcw, Dlapa:2020cwj,Henn:2020lye}).
Here we solve this problem by constructing the basis with a heuristic approach, which we
then validate by constructing the differential equation and observing the canonical form.
Even when the pure basis is known,
the construction of the analytic form of the differential equation is a technically challenging
procedure. We
employ the numerical sampling method of ref.~\cite{Abreu:2018rcw} implemented
over finite fields~\cite{vonManteuffel:2014ixa, Peraro:2016wsq}, which we show
can also be applied in cases where square roots must be taken in intermediate stages.
Integral reduction can then be performed numerically using, for example, standard public packages
\cite{vonManteuffel:2012np,Maierhoefer:2017hyi,Smirnov:2019qkx}
or modern unitarity based methods~\cite{Gluza:2010ws, Schabinger:2011dz, 
Ita:2015tya, Larsen:2015ped, Georgoudis:2016wff, 
Abreu:2017xsl, Abreu:2017hqn, Bendle:2019csk, Agarwal:2019rag}.
A further requirement for the application of the numerical sampling
approach of ref.~\cite{Abreu:2018rcw} is the symbol alphabet. 
With that in mind, here we show how the
full symbol alphabet can be constructed from a technically simpler computation
of cut differential equations.
We organize the symbol alphabet and observe that, despite the complex five-point one-mass
scattering kinematics, it can be written in a remarkably compact form.
We then obtain the symbols of the integrals from their differential equation
and show that the (extended) Steinmann relations \cite{Steinmann, Steinmann2, Cahill:1973qp,
Caron-Huot:2016owq, Dixon:2016nkn} follow from the structure of the differential equation.

The differential equation is also a useful tool to represent the master integrals
in terms of known sets of functions, which can then be used
for their efficient numerical evaluation. 
There also exist numerical approaches based on Monte-Carlo
integration~\cite{Smirnov:2015mct,Borowka:2017idc,Mandal:2018cdj,Capatti:2019ypt,Capatti:2019edf,
Runkel:2019yrs}, some of which have been used to supply two-loop integrals in
amplitude computations \cite{Borowka:2016ehy, Borowka:2016ypz, Borowka:2018anu,
Jones:2018hbb, Maltoni:2018zvp, Chen:2019fla}. However, because the efficiency and precision of
Monte-Carlo techniques are often limiting, analytic results are still desirable.
Obtaining such results from the differential equation can be practically difficult
in multi-scale applications such as five point integrals. Indeed, whilst the results would
be naturally written in term of multiple polylogarithms, the analytic continuation 
required to be able to use them all over phase-space can be challenging.
In this paper we apply the method of~\cite{Francesco:2019yqt}, where analytic
solutions to the differential equation are constructed on a 1-dimensional path in the form of a
collection of generalized power series. Such an approach trivializes the
integration step and analytic continuation is easily implemented by appropriate
choice of integration contours. As such, the integrals can easily be computed in not just
the Euclidean but also the physical regions with high numerical precision.
The approach of ref.~\cite{Francesco:2019yqt} has already been
applied in the computation of two-loop integrals for the QCD corrections to
Higgs+jet
production~\cite{Francesco:2019yqt,Bonciani:2019jyb,Frellesvig:2019byn}. Here we
apply it for the first time to five-point kinematics. We review the application of the method to a canonical differential equation and explain how to use it to compute boundary
conditions. Furthermore, we demonstrate the readiness of the method for LHC
physics in a number of ways, such as computing high-precision boundary conditions for
the integrals at hand in both Euclidean and physical regions and showing the
efficiency with various studies over physical phase space.

The main numerical and analytic results of the paper are provided in a set of
ancillary files. The definition of the pure master integrals is given in
\texttt{anc/*/pureBasis-*.m}. The alphabet is given in the files
\texttt{anc/alphabet.m}. The differential equations for the three integral
families are given in the files \texttt{anc/*/diffEq-*.m}. High precision
reference values are given in \texttt{anc/*/numIntegrals-*.m}. For convenience,
we provide an example of how to use these ancillary files in
\texttt{anc/usageExample.m} where we also generate the symbols of the master
integrals. 

The paper is structured as follows. First, in section
\ref{sec:scatKin} we describe important features of the kinematics relevant for
five-point one-mass scattering. In section \ref{sec:masterInt} we describe the
loop integrals which we compute.
Next, in section \ref{sec:DifferentialEquations}, we discuss our numerical
construction of the differential equations and thereby the basis of pure
integrals. Further, in section \ref{sec:AnalyticStructure}
we discuss  the symbol alphabet and the implications of the differential equations for the analytic
structures of scattering amplitudes. In section
\ref{sec:DEIntegration} we discuss the application of the generalized series
method to the solution of the differential equations. 
In section \ref{sec:numerics}, we study numerical evaluation of the
integrals in physical regions.
Finally, we summarize the
results of our work and discuss extensions in section \ref{sec:Conclusions}.

%% file: kinematics.tex

\section{Scattering kinematics}
\label{sec:scatKin}

The main result of this paper is a calculation of a basis of two-loop
integrals relevant for planar five-point scattering processes with a single
massive external leg. However, before we delve into that problem, we first
briefly discuss the kinematics of these processes and introduce some quantities
that will be relevant in the following sections.

The momenta of the scattering particles are labelled $p_i,$ $i=1,\ldots,5$, and
fulfil momentum conservation, $\sum_{i=1}^5 p_i =0$. 
Without loss of generality, we assume $p_1$ to be massive, 
i.e.~$p_1^2 \ne 0$, and the remaining ones to be
massless, $p_{i}^2=0$ for  $i=2,\ldots,5$. 
Out of these momenta,  we can form six independent Mandelstam variables
of the form  $s_{ij}=(p_i+p_j)^2$, which we choose to be
\begin{equation} \label{eqn:orderedInvariants}
\vec s= \{ \offShellScale{} \,,
s_{12}\,, s_{23}\,,s_{34}\,, s_{45}\,, s_{15}\}\,.
\end{equation}
For concreteness, in this paper we use the metric $g = \mathrm{diag}(+,-,-,-)$,
which we extend with further minus signs when working in $D$ dimensions.
These variables are not sufficient to characterize the kinematics of the scattering
process: there is an additional parity label which can be
captured by the parity-odd Levi-Civita contraction
\begin{equation}
\label{eq:tr5}
\trFive{} = 4 i \varepsilon_{\alpha\beta\gamma\delta} 
\,p_1^\alpha p_2^\beta p_3^\gamma p_4^\delta\,.
\end{equation}
Indeed, space-time parity inverts all spatial momentum components,
\begin{equation}\label{eq:parityTrans}
 P:\quad ( p_i^0, \vec p_i )\quad \rightarrow \quad ( p_i^0, -\vec p_i )\,,
\end{equation}
and, while Mandelstam variables are invariant,
$\trFive{}$ gains a sign under this transformation.

It is also useful to introduce Gram determinants when discussing kinematics of scattering
processes. They are given by the determinants of the
Gram matrix $G(q_1,\ldots,q_n)$, which we define as
\begin{equation}\label{eq:gramGen}
	G(q_1,\ldots,q_n)=2\,V^T(q_1,\ldots,q_n)\,g\,V(q_1,\ldots,q_n)
	=2 \, \{q_i\cdot q_j\}_{i,j\in\{1,\ldots,n\}}\,,
\end{equation}
where the factor of two is conventional and $V(q_1,\ldots,q_{n})$ is 
a $4\times n$ matrix whose columns are the vectors $q_i$.
It can be shown from the definition of the Gram matrix that
if the $q_i$ are linearly dependent then the Gram determinant
vanishes, and also that this determinant is invariant under shifts 
of any of the $q_i$ by any of the other momenta.
Returning to the discussion of five-point one-mass kinematics, we note that
the parity-odd $\trFive{}$ is related to the parity-even five-point Gram 
determinant through
\begin{equation}
	\label{eq:gram5}
	\Delta_5 =  \det G(p_1,p_2,p_3,p_4)=\det \{2\, p_i\cdot p_j \}_{i,j\in\{1,2,3,4\}}=
	\textrm{tr}_5^2\,.
\end{equation}
In other words, $\trFive{}$ is a square root of a polynomial
in the Mandelstam variables $\vec s$.
Two other square roots related to Gram determinants which are not perfect squares
are relevant for the scattering kinematics we are considering. 
The associated Gram determinants can be written in terms 
of the K\"all\'en function $\lambda(a,b,c)$:
\begin{align}
\label{eq:gram3}
\Delta_3&=-\det G(p_1, p_2+p_3)
= \lambda(\offShellScale{}, s_{23},s_{45}) \,,\\[2ex]
\label{eq:gram3NP}
\Delta_3^{\ncg}&=-\det G(p_1,p_3+p_4)
= \lambda(\offShellScale{},s_{25},s_{34})\,, \\[2ex]
\lambda(a,b,c)&=a^2+b^2+c^2-2ab-2ac-2bc\,,
\end{align}
where the minus sign is conventional. Our notation is explained by the fact that
$\Delta_3$ is naturally associated with a degeneration of the
kinematics which preserves the cyclicity of the momenta,
 $\{p_1,p_2,p_3,p_4,p_5\}\to\{p_1,p_2+p_3,p_4+p_5\}$, while 
$\Delta_3^{\ncg}$ is associated with a degeneration that does not preserve it,
$\{p_1,p_2,p_3,p_4,p_5\}\to\{p_1,p_3+p_4,p_2+p_5\}$.
Let us note that, by properties of the Gram determinant, other equivalent choices of the arguments of $G$ are possible.
For example, the choice in \eqref{eq:gram5} of all momenta but $p_5$ is purely conventional.

We finish this section with a brief comment on the analytic structure
of Feynman integrals, to which we will return later in the paper. They evaluate to functions
of the Mandelstam variables $\vec s$ with a complicated branch cut structure.
More precisely,
the integrals we compute have branch cuts starting at $\offShellScale{}=0$,
$s_{12}=0$, $s_{23}=0$, $s_{34}=0$, $s_{45}=0$ and $s_{15}=0$.
For each integral, we thus find it convenient to label different kinematic regions by the sign of the Mandelstam
invariants. We highlight two types of regions that we will return to 
in following sections.
First, the region where we are away from any branch cuts and where the integrals
evaluate to real numbers. This region is called the Euclidean region,
and in our case it corresponds to having  all Mandelstam variables negative.
For five-point one-mass kinematics, it is not a physical region (i.e., there is
no physical configuration of momenta that corresponds to values of Mandelstam
variables in the Euclidean region). Second, we consider regions
associated with the production of a massive vector boson 
in association with two jets in QCD. This physical process is 
a natural application of the one-mass five-point two-loop integrals.
We assign the massless momenta $p_i\,,i=2,\ldots,5$
to massless partons and the massive momentum $p_1$ to the vector boson,
which we assume to decay, e.g.~into a lepton pair. 
This implies that the momentum $p_1$ is timelike, i.e.,~$\offShellScale{}>0$.
Since any of the parton momenta may be in the initial state, we have
six different channels,
\begin{equation}
p_i+p_j \rightarrow p_1+p_k+p_l\,,
\end{equation}
where $i,j,k,l$ take distinct values in $\{2,3,4,5\}$
and to each channel corresponds a kinematic region.
In \tab{tab:regions}  we give the signs of the kinematic invariants
for each region.
We note that, since momenta corresponding to a physical scattering process must
have real components and $\det(g)=-1$, it follows from \eqref{eq:gramGen} that
$\Delta_5<0$ .

\begin{table}[]
\centering
 \begin{tabular}{| c | c | c | c | c | } 
\hline
	&  Initial State  	& $	>0	$ & $<0	$\\\hline
Euclidean 	& ---	& &	$ s_{12}, s_{23}, s_{34}, s_{45}, s_{15}, \offShellScale{} $\\ \hline
\multirow{6}{*}{V-production}&$2,3$ & $ s_{23},s_{45},s_{15},\offShellScale $&$ s_{12},s_{34} $ \\   
	 	&	$2,4$		& $ s_{15},\offShellScale 		$&$ s_{12},s_{23},s_{34},s_{45}   $ \\
	 	&	$2,5$		& $ s_{34},\offShellScale 		$&$ s_{12},s_{23},s_{45},s_{15}   $ \\ 
	 	&	$3,4$		& $ s_{12},s_{34},s_{15},\offShellScale $&$ s_{23},s_{45} 		  $  \\
	 	&	$3,5$		& $ s_{12},\offShellScale 		$&$ s_{23},s_{34},s_{45},s_{15}	  $	\\
	 	&	$4,5$		& $ s_{12},s_{23},s_{45},\offShellScale $&$ s_{34},s_{15}         $ \\\hline
  \end{tabular}
 \caption{Signs of ordered Mandelstam invariants in the Euclidean and the physical 
		phase space.}
 \label{tab:regions}
\end{table}

%% file: masters.tex

\section{Two-loop planar five-point one-mass integrals}
\label{sec:masterInt}

\begin{figure}
	\centering
	\begin{subfigure}{0.45\textwidth}\centering
		\includegraphics[scale=0.7]{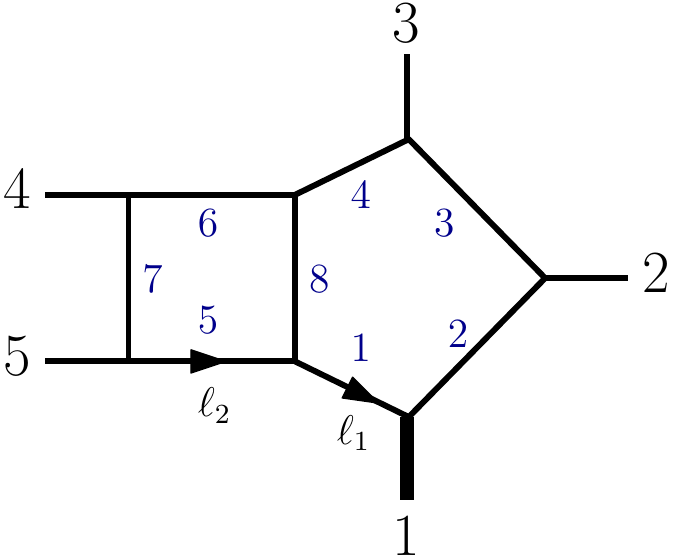}
		\caption{$I^{[\mzz]}[\vec\nu]$}
		\label{fig:mzz}
	\end{subfigure}
	\begin{subfigure}{0.45\textwidth}\centering
		\includegraphics[scale=0.7]{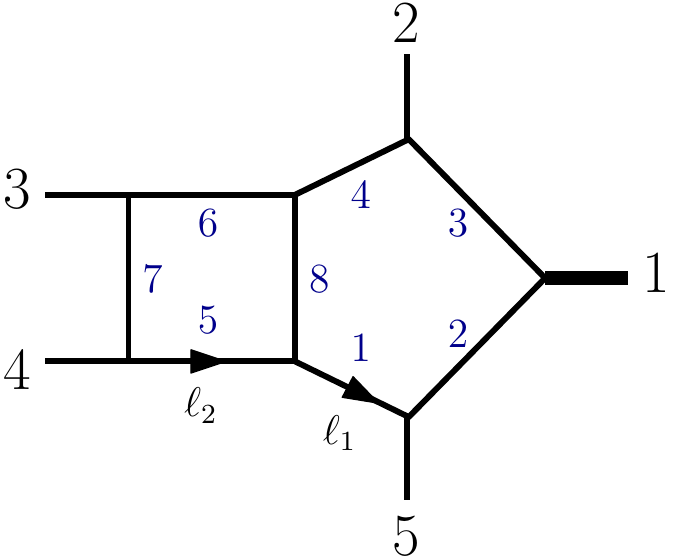}
		\caption{$I^{[\zmz]}[\vec\nu]$}
		\label{fig:zmz}
	\end{subfigure}\vspace{3mm}
	\begin{subfigure}{0.45\textwidth}\centering
		\includegraphics[scale=0.7]{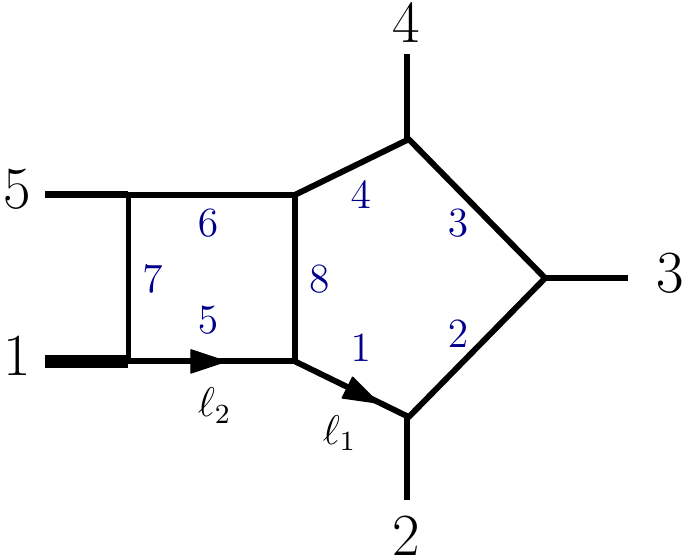}
		\caption{$I^{[\zzz]}[\vec\nu]$}
		\label{fig:zzz}
	\end{subfigure}
	\begin{subfigure}{0.45\textwidth}\centering
		\includegraphics[scale=0.7]{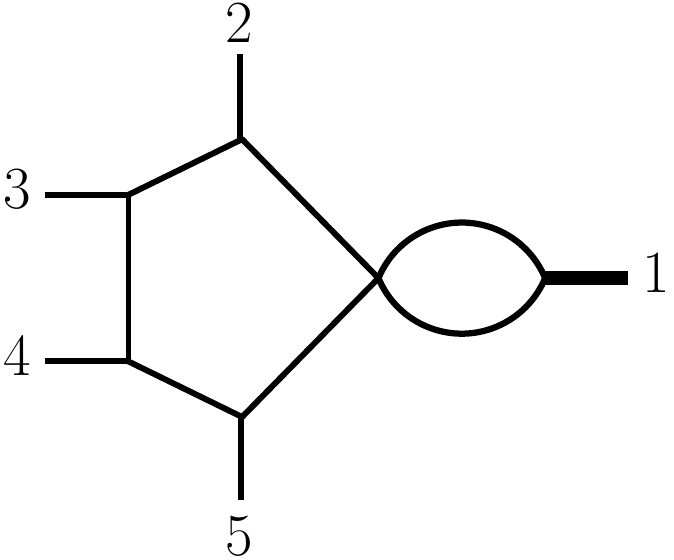}
		\caption{$I^{[\sqrd]}[\vec\nu]$}
		\label{fig:sqrd}
	\end{subfigure}
	\caption{Two-loop five-point one-mass topologies. The thick external line
	with label 1 denotes the massive external leg.}
	\label{fig_families_int}
\end{figure}

Planar five-point scattering amplitudes with a single massive external leg can
be written as a linear combination of Feynman integrals. These integrals form a
linear space spanned by a set of so-called `master integrals', which can be
generated by considering the four integral topologies depicted in
\fig{fig_families_int}.\footnote{\label{foot:relabellings} More precisely,
  to compute an amplitude we would also need to consider the topologies obtained
  by the relabelling $(2\leftrightarrow 5,3\leftrightarrow 4)$.} In this section
we establish our notation and briefly describe these linear spaces.

The topologies in \fig{fig_families_int} can be categorized as either
genuine two-loop, or `one-loop squared'. The integrals of the topology
$I^{[\sqrd]}$ of \fig{fig:sqrd}, all factorize into a product of two
one-loop integrals. Computing them is thus not a genuine two-loop problem, and
indeed they lack many of the features associated with multi-loop integrals (for
instance, there are no irreducible scalar products). Given that computing the
master integrals in this topology is a one-loop problem, we will not discuss
this topology further and in the remainder of this paper will choose instead to
discuss corresponding one-loop integrals.
In contrast, the integrals of the topologies in figs.~\ref{fig:mzz},
\ref{fig:zmz} and \ref{fig:zzz}, which all have a `penta-box' as top diagram,
are genuine two-loop integrals and the main result of this paper.
They differ by the position of the massive external leg, and we encode this in the
label for each topology. Precisely, the notation characterizes the mass
assignment for the three external legs attached to the pentagon subloop: they
can all be `zero mass' ($\zzz$), the middle leg can be massive ($\zmz$), or the
first leg can be massive ($\mzz$).
All other assignments are related to these choices by
relabelling of the kinematics, see footnote \ref{foot:relabellings}.
This notation is also used in the ancillary files accompanying this paper.

To each topology $f$ is associated an integral of the form 
\begin{equation}\label{eq:intNorm}
I^{[f]}[\vec \nu] =  e^{2\epsilon\gamma_E} \int 
\frac{d^D\ell_1}{i\pi^{D/2}}\frac{d^D\ell_2 }{i\pi^{D/2}} 
\frac{\rho_{9,f}^{-\nu_9}\,\,\rho_{10,f}^{-\nu_{10}}\,\,\rho_{11,f}^{-\nu_{11}}}
{\rho_{1,f}^{\nu_1}\,\,\rho_{2,f}^{\nu_2}\,\,
\rho_{3,f}^{\nu_3}\,\,\rho_{4,f}^{\nu_4}\,\,
\rho_{5,f}^{\nu_5}\,\,\rho_{6,f}^{\nu_6}\,\,
\rho_{7,f}^{\nu_7}\,\,\rho_{8,f}^{\nu_8}}\,,
\end{equation}
with $D=4-2\epsilon$, and we have included some normalization factors
that are conventional in dimensional regularization ($\gamma_E$ is the 
Euler-Mascheroni constant).
The explicit expression of the inverse propagators $\rho_{1,f}$, \ldots, $\rho_{8,f}$
can be read from the diagrams of \fig{fig_families_int}. We choose the 
so-called {irreducible scalar products}
$\rho_{9,f}$, $\rho_{{10},f}$, $\rho_{{11},f}$ as:
\begin{align}\begin{split}
	\rho_{9,\mzz}&= (l_1-p_5)^2\,,\quad
	\rho_{{10},\mzz}= (l_2+p_1)^2\,,\quad
	\rho_{{11},\mzz}= (l_2+p_1+p_2)^2\,, \\
	\rho_{9,\zmz}&= (l_2+p_5)^2\,,\quad
	\rho_{{10},\zmz}= (l_2+p_1+p_5)^2\,,\quad
	\rho_{{11},\zmz}= (l_1-p_4)^2\,, \\
	\rho_{9,\zzz}&= (l_1-p_1)^2\,,\quad
	\rho_{{10},\zzz}= (l_2+p_2)^2\,,\quad
	\rho_{{11},\zzz}= (l_2+p_2+p_3)^2\,.
\end{split}\end{align}
The set of powers $\vec\nu$ is a vector of integers, with the restriction
that $\nu_9,\nu_{10},\nu_{11}\leq0$ (i.e., irreducible scalar
products are not allowed to be in the denominator).

For a given topology $f$, each set of powers $\vec\nu$ defines an integral that is
a member of a linear space $Y^{[f]}$. In this paper, we compute a set of integrals
that form a basis of these spaces, that is the set of master integrals
associated with each topology. Any integral in $Y^{[f]}$ can be rewritten
as a linear combination of the master integrals using integration-by-parts (IBP)
identities \cite{Chetyrkin:1981qh}.
For each topology, the master integrals all have a subset of the propagators in
the top topology. The dimensions of the vector spaces can be determined in
several different ways, and we find
\begin{equation}\label{eq:dimTopo}
	\textrm{dim}\left(Y^{[\mzz]}\right)=74\,,\qquad
	\textrm{dim}\left(Y^{[\zmz]}\right)=75\,,\qquad
	\textrm{dim}\left(Y^{[\zzz]}\right)=86\,.
\end{equation}
Our choice of bases is given in the ancillary files \texttt{anc/f/pureBasis-f.m},
where \texttt{f} is to be replaced by the name of each topology
(a pictorial representation of the basis can be found in the files
\texttt{anc/f/graphs-f.m} which was generated using ref.~\cite{Georgoudis:2016wff}). 
We note
that the same integrals can appear in different topologies, and
there is a large overlap between these different spaces.

In writing the elements of these bases, we 
often make use of functions $\mu_{ij}$ that are obtained by contracting the 
components of the loop momenta beyond four dimensions, which
we denote $\ell^{(D-4)}_i$. Explicitly,
\begin{eqnarray}
\label{eq:mu}
\mu_{ij}= \ell^{(D-4)}_i \cdot \ell^{(D-4)}_j\,.
\end{eqnarray} 
These functions can also be written as polynomials in the $\rho_{i,f}$. The latter
representation is more convenient if one wants to rewrite integrals
defined with the help of these functions as members of the vector
spaces $Y^{[f]}$. It is given in the ancillary file \texttt{anc/determinants.m}.

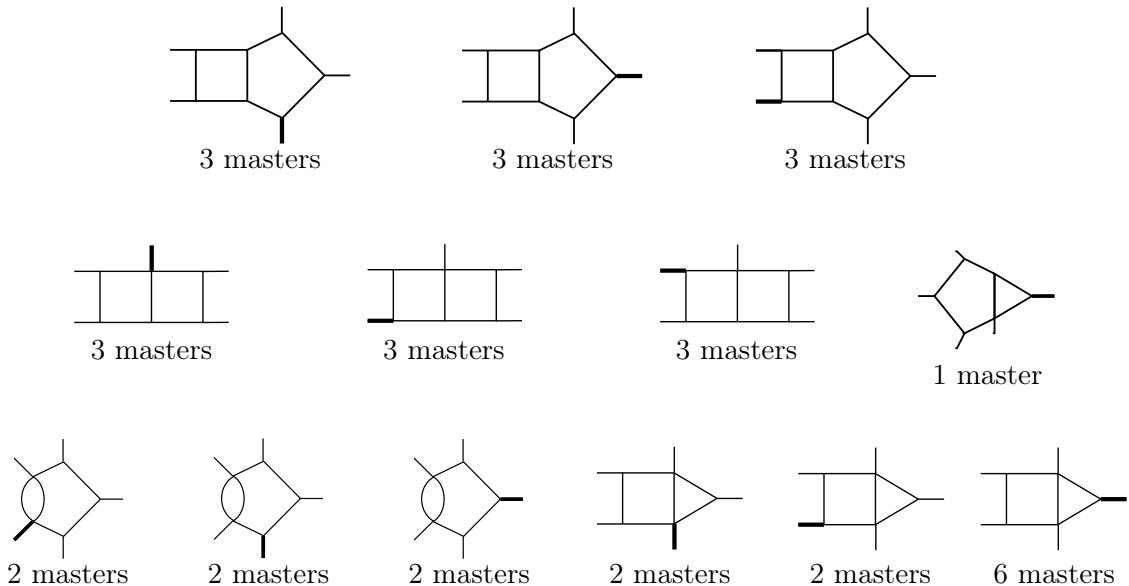
\begin{figure}[h] \begin{tikzpicture}[scale=1.1]
	  \node at (2.8,1){\includegraphics[scale=0.4]{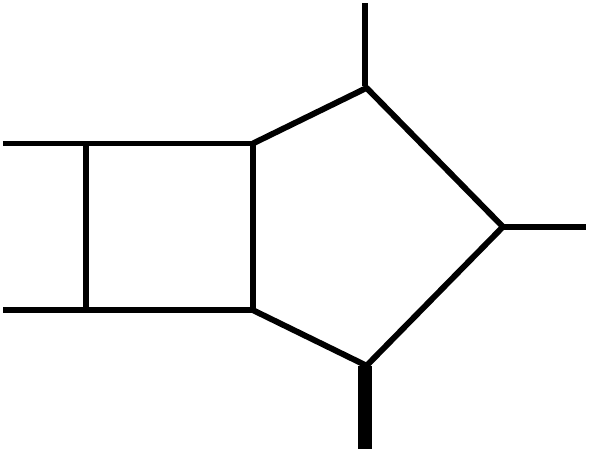}};
		\node at (2.8,0){3 masters}; \node at
		(6.3,1){\includegraphics[scale=0.4]{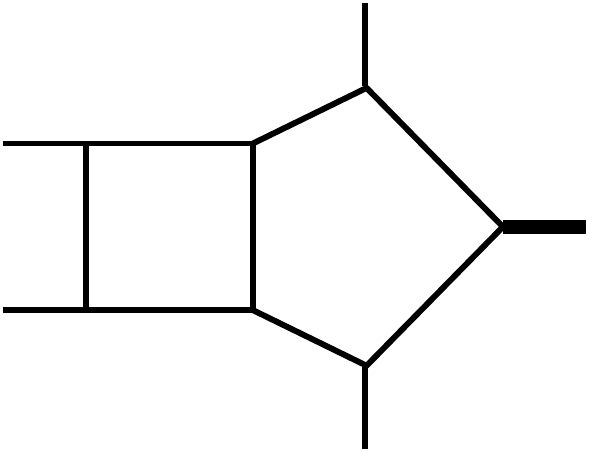}}; \node at
		(6.3,0){3 masters}; \node at
		(9.8,1){\includegraphics[scale=0.4]{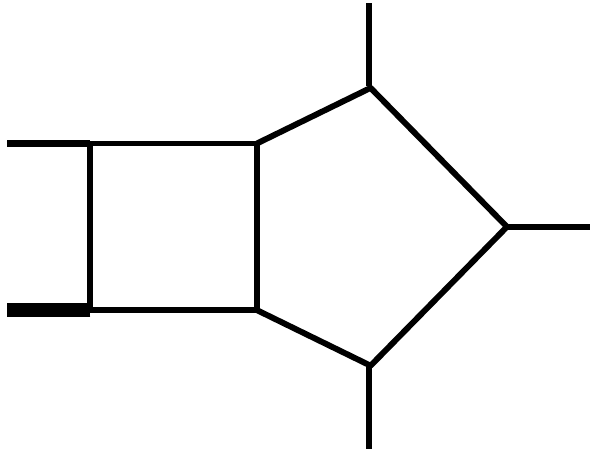}}; \node
		at (9.8,0){3 masters};
	\node at (1.5,-1.5){\includegraphics[scale=0.4]{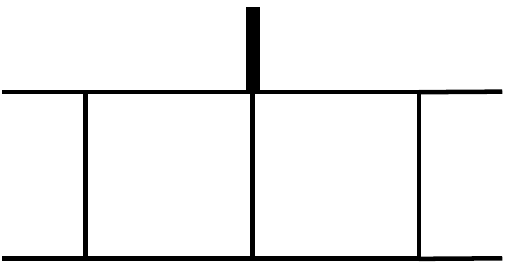}};
		\node at (1.5,-2.3){3 masters}; \node at
		(5,-1.5){\includegraphics[scale=0.4]{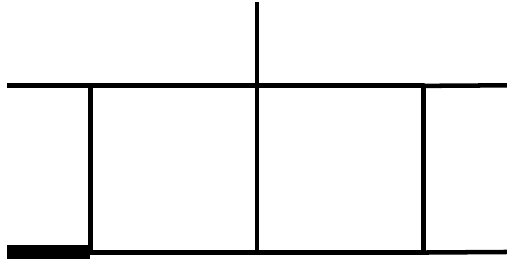}}; \node
		at (5,-2.3){3 masters}; \node at
		(8.5,-1.5){\includegraphics[scale=0.4]{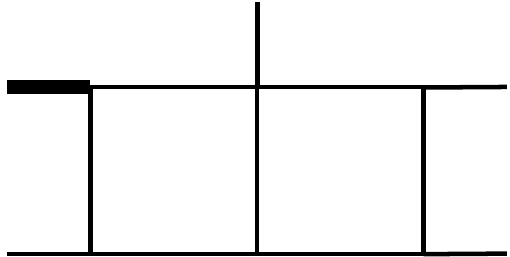}}; \node
		at (8.5,-2.3){3 masters};
    \node at
		(11.5,-1.7){\includegraphics[scale=0.35,
		angle=180]{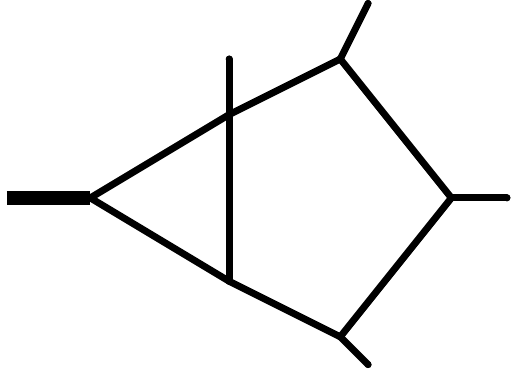}}; \node at (11.5,-2.6){1
		master};    
	\node at (0.5,-4.1){\includegraphics[scale=0.35]{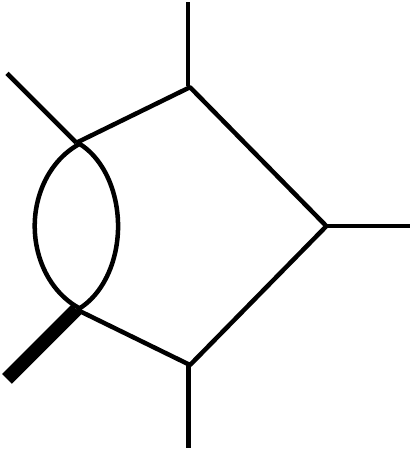}};
	\node at (0.5,-5.0){2 masters};
  \node at (2.9,-4.1){\includegraphics[scale=0.35]{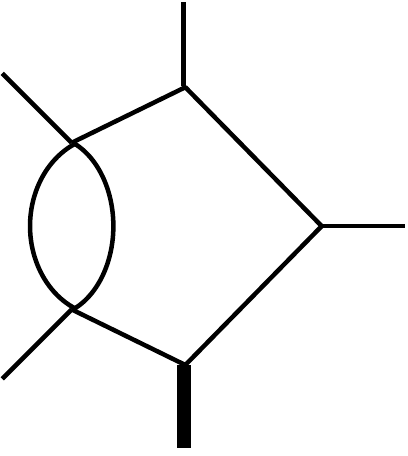}};
	\node at (2.9,-5.0){2 masters};
  \node at (5.3,-4.1){\includegraphics[scale=0.35]{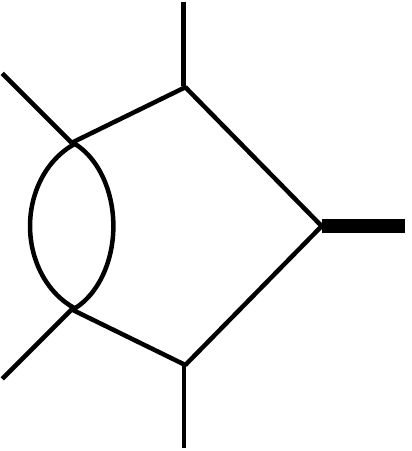}};
	\node at (5.3,-5.0){2 masters};
  \node at (7.7,-4.1){\includegraphics[scale=0.4]{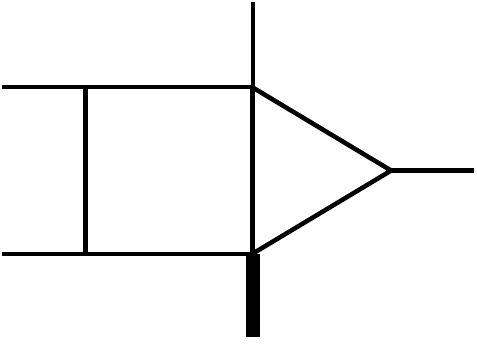}};
	\node at (7.7,-5.0){2 masters};
  \node at (10.1,-4.1){\includegraphics[scale=0.4]{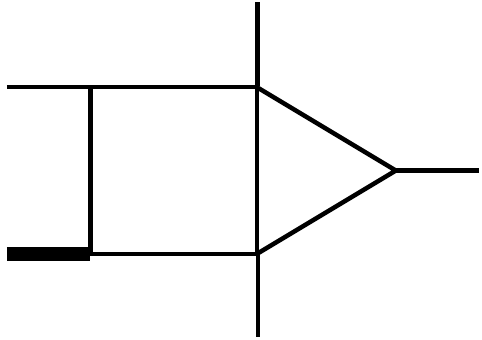}};
	\node at (10.1,-5.0){2 masters};
	\node at (12.3,-4.1){\includegraphics[scale=0.4]{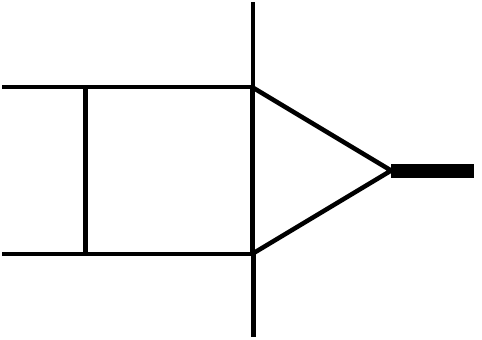}};
	\node at (12.3,-5.0){6 masters};
  \end{tikzpicture}
  \caption{Propagator structures of two-loop five-point master integrals.} 
  \label{fig_master_int}
\end{figure}

While in this paper we compute for the first time the full set of master
integrals required for two-loop five-point planar amplitudes, some
of those master integrals also appear in other amplitudes. In particular,
integrals associated with Feynman diagrams with four external legs or less
appear in four-point processes with two external masses
and have been previously computed \cite{Henn:2014lfa, Gehrmann:2018yef}. We will thus pay particular 
attention to integrals corresponding to diagrams with five external legs.
They are depicted in \fig{fig_master_int}, where we also give the number
of master integrals supported on their respective propagator structures.
Our choice of master integrals for these topologies, which we will discuss in the next
section, is given in 
appendix~\ref{pure_pentagons} as well as in the ancillary files, as was
already mentioned above. Finally, we also note that a full set of master integrals
for topology $I^{[\mzz]}[\vec \nu]$ has already been computed 
previously \cite{Papadopoulos:2015jft}.

%% file: differentialEquation.tex

\section{Semi-numerical construction of differential equation and pure basis}
\label{sec:DifferentialEquations}

One of the most effective approaches for computing Feynman integrals is 
solving the differential equation they satisfy
\cite{Kotikov:1990kg,
Kotikov:1991pm, Bern:1993kr, Remiddi:1997ny, Gehrmann:1999as,Henn:2013pwa}. 
Nevertheless, for complicated enough integrals such as the ones we are considering
in this paper, obtaining the differential equation can be challenging in itself.
In this section we discuss how we constructed the differential equations
required to compute the master integrals of topologies 
$I^{[\mzz]}$, $I^{[\zmz]}$ and $I^{[\zzz]}$ of \fig{fig_families_int}.
Our approach is based on numerical evaluations and builds on the one 
presented in \cite{Abreu:2018rcw}.

Before we discuss the details of our approach, we make
some general comments on differential equations for Feynman integrals to set up
our notation.
Let ${\bf I}$ be a vector of master integrals associated with a given 
topology. It is clear that the derivatives of the master integrals
are part of the same topology, and we can thus reduce them to the basis
of integrals in ${\bf I}$. We note that obtaining the IBP relations required for this
step is often the bottleneck in constructing the differential equations.
In full generality, the vector ${\bf I}$ fulfils a
differential equation
\begin{eqnarray}
\label{eqn:DE}
{\rm d} {\bf I} = \, {\bf M} \,{\bf I}\,,
\end{eqnarray}
where the connection {\bf M} is a matrix of differential forms depending 
on the dimensional regulator $\epsilon=(4-D)/2$. 

A refinement of the differential equation approach to the calculation of
Feynman integrals was proposed in \cite{Henn:2013pwa}: when Feynman integrals
evaluate to multiple polylogarithms, the solution
to the differential equation is greatly simplified if a basis of so-called
`pure' integrals is chosen. Indeed, in this basis the connection ${\bf M}$
in the differential equation \eqref{eqn:DE} takes a particularly simple form.
The regulator $\epsilon$ factorizes and the connection ${\bf M}$ is a total 
derivative  of a (singular) potential that depends logarithmically on kinematic 
expressions. That is,
\begin{eqnarray}
\label{epsFactorizedDE}
{\bf M} = \epsilon \sum_{\alpha}\,M_\alpha \,{\rm d}\log{(W_\alpha)} \,,
\end{eqnarray}
where the entries of the matrices $M_\alpha$ are rational numbers. The functions
$W_\alpha\in \mathcal{A}$ are known as the `letters' of the so-called `(symbol) alphabet'
$\mathcal{A}$ associated to the integrals. We will return to these notions in section
\ref{sec:AnalyticStructure}.

In \cite{Abreu:2018rcw}, a numerical method of constructing the differential
equation was introduced for the case of a pure basis of integrals with a known
symbol alphabet. The approach requires only the solution of small linear systems,
taking as input as many numerical evaluations of the differential equation as
there are letters. Since only numerical evaluations are required, all IBP
relations can be computed numerically, bypassing the often prohibitive
complexity of intermediate analytic expressions.
The non-trivial aspects of this method are the construction of
the pure basis and of the symbol alphabet, which are both as yet unknown for 
five-point one-mass two-loop processes. In this section, we will
address these points. After a brief description of
how we numerically sample differential equations, we discuss our construction of the 
bases of master integrals in section \ref{sec:pureBasis} and then, in
section \ref{sec:AlphabetConstruction}, we construct the symbol alphabet
from analytic cut differential equations. Crucially, using the data
from the numerical evaluations allows to target the simplest
cut differential equations that are required to obtain the full alphabet.

\subsection{The random direction differential equation}

Our approach to constructing differential equations for Feynman integrals
is based on a numerical evaluation of the differential equation, where
the Mandelstam variables $\vec s$ and the regulator $\epsilon$ take
numerical values. To this end, we introduce a `directional' partial derivative
\begin{equation}
     \vec{c} \cdot \frac{\partial}{\partial \vec{s}\,} {\bf I} =  
     {\bf C}(\epsilon, \vec{s}\,) \, {\bf I}.
     \label{eq:DirectionalDerivative}
\end{equation}
The vector $\vec{c}$ specifies a direction in the kinematic space,
and the operator $\vec{c} \cdot \frac{\partial}{\partial \vec{s}\,}$
replaces the total derivative $\mathrm{d}$. In contrast to the connection ${\bf
M}$, the matrix ${\bf C}(\epsilon, \vec{s}\,)$ is an algebraic function of
the kinematic data which can be easily evaluated numerically.
Nevertheless, for appropriate choices of $\vec{c}$, it is still sensitive to all
features of the connection. For this to be the case, the vector $\vec{c}$ must
not be chosen in any special direction. We thus fix it to a random
direction by making a random numerical choice for its components.

Let us now discuss the case of a pure basis. Then ${\bf C}(\epsilon, \vec{s}\,)$
takes a very specific form,
\begin{equation}
    {\bf C}(\epsilon, \vec{s}\,) = \epsilon\sum_{\alpha}\,M_\alpha \, 
    \vec{c} \cdot \frac{\partial}{\partial \vec{s}\,} \log{(W_\alpha)}.
    \label{eq:DirectionalDerivativeAnsatz}
\end{equation}
The factorization of the regulator can be checked by evaluating 
${\bf C}(\epsilon, \vec{s}\,)$ at different values of $\epsilon$. 
Furthermore, if the letters ${W_\alpha}$ are known, we can fix all of the 
$M_\alpha$ by evaluating ${\bf C}(\epsilon, \vec{s}\,)$
for as many values of $\vec s$ as there are symbol letters.

When numerically evaluating the matrix ${\bf C}(\epsilon, \vec{s})$, we find it
convenient to perform operations in a finite field of large cardinality. This
approach has many advantages, such as removing all issues related to loss of
precision in algebraic operations. However, one might worry that the presence of
square-roots might render the numerical
evaluation of the matrix ${\bf C}(\epsilon, \vec{s}\,)$ in a finite field
impossible. While this is true in general, it turns out not to be a problem in 
practical applications.
Indeed, it is a fact of number theory that for a finite field of cardinality
$p$, where $p>2$, $(p+1)/2$ of the elements of a finite field are perfect
squares, or more precisely `quadratic residues' \cite{hardy1979introduction},
and there exist completely
general algorithms which allow for the systematic computation of the square root 
in the finite field.\footnote{These algorithms, such as the
Tonelli-Shanks algorithm~\cite{tonelli1891bemerkung, shanks1973five},
are commonly available in modern computer-algebra systems.} 
This fact can be easily understood, as one can enumerate the quadratic residues.
Specifically, the set of distinct perfect squares is given by
\begin{equation}
    0^2, \, \, 1^2, \, \, 2^2, \, \, \ldots \, \, , \, \, \left( \frac{p-1}{2} \right)^2.
    \label{eq:QuadraticResidueList}
\end{equation}
These are quadratic residues by construction, so it remains to prove that they
are distinct and complete. Now, consider two different elements of the finite
field $s$ and $r$ which square to the same number. That is,
\begin{equation}
  s^2 = r^2 \quad (\mathrm{mod} \, \, p).
\end{equation}
It is clear that the equation is solved by $r = s$ and the partner solution $r =
p-s$. If we now consider \eqn{eq:QuadraticResidueList}, we see that no entries
are partners of one another, but this would not be true if we added any other
residue. Hence the set is distinct, complete and manifestly contains $(p+1)/2$
elements.
This observation implies that in practice we can take the square root
approximately 50\% of the time. As such, to avoid any issues related to the fact
that the square root of certain numbers cannot be represented in a given finite
field, we simply veto the randomly chosen points in which the
relevant square roots (see eqs.~\eqref{eq:tr5}, \eqref{eq:gram3} 
and \eqref{eq:gram3NP}) are not perfect squares.

\subsection{Constructing pure master integrals}
\label{sec:pureBasis}

Despite much progress in recent years \cite{Chicherin:2018old,Henn:2014qga,Lee:2014ioa,
Prausa:2017ltv,Gituliar:2017vzm,Meyer:2017joq,Meyer_2018,
Wasser:2018qvj, Abreu:2018rcw, Dlapa:2020cwj,Henn:2020lye}, 
which includes the development of automated approaches, the construction
of a pure basis for multi-scale dimensionally regulated Feynman integrals is not
yet a fully understood problem. Furthermore, for five-point integrals,
four-dimensional analyses of the integrands are often not enough,
see e.g.~\cite{Abreu:2019rpt,Chicherin:2018old}.
In this section we discuss how we constructed our bases of pure master integrals.
Our approach is based on constructing educated guesses for pure integrals,
and then checking numerically that $\epsilon$ factorizes in the matrix 
${\bf C}(\epsilon, \vec{s}\,)$. Strictly speaking, 
this does not imply that we have a pure basis,
which would also require that only \dlog{} forms appear in the 
connection. We will see in the next section that this is the case for the bases
we construct in this section.

For the integrals we are concerned with, pure bases are known for all 
integrals with four or fewer external legs 
\cite{Henn:2014lfa,Gehrmann:2015ora}.\footnote{For some low-point topologies we 
choose alternative basis integrals for technical convenience, giving 
preference to pure integrals without doubled propagators.}
The five-point sectors for which we need to find pure integrals
are depicted in \fig{fig_master_int}
and can be grouped into two sets: those where the number of master 
integrals is unchanged in the limit where the mass goes to zero, and those where
it is not. 
We find that for those with the same master count---all penta-boxes, 
double-boxes, penta-bubbles and all but one triangle-box---pure master 
insertions can be constructed in the same way as in 
\cite{Abreu:2018rcw,Abreu:2018aqd}.
For each such topology, we can separate the pure masters into those
that are even and those that are odd under the parity 
transformation of eq.~\eqref{eq:parityTrans}.
Educated guesses for pure even integrals can be motivated from a
four-dimensional analysis of the integrand as we now illustrate in an example.
Consider the integrand of the penta-box of \fig{fig:zzz}
for $D=4$,
\begin{equation}\label{eq:dlogpb}
  \mathcal{I}^{(4)}_{\textrm{pb}}=\mathcal{N}_{\textrm{pb}}
  \frac{d^4\ell_1}{\ell_1^2(\ell_1+p_2)^2(\ell_1+p_{23})^2(\ell_1+p_{234})^2}
  \frac{d^4\ell_2}{\ell_2^2(\ell_1-\ell_2)^2(\ell_2+p_{234})^2(\ell_2-p_{1})^2}\,,
\end{equation}
where $p_{ij}=p_i+p_j$, $p_{ijk}=p_i+p_j+p_k$ and we have introduced a loop-momentum dependent factor $\mathcal{N}_{\textrm{pb}}$ that we should fix
such that the integrand integrates to a pure function. It
has been conjectured that having this is equivalent to having an integrand that is a
\dlog{} form  with unit leading singularity (see e.g.~\cite{ArkaniHamed:2010gh}), 
and we will now see how this can be achieved for this example.
In this case we can proceed with a
loop-by-loop analysis. 

We first recall that in strictly 4 dimensions,
the integrands of one-loop box integrals can be written in terms of a \dlog{}
form,
\begin{equation}
  \mathcal{I}^{(4)}_{\textrm{box}}=\mathcal{R}_{\textrm{box}}
  d\log(g_1)\wedge d\log(g_2)\wedge d\log(g_3)\wedge d\log(g_4)
\end{equation}
where $\mathcal{R}_{\textrm{box}}$ and the $g_i$ depend
on the configuration of masses of the box integral. For our purposes, the form
of the $g_i$ is irrelevant, but that of the $\mathcal{R}_{\textrm{box}}$ is not.
As the coefficients of \dlog{} forms, they are known as `leading singularities'.
They depend only on external kinematics and for one-loop box integrals are given
by modified Cayley determinants associated with each integral (see
e.g.~\cite{Abreu:2017ptx}). Let us label the external legs of these box integrals
cyclically by $q_i$ and take $s = (q_1+q_2)^2$ and $t = (q_2+q_3)^2$. We will be particularly interested in the case of a
box with a single massive external leg (say $q_1^2\neq0$), which we call
$\textrm{b1m}$, and the case with three massive external legs (say $q_4^2=0$),
which we call $\textrm{b3m}$. For those cases
\begin{equation}
  \mathcal{R}_{\textrm{b1m}}=\frac{1}{s t}
  \qquad \mathrm{and} \qquad
  \mathcal{R}_{\textrm{b3m}}=\frac{1}{s t - q_1^2 q_3^2}\,.
\end{equation}
Note that the external legs $q_1$ and $q_3$ are diagonally opposed to each
other. It is clear that if we normalize the boxes by
$\mathcal{R}^{-1}_{\textrm{box}}$ we obtain a \dlog{} integrand with unit
leading singularity, and indeed these correspond to pure functions.

Let us now return to the integrand in eq.~\eqref{eq:dlogpb}, and focus
on the loop momentum $\ell_2$. Since it is nothing but the integrand
of a one-loop box with three external massive legs, it follows from the one-loop examples
we just discussed that it can be brought into the \dlog{} form
\begin{align}
  \begin{split}
  &\frac{d^4\ell_2}{\ell_2^2(\ell_1-\ell_2)^2(\ell_2+p_{234})^2(\ell_2-p_{1})^2}=\\
  &\qquad\qquad\frac{1}{s_{15}(p_1-\ell_1)^2-\offShellScale(l_1+p_{234})^2}
  d\log\omega_1\wedge d\log\omega_2\wedge d\log\omega_3 \wedge d\log\omega_4\,,
\end{split}
    \label{eq:PentaboxSubDlog}
\end{align}
where again the form of the $\omega_i$ is immaterial for our discussion.
Now, if we choose $\mathcal{N}_{\textrm{pb}}$ to be proportional to the inverse of
the leading singularity in \eqn{eq:PentaboxSubDlog}, then 
the integrand of \eqn{eq:dlogpb} factorizes into $\ell_2$ independent
propagators and a \dlog{} piece. Then, we can again
notice that the  $\ell_2$-independent propagators are those of a
one-loop box with a single massive leg, which we know has a \dlog{}
representation. Explicitly,
\begin{align}\label{eq:dlog2mh}
\begin{split}
  &\frac{d^4\ell_1}{\ell_1^2(\ell_1+p_2)^2(\ell_1+p_{23})^2(\ell_1+p_{234})^2}=\\
  &\qquad\qquad\frac{1}{s_{23}s_{34}}
  d\log\bar\omega_1\wedge d\log\bar\omega_2\wedge d\log\bar\omega_3 \wedge d\log\bar\omega_4\,.
\end{split}
\end{align}
In summary, by choosing
\begin{equation}\label{eq:pbeven}
  \mathcal{N}_{\textrm{pb}}=\epsilon^4\,s_{23}\,s_{34}\,
  \left(s_{15}(p_1-\ell_1)^2-\offShellScale(l_1+p_{234})^2\right)
\end{equation}
the integrand in \eqn{eq:dlogpb} can be written in a \dlog{} form with unit
leading singularity (the factor of $\epsilon^4$ is purely conventional). 
This four-dimensional argument is not sufficient to claim that the dimensionally
regulated integral is pure, 
but we view this analysis as a way to construct an educated
guess for a pure basis which we can later check.

Let us now discuss the remaining master integrals for topologies with the same
master count as in the massless case, but which are odd under parity. These
integrals involve numerator insertions that are written in terms of the
$\mu_{ij}$ defined in eq.~\eqref{eq:mu} and their parity properties follow from
the parity-odd factor $\trFive$ in the normalization.
We find that the na\"ive generalization of the odd integrands from the massless
to the massive case gives pure integrals: that is, we use the same
integrands, but the expressions now implicitly depend on $\offShellScale$.
While these integrands vanish in strictly four dimensions and 
 can thus not be captured by a four-dimensional analysis,\footnote{However, 
a $D$-dimensional analysis of the integrand can be performed, 
see e.g.~\cite{Abreu:2019rpt,Chicherin:2018old} for examples.}
they are natural objects to consider. A detailed analysis of why this is the
case is beyond the scope this paper, so we only suggest motivations.
First, they can be used to shift the dimension of the integral, 
and purity of Feynman integrals depends on which dimensions they are computed in.
Second, they are related to (generalized) Gram determinants, 
and thus vanish at special configurations of the loop-momenta. This naturally
means that they remove maximal codimension residues of the integrand, helping to
construct \dlog{} forms with unit leading singularity. A further benefit is
that, as these integrands vanish in exactly four dimensions, they lead to
integrals whose Laurent series around $\epsilon=0$ usually starts later than
their even counterparts. As an illustration, for the penta-box 
of \fig{fig:zzz} we can construct two odd pure integrals with the numerators
\begin{equation}\label{eq:pbodd}
  \mathcal{N}^{(1)}_{\textrm{pb,odd}}=\epsilon^4\,s_{15}\,\trFive\,\mu_{12}\,,\quad
  \textrm{and}\,\qquad
  \mathcal{N}^{(2)}_{\textrm{pb,odd}}=\epsilon^4\,\frac{1-2\epsilon}{1+2\epsilon}
  \,\trFive\,(\mu_{11}\mu_{22}-\mu_{12}^2)\,,
\end{equation}
and while the integral obtained from eq.~\eqref{eq:pbeven} starts at order
$\epsilon^{0}$, the ones obtained from eq.~\eqref{eq:pbodd} start at order
$\epsilon^5$. This has clear advantages when using these integrals for evaluating
two-loop amplitudes.

\begin{figure}
  \centering
  \begin{subfigure}{0.45\textwidth}\centering
    \includegraphics[height=3cm]{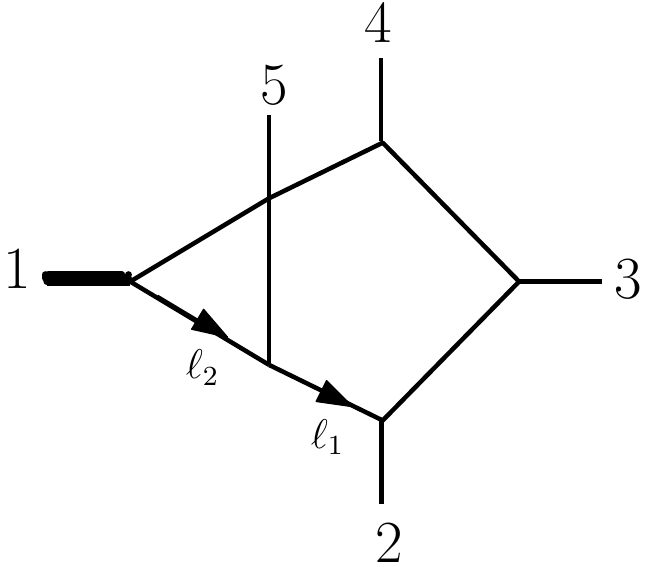}
    \caption{}
    \label{fig:pentaTri}
  \end{subfigure}
  \begin{subfigure}{0.45\textwidth}\centering
    \includegraphics[height=3cm]{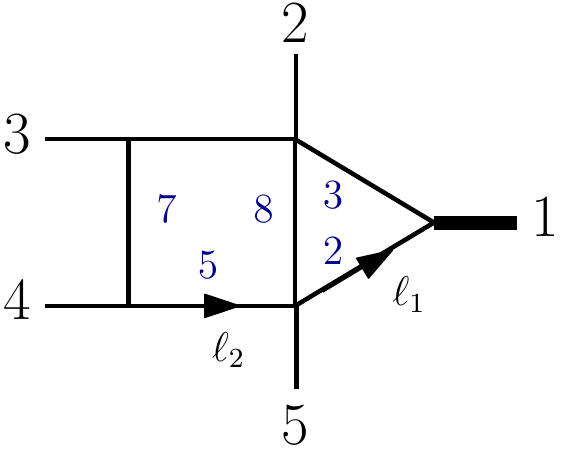}
    \caption{}
    \label{fig:badTB}
  \end{subfigure}
  \caption{Topologies with more masters than in the limit $p_1^2\to0$.}
  \label{fig:badTopo}
\end{figure}

Let us now discuss the two topologies that cannot be understood
by simple generalization of the pure basis of the massless five-point 
two-loop integrals, see \fig{fig:badTopo}.
The first is the penta-triangle topology of \fig{fig:pentaTri}.
Despite the fact that it does not appear in the massless case, this is
in fact a simple case to solve.
We require a single pure integral, and can use a logic similar to the one discussed above
for odd integrals. We recall that, with an appropriate normalization,
a triangle in $4-2\epsilon$ dimensions and a pentagon in $6-2\epsilon$
are pure. As we already hinted at above, the
latter can be represented by a $\mu^2$ insertion
on the $4-2\epsilon$ pentagon, where $\mu$ denotes the $(D-4)$-dimensional
component of the pentagon's loop momentum.
A natural educated guess for a pure integrand is then to take as a numerator
\begin{equation}
   \mathcal{N}_{\textrm{pt}}= \epsilon^4 \trFive{} \mu_{11}.
\end{equation}
The validity of this guess can be verified in the usual way.

The final and most challenging case we need to address 
is the triangle-box with a massive 
leg on the triangle side of \fig{fig:badTB}. 
The main difficulty lies in the fact that we must construct six pure 
integrals with this set of propagators.
As for all other triangle-box topologies,
two pure insertions can be obtained as simple generalizations of the
massless case,
\begin{align}\begin{split}
\mathcal{N}^{(1)}_{\textrm{tb}}=\epsilon^4 s_{34}\sqrt{\Delta_3}\,,\qquad\qquad
\mathcal{N}^{(2)}_{\textrm{tb}}=\epsilon^3\mu_{22}\text{tr}_5\frac{1}{\rho_8}\,. 
\end{split}\end{align}
Two further integrands can be constructed using the \dlog{}
logic that was used to build \eqn{eq:pbeven}. Let us consider
the triangle sub-loop in \fig{fig:badTB}, and the one-loop
IBP relation
\begin{align}\label{eq:oneLIBP}
  \begin{split}
  (\offShellScale{} s_{45}) \, \, \eqnDiag{\includegraphics[scale=0.4]{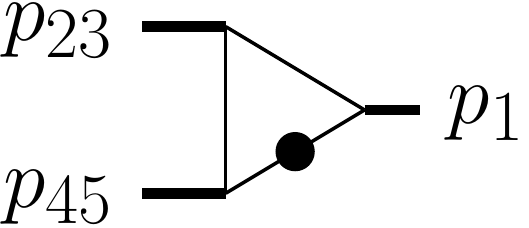}}
  +& \frac{D-4}{2}(\offShellScale{}-s_{23}+s_{45}) \,\, \eqnDiag{\includegraphics[scale=0.4]{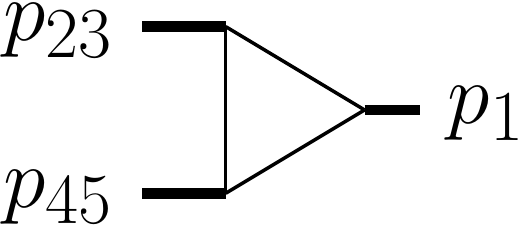}} \\
  = (D-&3)  \left(\eqnDiag{\includegraphics[scale=0.4]{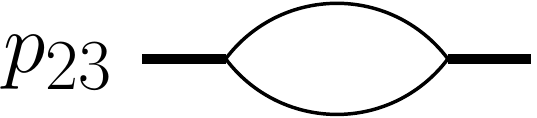}} -
  \eqnDiag{\includegraphics[scale=0.4]{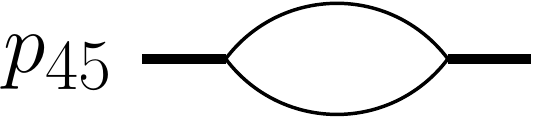}} - 
  \eqnDiag{\includegraphics[scale=0.4]{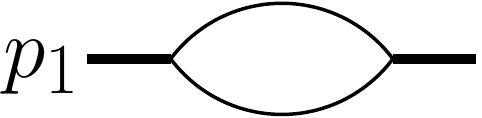}}\right),
  \end{split}
\end{align}
together with the same relation obtained by the exchange $p_{23} \leftrightarrow p_{45}$.
It can be easily shown that a one-loop bubble in $D=4-2\epsilon$ dimensions multiplied by
$(D-3)$ is related by a simple numerical factor to a $2-2\epsilon$ bubble normalized by its scale. 
The latter is known to be pure, and thus to have an integrand which is a \dlog{} form with unit leading
singularity.
After replacing the box sub-loop by its \dlog{} form, equivalent to that
in equation~\eqref{eq:dlog2mh},
we can then formally replace the triangle sub-loop in \fig{fig:badTB}
by the left-hand side of eq.~\eqref{eq:oneLIBP} (or its equivalent under
$p_{23} \leftrightarrow p_{45}$) to obtain two candidate numerators that
correspond to pure integrals:
\begin{align}\begin{split}
&\mathcal{N}^{(3)}_{\textrm{tb}}= \epsilon^4s_{34}\left( 
(\offShellScale{} - s_{23} + s_{45}) 
- \frac{1}{\epsilon}\frac{\offShellScale{}  s_{45}}{\rho_2}\right) \,,\\
&\mathcal{N}^{(4)}_{\textrm{tb}}= \epsilon^4s_{34}\left(
(\offShellScale{} - s_{45} + s_{23}) 
- \frac{1}{\epsilon}\frac{\offShellScale{} s_{23}}{\rho_3}\right)\,.
\end{split}\end{align}
There are two further pure integrals, for which we did not built educated
guesses. Instead, we rely on the fact that we can easily test if $\epsilon$
factorizes in the differential equation by using simple numerical evaluations.
Combined with the fact that the system of differential equations can be further
simplified by imposing that certain propagators are set to zero
(see e.g.~\cite{Ita:2015tya,Larsen:2015ped,Zeng:2017ipr,Abreu:2018rcw}),
we obtain a very efficient method of constructing the remaining two insertions
by requiring that $\epsilon$ factorizes in the differential equation.
With this approach we constructed the following other two numerators:
\begin{align}\begin{split}
&\mathcal{N}^{(5)}_{\textrm{tb}}=\epsilon^4\left(s_{34} (\offShellScale + s_{23} - s_{45}) 
+ \frac{1}{\epsilon}\offShellScale s_{34}\frac{(\ell_1-p_4)^2}{\rho_2}+
\frac{1}{\epsilon}s_{15} s_{34} \frac{\rho_7}{\rho_5}\right)\,,\\
&\mathcal{N}^{(6)}_{\textrm{tb}}=\epsilon^3 \mu_{12} \text{tr}_5\frac{1}{\rho_8}\,.
\end{split}\end{align}

As noted throughout this section, at this stage we cannot yet determine if the
integral we have chosen for the five-point topologies of
\fig{fig_master_int} are pure. We can only check that $\epsilon$ factorizes
in the matrix ${\bf C}(\epsilon, \vec{s}\,)$ of
eq.~\eqref{eq:DirectionalDerivative}, which is a necessary but not sufficient
condition for the master integrals to be pure. We have collected all
the integrands for the diagrams in \fig{fig_master_int} in 
appendix~\ref{pure_pentagons}.

The bases we have constructed through this procedure for the topologies in
figs.~\ref{fig:mzz}, \ref{fig:zmz} and \ref{fig:zzz} can be found in
\texttt{anc/f/pureBasis-f.m}, with \texttt{f}=\texttt{mzz}, \texttt{zmz}, 
\texttt{zzz}. We also include the basis for the one-loop integrals in 
\texttt{anc/1loop/pureBasis-1loop.m}.

\subsection{Analytic form of differential equations}
\label{sec:AlphabetConstruction}

We now discuss how we obtain the analytic form of the differential equations
through an ansatz procedure. Specifically, we work with an ansatz consistent
with the assumption that our bases of master integrals for the topologies
in \fig{fig_families_int} are pure. The structure of the ansatz is that the
matrices ${\bf C}(\epsilon, \vec{s}\,)$ take the form given in
eq.~\eqref{eq:DirectionalDerivativeAnsatz}. In our case, we take the ansatz as a
working assumption. By verifying the ansatz with an overconstraining set of
numerical data, we verify that our bases are indeed pure. Working with this
assumed ansatz, in order to completely determine the differential equations we
need to determine the letters $W_{\alpha}$ and the associated matrices
$M_\alpha$. At this stage, even the number of letters, i.e., the dimension of
the symbol alphabet, is unknown.

Let us consider this dimensionality question for any differential
equation \eqref{eqn:DE} where the basis~$\bf{I}$ is pure and of dimension $n$.
Given the ability to numerically evaluate the directional derivative matrix
${\bf C}(\epsilon, \vec{s}\,)$ of eq.~\eqref{eq:DirectionalDerivativeAnsatz},
one can easily determine the dimension of the alphabet relevant for the basis.
First note that, as the directional derivative matrix ${\bf C}(\epsilon, \vec{s}\, )$
is considered in a random direction, its entries span a vector space which
is equivalent to the one spanned by the alphabet. 
Therefore it is sufficient to count the number of
entries of ${\bf C}(\epsilon, \vec{s}\,)$ which are linearly independent.
Numerically this can be easily achieved by sampling the directional derivative
matrix. We begin by flattening the $n \times n$ matrix ${\bf
C}(\epsilon, \vec{s}\,)$ into a single vector of length $n^2$. In a finite
field of large cardinality, we now generate
random phase space points $\vec{s}\,^{(k)}$, $k = 1,\ldots, m$, and fix
$\epsilon=\epsilon_0$. Now, we evaluate
our vector on these points to construct a new, finite-field valued, $m \times
n^2$ matrix $\mathcal{C}_l(\epsilon_0, \vec{s}\,^{(k)})$ whose rows
are the flattened matrices ${\bf C}(\epsilon_0, \vec{s}\,^{(k)})$. The indices
of this matrix are $k$ and $l$.
Given this construction, the rank of the matrix $\mathcal{C}_l(\epsilon_0,
\vec{s}\,^{(k)})$ is bounded from above by both the number of rows $m$ and the
dimension of the alphabet itself, therefore
\begin{equation}
  {\rm rank}\,\,\mathcal{C}_l(\epsilon_0, \vec{s}\,^{(k)}) = \mathrm{min}[ m, \dim(\mathcal{A}) ].
  \label{eq:DimensionCounting}
\end{equation}
This follows as linear relations between the columns of the evaluation matrix
are inherited from linear relations between the entries of the directional
derivative matrix. Equation~\eqref{eq:DimensionCounting} implies that if we sequentially raise $m$ and
find that the rank stops increasing, then we will have identified the dimension of the alphabet.

To apply this approach to the integrals we are interested in, we note that
we want to construct four different differential equations, one for each genuine two-loop 
topology of \fig{fig_families_int} and one for the one-loop five-point one-mass topology.
By direct application of the above steps we find that
\begin{align}\begin{split}
  \dim\left(\mathcal{A}^{[\mzz]}\right)=38\,,\quad
  &\dim\left(\mathcal{A}^{[\zmz]}\right)=48\,,\quad
  \dim\left(\mathcal{A}^{[\zzz]}\right)=49\,,\\
  &\dim\left(\mathcal{A}^{[\rm{1-loop}]}\right)=30\,.
\end{split}\end{align}
Another perhaps more interesting number is the dimension of the union of the
alphabets, corresponding to all master integrals. We thus flatten the matrices
${\bf C}^{[\mzz]}(\epsilon_0, \vec{s}\,^{(k)})$, ${\bf C}^{[\zmz]}(\epsilon_0, \vec{s}\,^{(k)})$,
${\bf C}^{[\zzz]}(\epsilon_0, \vec{s}\,^{(k)})$ and 
${\bf C}^{[\rm{1-loop}]}(\epsilon_0, \vec{s}\,^{(k)})$ into four different
vectors, and then join them together to form one larger vector.
We then determine the dimension of the full alphabet by computing the rank
of the matrix constructed with these vectors evaluated at successive random values
$\vec{s}\,^{(k)}$. We find that the dimension of the union of the four alphabets is 55.

We are now left with the task of determining the set of letters that we need to 
express the four differential equations, i.e.~a basis of the 55 dimensional 
alphabet. Once again, we can use 
the numerical data we have collected. We consider the matrices
${\mathcal C}^{[\mzz]}_l(\epsilon_0, \vec{s}\,^{(k)})$, 
${\mathcal C}^{[\zmz]}_l(\epsilon_0, \vec{s}\,^{(k)})$,
${\mathcal C}^{[\zzz]}_l(\epsilon_0, \vec{s}\,^{(k)})$ and
${\mathcal C}^{[\rm{1-loop}]}_l(\epsilon_0, \vec{s}\,^{(k)})$
and row reduce them.
For each non-zero row in the row-reduced echelon form, the index of the leading
non-zero column labels an independent basis element.
By appropriately ordering the columns, one can prioritize different
elements in this basis search. We first choose to determine as many letters
as possible from the one-loop differential equation, since these are trivial
to obtain in analytic form.
This leaves 25 letters to be
determined. To determine those, we first note that each column of the matrices
$\mathcal{C}^{[f]}$ corresponds to the coefficient of an integral $i$ in the differential
equation of another integral $j$. We choose to prioritize the columns for which
$(i,j)$ share as many propagators as possible, essentially organizing
the matrix into increasingly `off-shell' blocks. This organization leads us
to an important observation: a basis of symbol
letters can be found in the maximal and next-to-maximal cut differential
equations for sectors with 6 or 7 propagators (this statement is true for the full
55 letters, not just for the 25 that are new at two loops). 
Whilst this observation is
theoretically interesting, it is also of immediate practical consequence. Cut
differential equations are technically much easier to construct analytically, 
especially when using IBP-reduction methods tailored for the presence of
unitarity cuts \cite{Abreu:2018rcw, Zeng:2017ipr, Bosma:2017hrk}. Alternatively,
public Laporta-based IBP programs such as KIRA \cite{Maierhoefer:2017hyi}
can be used to compute the relevant cut IBP relations.
Guided by the numerical differential equations, we have thus reduced the
problem of determining the symbol alphabet to the calculation of a few
trivial one-loop or \emph{cut} two-loop differential equations. We also note that
by checking that these trivial differential equations are pure we
prove that the bases we have chosen in the previous section is indeed pure.

Armed with the basis of the symbol alphabet extracted from the cut differential
equations, the only missing ingredients to obtain the analytic form of the
directional differential equation matrix ${\bf C}(\epsilon, \vec{s}\,)$ are the
matrices $M_\alpha$ in eq. \eqref{eq:DirectionalDerivativeAnsatz}. These
can be constructed by reusing the set of numerical evaluations of the
directional differential equation matrix. First, we compute the
$\dim(\mathcal{A}) \times \dim(\mathcal{A})$ matrix of evaluations of the
``random directional'' \dlog{}s
\begin{equation}
\mathcal{W}_{\alpha k} = \vec{c} \cdot \left[  \frac{\partial}{\partial \vec{s}} \log( W_\alpha )\right]|_{\vec{s} = \vec{s}^{(k)}}.
\end{equation}
This matrix is invertible as the random directional \dlog{}s are independent by construction.
This allows us to explicitly compute the coefficient matrices through
\begin{equation}
  M_{\alpha} = \sum_k \frac{1}{\epsilon_0} \mathcal{W}^{-1}_{\alpha, k} {\bf C}(\epsilon_0, \vec{s}^{(k)}).
\end{equation}
As expected from previous experience \cite{Abreu:2018rcw}, the rational numbers
involved are easily reconstructed from their image in a single finite field of
cardinality $\mathcal{O}(2^{31})$.

The differential equations we have constructed in this way can be found
in the ancillary files
\texttt{anc/f/diffEq-f.m}, for \texttt{f}=\texttt{mzz}, \texttt{zmz}, \texttt{zzz}
or  \texttt{1loop}. They are written in terms of the alphabet that can
be found in \texttt{anc/alphabet.m}, whose construction will be described in
the next section.

%% file: analyticStructure.tex

\section{Analytic structure of planar five-point one-mass scattering at two loops}
\label{sec:AnalyticStructure}

The differential equations satisfied by the master integrals that we have
constructed in the previous section contain a lot of information about the
analytic structure of not just the integrals, but also the scattering
amplitudes they appear in. In this section, we explore this structure with the
help of the `symbol' \cite{Goncharov:2010jf} which can be constructed with
minimal effort from a canonical differential equation. Let us review some basic
concepts to set up our notation. Consider the $\epsilon$ expansion of the master
integrals. At each order in $\epsilon$, the master integrals are computed by
integrating the previous order with respect to a kernel that is fixed by the
connection matrix ${\bf M}$ in eq.~\eqref{epsFactorizedDE}. In particular, the
kernel is given by linear combinations of \dlog{} forms. More precisely, we have
\begin{eqnarray}
\label{eqn:solExpansion}
		{\bf I} = \sum_{i=0} {\bf I}^{(i)} \epsilon^i \,, \qquad 
		{\bf I}^{(i+1)} = \int \sum_{\alpha} M_\alpha  \mathrm{d}\log(W_\alpha) \, {\bf I}^{(i)} \,,
\end{eqnarray}
where we have used the fact that we normalize our master integrals to have no poles in $\epsilon$.
As ${\bf I}^{(0)}$ lives in the kernel of the derivative, it has to be a constant vector. The vector ${\bf I}^{(n)}$ is a function built from
$n$ iterated integral over a series of \dlog{} kernels. That is,
\begin{equation}
{\bf I}^{(n)} = \sum_{\alpha_1, \ldots, \alpha_n} {\bf c}_{\alpha_1, \ldots, \alpha_n} 
\int \mathrm{d}\log W_{\alpha_1} \cdots \mathrm{d} \log W_{\alpha_n}.
\end{equation}
The number of iterated integrations is called the \emph{weight} of the function.
To explicitly obtain the functions ${\bf I}^{(n)}$ we must specify the integration
contour. However, a great deal
of analytic information can be understood from the integrand alone. To this end,
it is common to introduce the notion of a symbol, which captures the integrand
information. The symbol is simply a vector in the tensor product space of the letters
\begin{equation}
S[{\bf I}^{(n)}] = \sum_{\alpha_1, \ldots, \alpha_n} 
{\bf c}_{\alpha_1, \ldots, \alpha_n} \left[ W_{\alpha_1}, \cdots, W_{\alpha_n}\right],
\end{equation}
where the length of the tensor equals the weight of the function. Note that the
fact that the differential equation is in canonical form naturally ties the
order in the Laurent expansion with the weight of the functions, see
eq.~\eqref{eqn:solExpansion}. It is clear that the symbol is controlled by the
differential equation. In particular, the tensors ${\bf c}$ are computed from
the products of the matrices $M_{\alpha}$ in eq.~\eqref{epsFactorizedDE} and
control which tensor products appear in the symbol of the integrals. In the case
where ${\bf I}$ is a vector of Feynman integrals, there is a constraint on the
symbol known as the first-entry condition \cite{Gaiotto:2011dt}. In our case, it
states that ${\bf c}_{\alpha_1, \ldots, \alpha_n}=0$ if $W_{\alpha_1}\notin\vec
s$, where we already use the fact that the Mandelstam variables $\vec s$ are
part of our alphabet. As we will see in section \ref{sec:strucSymb}, this proves
to be a very strong constraint, which almost fully constrains the initial
condition at weight 0.

In this section, we will first discuss how to construct a simple set of symbol letters
in order to simplify the form of the differential equations (and thus of the
symbol). We will then discuss some properties of the symbols of the master
integrals.

\subsection{Choosing letters}
\label{sec:ChoosingLetters}

It is clear that the choice of symbol letters is not unique: their logarithms
generate a vector space, and any basis of that space is equivalent.
In section \ref{sec:AlphabetConstruction}, we discussed how to extract 
a complete set of letters from cut differential equations. However,
what we na\"ively obtain from the differential equations might not
be the most convenient choice of alphabet. We now discuss some steps we have
taken to simplify the alphabet and attempt to choose letters that make manifest 
some analytic properties of the integrals.

Let us start from a pure differential equation whose connection takes the form
of eq.~\eqref{epsFactorizedDE}.
A first step is to take an independent set of irreducible factors of the
\dlog{} forms in the differential equation as letters, 
but this can be practically difficult. The issue finds its origin in the square
roots in the problem---in our case, the Gram determinants of
eqs.~\eqref{eq:gram5}, \eqref{eq:gram3} and \eqref{eq:gram3NP}. As observed in
the literature~\cite{Heller:2019gkq, Bourjaily:2019igt}, expressions involving
square roots cannot be factorized uniquely, meaning that elucidating multiplicative
relations between candidate letters is analytically challenging. Furthermore,
for letters involving these square roots, it is not a priori clear what the most
compact and/or physically relevant basis is.

To combat these difficulties, we employ a numerical sampling approach, which
uncovers multiplicative relations between letters even in the presence
of square roots. 
Consider a set of functions $\Omega = \{\Omega_i\}_{i=1,\ldots, N}$ as new
candidate letters. We want to know if they live in the alphabet, and if there are
any multiplicative dependencies between them. 
To answer these questions, we construct the list
\begin{equation}
    L(\vec{s}\,) = \{\log(|\Omega_1|), \ldots, \log(|\Omega_N|), \log(|W_1|), \ldots, \log(|W_n|)\}.
\end{equation}
That is, we take the list $\Omega$, append the alphabet and take the logarithm
of the absolute value of each element. All multiplicative relations between the 
elements of $\Omega$ and the alphabet now become linear relations between the 
elements of $L(\vec{s})$. The absolute value
plays the role of throwing away any sign information which is not relevant for
symbol letters. Similar to the algorithmic construction of the alphabet, all
linear relations between the elements of $L(\vec{s})$ can be extracted by
constructing the square matrix $L_i(\vec{s}^{(k)})$ from $n+N$ randomly chosen
values of $\vec{s}$ (the indices $i$ and $k$ denote the entries of the matrix). 
As the matrix is not large, all practical questions of
numerical stability are avoided using high precision floating point arithmetic.
Having constructed $L_i(\vec{s}^{(k)})$, we can now use similar techniques to
section \ref{sec:AlphabetConstruction}. Firstly, we can easily check if all
elements of $\Omega$ indeed live in the alphabet as this implies that
$\mathrm{rank}(L_i(\vec{s}^{(k)}))=n$. Secondly, by ordering the elements of
$L(\vec{s})$ to put preferred elements first, a new basis of the alphabet is
algorithmically picked out by reading the linearly independent columns from the
row reduced form of $L_i(\vec{s}^{(k)})$.
This approach allows us, with no explicit rationalization of the kinematics, to
easily construct alternative bases of the alphabet, prioritizing
the letters with the properties we find most important.

With this technique in hand, we can easily find a set of symbol letters from
analytically factorizing those found in the differential equation.
We favour letters with lower mass dimension. It is nevertheless clear that the
non-uniqueness of the factorization still remains a barrier to simplicity.
To proceed, we rely on an observation made in reference \cite{Heller:2019gkq},
where it was pointed out that one can construct
candidate symbol letters involving a single square root from knowledge of the
polynomial part of the alphabet and the square root alone. Employing this method
we find that it generates a large number of letters which live in the alphabet,
but crucially many are new representations with lower mass dimension. Following
these steps, we obtain a sufficient set of letters with one square root whose
mass dimension is no greater than four.

The final step in our organization procedure is to choose the alphabet to have
manifest behavior with respect to changing the signs of the square roots. 
The reason for this choice is that Feynman integrals are invariant under
this change, but this invariance might be broken by the 
normalizations introduced when constructing a pure basis (see for instance
the distinction between even and odd integrals in section \ref{sec:pureBasis}).
It is clear that the operations of flipping each sign compose to form a group, which is
known in the mathematics literature as a `Galois group'.\footnote{Mathematically, the
  Galois group arises when considering field
  extensions~\cite{dummit2004abstract}. Here we are implicitly working in the
  field of rational functions of Mandelstam variables extended by the addition
  of the square roots in eqs.~\eqref{eq:tr5}, \eqref{eq:gram3} and \eqref{eq:gram3NP},
  denoted by $\mathbb{Q}(\vec{s}, \sqrt{\Delta_3},
  \sqrt{\Delta_3^{\ncg}}, \sqrt{\Delta_5})$. This field has a privileged set of
  field automorphisms---those that reduce to the identity on the underlying
  field $\mathbb{Q}(\vec{s})$. These automorphisms form a group under
  composition, the Galois group. Beyond square roots, these concepts
  generalize to more complicated radicals, such as those found
  in~\cite{Bourjaily:2018aeq}.
} By choosing each letter to map to themselves, or their reciprocal, under each
element of the group, we ensure that the \dlog{}s form an irreducible representation 
of the group, and that, consequently, so will the symbols.

\subsection{The symbol alphabet}
\label{sec:AlphabetResults}

With the procedure described in the previous section we are able to 
construct an alphabet whose letters have low mass dimension,
and with manifest properties under the Galois group associated to the 
square roots in the problem. 
As noted in section \ref{sec:masterInt}, the set of master integrals
we compute is not sufficient for two-loop planar five-point one-mass
amplitudes, as we also require the integrals obtained by the
exchange $(2\leftrightarrow 5,3\leftrightarrow 4)$ of the external legs.
To obtain the symbol relevant for the amplitude, we complete
the letters by including their image under this transformation. This
increases the size of the alphabet from 55 to 58. In this section
we present the alphabet of the amplitude.

We split the letters into two main sets: those that do and those that do not
appear in the master integral symbols up to weight four (after imposing the
first-entry condition discussed at the start of this section), which is the
weight of the contributions that are relevant for two-loop amplitudes. We will
first list the 49 `relevant' letters, which we organize according to their
simplicity and transformation properties under the Galois group.
The remaining 9 letters are `irrelevant', in that they do not turn up in the
symbols of the integrals up to weight 4.
Each set we present is
closed under the $(2\leftrightarrow 5,3\leftrightarrow 4)$ exchange.
In the following, we often choose representations of the letters which help to
manifest the soft limits in which they vanish. In the ancillary files
\texttt{anc/alphabet.m} we present the alphabet written explicitly in terms of
independent Mandelstam variables.

We start with letters that are invariant under the Galois group.
The first set consists of the letters corresponding to the Mandelstam
variables that are allowed in the first entry of the symbol,
\begin{align}
  \begin{split}
  \{W_1, \ldots, W_6\} &= \{ \offShellScale{}, \, s_{34}, \, s_{12}, 
  \, s_{15}, \, s_{23}, \, s_{45}\}\,.
  \end{split}
\end{align}
The next two sets are again invariant under the Galois group
and of mass dimension two.
They are either two-particle invariants or simple differences
of Mandelstam variables
\begin{align}
  \begin{split}
  \{W_7, \ldots, W_{13}\} &= \{ 2\,p_2 \cdot p_5, \, 2\, p_1 \cdot p_2, \, 2\, p_1  \cdot p_5,
   \, 2\, p_1 \cdot p_3, \, 2\, p_1 \cdot p_4,\\ 
  &\qquad2 \,p_2 \cdot p_4, \, 2 \,p_3 \cdot p_5\}\,,\\
  \{W_{14}, \ldots, W_{21}\} &= \{ 2\,p_2\cdot(p_3+p_4),\, 2\,p_5\cdot(p_3+p_4),\, 2\,p_2\cdot(p_4+p_5),
  \, 2\,p_5\cdot(p_2+p_3),\\
  &\qquad 2\,p_3\cdot(p_1+p_2),\,2\,p_4\cdot(p_1+p_5),\,2\,p_3\cdot(p_1+p_5),\,2\,p_4\cdot(p_1+p_2) \}.
  \end{split}
\end{align}
We then list invariant letters that are slightly more complicated polynomials of the 
Mandelstam variables $\vec s$, now of mass dimension four.
We separate a set that depends on four-point kinematics,
\begin{align}\label{eq:invDim4}
  \begin{split}
  \{W_{22}, \ldots, W_{30}\} =& \{ \trp(1\,2\,1\,5), \, \trp(1\,2\,1\,3),  
  \,\trp(1\,5\,1\,4), \,\trp(1\,2\,1\,4), \,\trp(1\,5\,1\,3),  \\
  &\trp(1\,2\,1\,[4+5]), \, \trp(1\,5\,1\,[2+3]), \\
  &\trp( [2 + 3] \, 4 \, [2 + 3] \, 1),\,
  \trp( [4 + 5] \, 3 \, [4 + 5] \, 1)\}\,,
  \end{split}
\end{align}
from a set that depends on five-point kinematics
\begin{align}
  \begin{split}
  \{W_{31}, W_{32}\} &= \{ 
  \trp( 1 \, 2 \, 3 \, 4) - \trp( 1 \, 2 \, 4 \, 5), \,
  \trp( 1 \, 5 \, 4 \, 3) - \trp( 1 \, 5 \, 3 \, 2)\}\,.
  \end{split}
\end{align}
Here we have introduce $\trp(i_1 \ldots i_n)$, which is defined as
\begin{equation}
  \mathrm{tr}_{\pm}(i_1 \ldots i_n)
  = \mathrm{tr}\left( \left[\frac{1\pm \gamma_5}{2}\right] \slashed{p}_{i_1} 
  \ldots \slashed{p}_{i_n} \right),
\end{equation}
and, in the case $n=4$, gives
\begin{equation}
  \mathrm{tr}_{\pm}(i \, j \, k \, l)=
  2\left(
  (p_i\cdot p_j) (p_k\cdot p_l) - (p_i\cdot p_k) (p_j\cdot p_l)
  + (p_i\cdot p_l) (p_j\cdot p_k)
  \pm i \varepsilon^{\mu \nu \rho \sigma} p_i^\mu p_j^\nu p_k^\rho p_l^\sigma
  \right).
\end{equation}
This object is manifestly multilinear in the external momenta and manifestly
vanishes in the limit where any of the involved momenta go to zero.
We note that this object is chiral if the vectors $p_i$, $p_j$, $p_k$ and $p_l$ are 
linearly independent, as in this case $\mathrm{tr}_{\pm}(i\,j\,k\,l)$ depends on $\trFive$. 
If this is not the case, as in eq.~\eqref{eq:invDim4}, then
$\mathrm{tr}_{+}(i\,j\,k\,l)=\mathrm{tr}_{-}(i\,j\,k\,l)$
is invariant under the Galois group action associated
with the flip of the sign of $\trFive$.

We next list some letters that are not invariant under the Galois group
associated to the square roots in the problem. Two sets depend on the three-point Gram determinants $\Delta_3$
and~$\Delta_3^{\ncg}$, and already arise in one loop integrals \cite{Abreu:2017mtm}. The
first is associated to three-mass triangle integrals whilst the second is
associated to the two-mass hard box,
\begin{align}
  \begin{split}
    \{W_{33}, \ldots, W_{36}\} &= \bigg\{ \frac{s_{12} + s_{13} + \sqrt{\Delta_3}}{s_{12} + s_{13} - \sqrt{\Delta_3}}, \frac{s_{14} + s_{15} + \sqrt{\Delta_3}}{s_{14} + s_{15} - \sqrt{\Delta_3}}, \\
    &\qquad \frac{s_{12} + s_{15} + \sqrt{\Delta_3^{\ncg}}}{s_{12} + s_{15} - \sqrt{\Delta_3^{\ncg}}}, \frac{ s_{14} + s_{13}+ \sqrt{\Delta_3^{\ncg}}}{ s_{14} + s_{13} - \sqrt{\Delta_3^{\ncg}}}\bigg\},\\[1.0ex]
  \{W_{37}, W_{38}, W_{39}\} &= \bigg\{ \frac{s_{12} - s_{13} + \sqrt{\Delta_3}}{s_{12} - s_{13} - \sqrt{\Delta_3}},  \frac{s_{15} - s_{14} + \sqrt{\Delta_3}}{s_{15} - s_{14} - \sqrt{\Delta_3}}, \frac{s_{12} - s_{15} + \sqrt{\Delta_3^{\ncg}}}{s_{12} - s_{15} - \sqrt{\Delta_3^{\ncg}}}\bigg\}\,.
    \end{split}
\end{align}
A set involves the Levi-Civita contraction $\trFive$,
\begin{align}
  \begin{split}
  \{W_{40}, \ldots, W_{46}\} &= \bigg\{
  \frac{\trp(2\,3\,4\,5)}{\trm(2\,3\,4\,5)}, \,
  \frac{\trp(1\,2\,3\,4)}{\trm(1\,2\,3\,4)}, \,
  \frac{\trp(1\,5\,4\,3)}{\trm(1\,5\,4\,3)}, \,
  \frac{\trp(4\,5\,1\,2)}{\trm(4\,5\,1\,2)}, \\
  &\quad \quad
  \frac{\trp(3\,2\,1\,5)}{\trm(3\,2\,1\,5)}, \,
  \frac{\trp(1\,2\,4\,3)}{\trm(1\,2\,4\,3)}, \,
  \frac{\trp(1\,5\,3\,4)}{\trm(1\,5\,3\,4)} \bigg\}.
  \end{split}
\end{align}
A single `relevant' letter involves two square roots,
\begin{align}
  \begin{split}
  W_{47} &= \frac{\Omega^{--} \, \Omega^{++}}{\Omega^{+-} \, \Omega^{-+}}\,, \quad 
  \mathrm{where} \quad \Omega^{\pm \pm} = 
  s_{12} s_{15} - s_{12} s_{23} - s_{15} s_{45} \pm s_{34} \sqrt{\Delta_3} \pm \trFive{}.
    \end{split}
\end{align}
Finally, two of the square-roots themselves are `relevant' letters
\begin{align}
  \begin{split}
  \{W_{48}, W_{49}\} &= \{ \sqrt{\Delta_3}, \trFive{} \}.
  \end{split}
\end{align}
While these two letters are clearly not invariant under the Galois group, 
their contribution to the symbol is, since only
the logarithm of their absolute value is relevant.

Beyond these `relevant' letters, there are also 9 `irrelevant' letters which do
not appear in the symbol up to weight 4. 
They can be organized in a
similar way as above. There are four letters which are invariant under the
action of the Galois group,
\begin{align}
  \begin{split}
	W_{50}&=\sqrt{\Delta_3^{\ncg}}\\[1.0ex]
  W_{51} &= \trp(1 \, 3 \, 1 \, 4), \\[1.0ex]
  \{W_{52}, W_{53}\} &= \{\trp(2 \, 1 \, [1+5] \, 4 \, [1+5] \, 1), \trp(5 \, 1 \, [1+2] \, 3 \, [1+2] \, 1)\},\\
  \end{split}
\end{align}
where in the last set we make use of a six index $\trp$. The remaining five
have non-trivial properties under the Galois group and are given by
\begin{align}
  \begin{split}
  &W_{54} = \frac{s_{13} - s_{14} + \sqrt{\Delta_3^{\ncg}}} {s_{13} - s_{14} - \sqrt{\Delta_3^{\ncg}}}\,,\\[1.0ex]
  &\{ W_{55}, W_{56} \} = \left\{\frac{\trp(1 \, 5 \, 3 \, [1+2])}{\trm(1 \, 5 \, 3 \, [1+2])}, 
  \frac{\trp(1 \, 2 \, 4 \, [1+5])}{\trm(1 \, 2 \, 4 \, [1+5])} \right\}\,, \\[1.0ex]
  &W_{57} = \frac{\trp(1\, 3\, 2\, 4)}{\trm(1\, 3\, 2\, 4)} \, 
  \frac{\trm(1 \, 4 \, 5 \, 3)}{\trp(1 \, 4 \, 5 \, 3)}, \\[1.0ex]
  &W_{58} = \frac{\tilde{\Omega}^{--} \, \tilde{\Omega}^{++}}{\tilde{\Omega}^{+-} \, \tilde{\Omega}^{-+}}\,,
  \quad \mathrm{where} \quad \tilde{\Omega}^{\pm \pm} = 
  s_{12} s_{13} - s_{12} s_{25} - s_{13} s_{34} \pm s_{45} \sqrt{\Delta_3^{\ncg}} \pm \trFive{}.
  \end{split}
\end{align}

\begin{figure}[]
\begin{center}
\includegraphics[width=4cm]{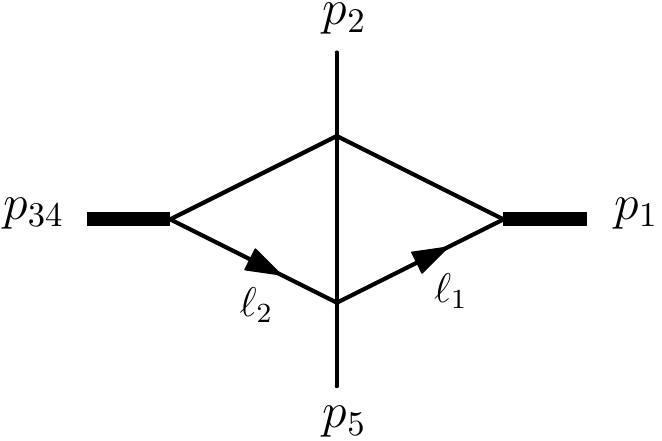}
\caption{Triangle-triangle topology with five master integrals. The scalar
integral, normalized as in eq.~\eqref{eq:normBadSlashed} is the only integral
that depends on letters $\{W_{35}, W_{36}, W_{39}\}$ at weight 4.}
\label{fig:2massSlashed}
\end{center}
\end{figure}

Let us make a number of comments on the symbol alphabet. 
First, all 30 letters that appear in the one-loop
alphabet are `relevant' letters at two-loops. Specifically, the one-loop alphabet
is comprised of
\begin{align}
    A_{1-\mathrm{loop}} = 
    \{&W_1, \ldots,  W_9, W_{12}, \ldots W_{15}, W_{18}, W_{19}, W_{22}, \ldots, W_{24}, W_{33}, W_{34}, W_{37}, W_{38}, W_{40}, \nonumber\\
      &W_{43}, \ldots, W_{49}\}.
\end{align}
Second, the letters $W_{30}$, $W_{53}$ and $W_{55}$ do not appear in the
presented integrals, but at amplitude level. 
Third, we comment on the relevant letters depending on $\sqrt{\Delta_3^{\ncg}}$,
$\{W_{35}, W_{36}, W_{39}\}$. Up to weight 4, these appear in a single master integral,
the scalar integral associated with the topology in \fig{fig:2massSlashed}
normalized with
\begin{equation}\label{eq:normBadSlashed}
  \mathcal{N}=\epsilon^4 \sqrt{\Delta_3^{\ncg}}\,.
\end{equation} 
This integral is first non-zero at weight 4. We note that $\sqrt{\Delta_3^{\ncg}}$
is also a letter, but it does not appear in any of the master integrals at weight
4 and as such is part of the `irrelevant' letters.
Finally, we note that only a small number of symbol letters cannot be determined from maximally-cut
differential equations. Specifically we find that, at amplitude level, the only letters that first appear
at the next-to-maximal-cut level are $W_{54}, W_{56}$ and $W_{58}$.
Remarkably, this implies that \emph{all `relevant' letters can be determined from the
maximal-cut differential equations}.

\subsection{Structure of symbols of master integrals}
\label{sec:strucSymb}

Having discussed the symbol alphabet, which describes the possible entries in
the symbol, it remains to discuss the patterns of letters which turn up in
practice in the master integrals.

We already discussed the first-entry condition at the beginning of this section.
Here we will illustrate how strong this condition is by showing how it determines
the weight 0 value of the integrals. It is clear from the definition of the
symbol that, at weight one, we have
\begin{equation}
  S[{\bf I}^{(1)}] =\sum_{\alpha} [W_\alpha]\,M_\alpha\, {\bf I}^{(0)},
\end{equation}
where ${\bf I}^{(0)}$ is a vector of rational numbers (of weight 0).
The first entry condition states that $S[{\bf I}^{(1)}]$ should not
contain $[W_\alpha]$ if $\alpha>6$. This means that the vector
${\bf I}^{(0)}$ must be in the kernel of the matrices $M_\alpha$ with $\alpha>6$,
that is, it lives in the intersection of the nullspaces of theses matrices.
Constructing such a vector is a simple linear algebra exercise.
Remarkably, the intersection of the nullspaces has dimension 1, which means
that ${\bf I}^{(0)}$ is fully determined by this exercise, up to an overall normalization
that any nullspace calculation is obviously blind to.
This is consistent with the fact that the differential equation is homogeneous
in ${\bf I}$.
As an example of how to use the differential equations in our ancillary files,
we implemented this calculation in a \texttt{Mathematica} function that can 
be found in \texttt{anc/usageExample.m} and allows to compute the symbols
of all the master integrals.

\begin{figure}[]
\begin{center}
\includegraphics[width=4cm]{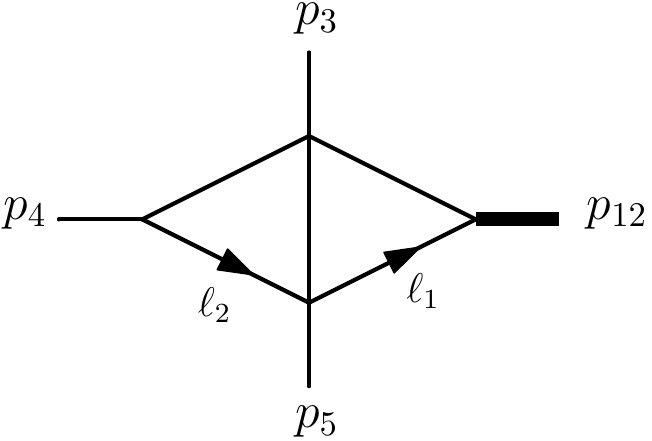}
\caption{Triangle-triangle topology with two master integrals. The scalar
integral, normalized by $\eps^4(s_{34}+s_{45})$, is the only integral that
depends on $W_{21}$ at weight 4. Its image under $(2\leftrightarrow 5,3\leftrightarrow 4)$
 is the only integral that depends on $W_{20}$ at weight 4.}
\label{fig:1massSlashed}
\end{center}
\end{figure}

Beyond the first entry, we also find that the letters that appear in the 
second entry of the symbols are highly constrained. 
Given the form of the differential equation, the weight-two symbols
fully determine the first two entries of any symbol tensor at any weight.
We find that,
at weight two, the symbols of all the master integrals required for planar
two-loop five-point one-mass amplitudes correspond to the (weight two) 
symbols of one-loop boxes and triangles which preserve the cyclic ordering of the
external legs. 
This fact was already observed in the massless
case \cite{Gehrmann:2018yef}, and is well understood at one-loop 
\cite{Abreu:2017enx,Abreu:2017mtm}. We stress that this is a non-trivial constraint
on the symbols: simply imposing that the symbol-tensors correspond to the symbol
of a function (i.e., that it is `integrable' \cite{Goncharov:2010jf}) would
allow letters
$\{W_{1}, \ldots W_{6}, W_{8}, W_{9}, W_{14}, W_{15}, W_{18}, \ldots,W_{22}, W_{33}, W_{34},
W_{35}, W_{39}\}$ to appear at weight two. We find that letters 
$\{W_{20}, W_{21},W_{35}, W_{39}\}$ do not appear. Interestingly, these
letters first appear at weight four, and are associated with only two topologies:
$W_{21}$ appears in the scalar integral of \fig{fig:1massSlashed} normalized
to $\eps^4(s_{34}+s_{45})$ and $W_{20}$ in its image under $(2\leftrightarrow 5,3\leftrightarrow 4)$,
and $\{W_{35}, W_{39}\}$ appear in the integral of \fig{fig:2massSlashed}
with the normalization in eq.~\eqref{eq:normBadSlashed} that we have already 
discussed.

\begin{figure}
  \centering
  \includegraphics[width=0.3\textwidth]{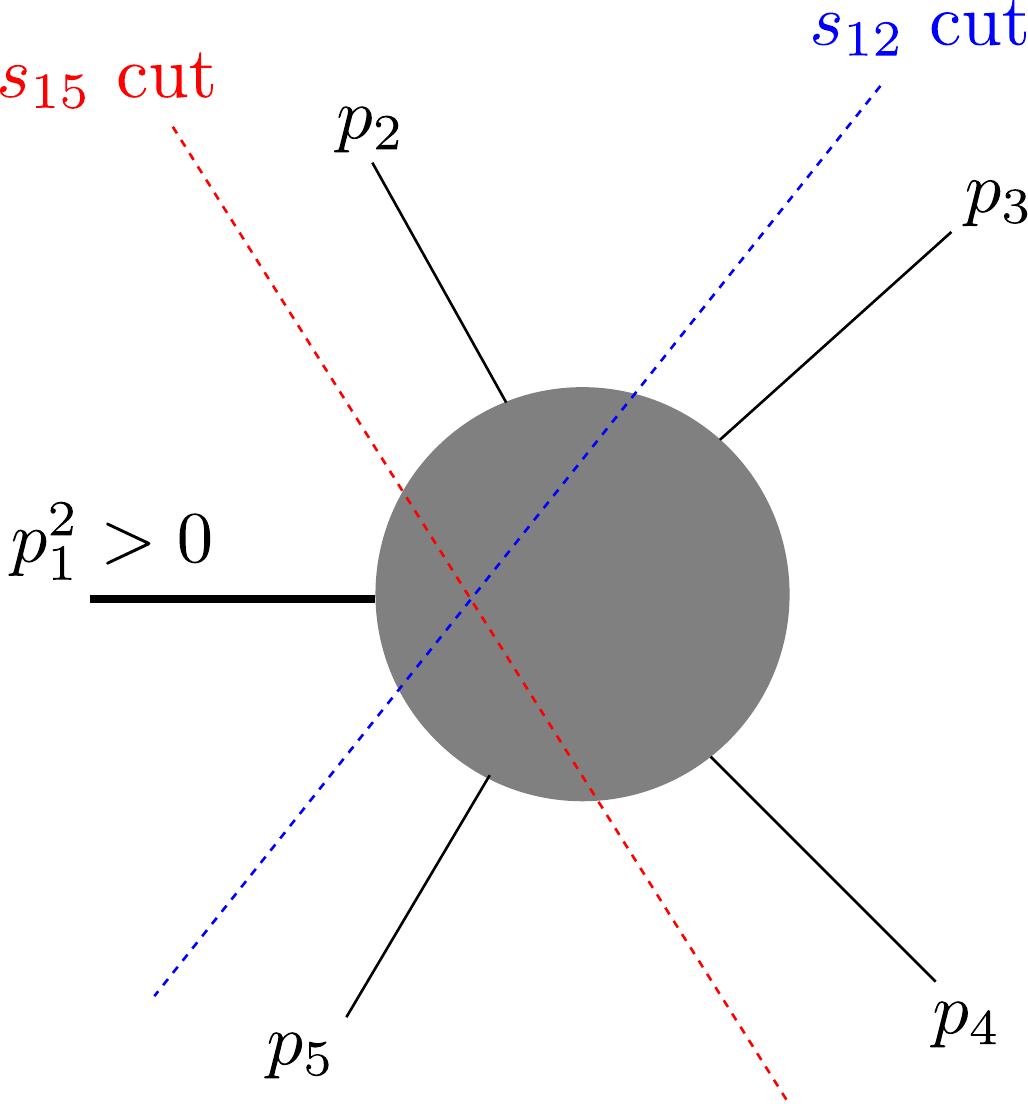}
  \caption{Illustration of Steinmann relations: the cuts in the $s_{12}$ and $s_{15}$
channels overlap and are therefore incompatible.}
  \label{fig:steinmann}
\end{figure}

It is also interesting to contrast other constraints on the symbol
alphabet with the structure of the differential equations. For instance,
the Steinmann relations \cite{Steinmann, Steinmann2,
Cahill:1973qp, Caron-Huot:2016owq, Dixon:2016nkn}
state that there is no double discontinuity associated with 
overlapping channels. In our case, this means that there should be no
double discontinuity associated with the $s_{12}$ and $s_{15}$
channels.\footnote{Due to subtleties with massless particles, we 
consider only channels involving at least 3 adjacent massless external particles, or two
adjacent external particles where at least one particle is massive.} 
Consistent with this expectation, we observe that letters 
$W_3$ and $W_4$ never appear consecutively in any symbol tensor.
Our results also confirm a stronger version of the constraint,
known as the `extended Steinmann relations',
which states that the two letters cannot appear in the $n$-th and $(n+1)$-th 
letters in a symbol tensor for any $n$. Indeed, we find that the matrices
$M_{3}$ and $M_{4}$ satisfy the relations
\begin{equation}
  M_{3} \, M_{4} = M_{4} \, M_{3} = 0,
  \label{eq:diffEq-steinmann}
\end{equation}
which implies that the extended Steinmann relations will be satisfied at all weights. 
We thus see that the structure of the differential equations naturally
encodes the extended Steinmann relations.

In addition to (extended) Steinmann relations, we have empirically observed more
`forbidden pairs' of symbol letters which never appear consecutively in the
symbols, by looking for pairs $(i,j)$ such that 
\begin{equation}\label{eq:matForbidden}
  M_iM_j=M_jM_i=0.
\end{equation}
We find many such pairs. 
We can however restrict them by demanding that the set of conditions be closed under the exchange
$(2\leftrightarrow 5,3\leftrightarrow 4)$ so that they are conditions on the symbol
of the planar amplitudes, and furthermore impose that $i\leq 6$,
that is $W_i$ is a letter that can appear in the first entry, and $j$ appears
in the second entry of at least one master integral.
Under these conditions, we find 
four pairs of forbidden letters (besides the pair $(3,4)$ which we already discussed):
\begin{align}\begin{split}
  (W_{3}, W_{18}) &= \big(s_{12}, \, 2\,p_4\cdot(p_1+p_5)\big)\,, 
  \quad (W_{4}, W_{19}) = \big(s_{15}, \, 2\,p_3\cdot(p_1+p_2)\big)\,,\\
  (W_{3}, W_{34}) &= \left(
  s_{12}, \, \frac{s_{14} + s_{15} + \sqrt{\Delta_3}}{s_{14} + s_{15} - \sqrt{\Delta_3}}\right)\,,
  \quad
  (W_{4}, W_{33}) = \left(
  s_{15}, \, \frac{s_{13} + s_{12} + \sqrt{\Delta_3}}{s_{13} + s_{12} - \sqrt{\Delta_3}}
  \right) .
  \label{eq:extraForbiddenPairs}
\end{split}\end{align}
We stress that, given that these pairs satisfy eq.~\eqref{eq:matForbidden},
these letters cannot appear next to each other for any symbol tensor and at any weight.
We leave it to future work to elucidate the nature of these extra Steinmann-like relations.
%

%% file: seriesSolution.tex

\section{Series solution of the differential equations}
\label{sec:DEIntegration}

In this section we discuss our approach to solving the differential equations
constructed previously, which follows the strategy proposed 
in \cite{Francesco:2019yqt}.
That is, we solve the differential equation along a path connecting a known
boundary point and a target point, and the solution is written in terms of
univariate generalized power series. After discussing how to construct the
solution along a path, we discuss analytic continuation around the different
branch-points, the determination of the boundary values, and the estimation
of the numerical precision of our solutions.

\subsection{Series solution along a path}
\label{sec:localSolution}

Our approach to evaluate the master integrals is to
solve their differential equations with
generalized power series~\cite{Francesco:2019yqt}. In
this method, the system of partial differential equations~\eqref{eqn:DE} 
is integrated along a one-dimensional path connecting two fixed points in the 
space of the Mandelstam variables. 
For concreteness, we focus our discussion in the case where  ${\bf I}$ is 
a vector of pure integrals, that is where
the connection ${\bf M}$ takes the form of eq.~\eqref{epsFactorizedDE}.
In the following, the univariate path will be parametrized by $t$ and for convenience we will
take it to be the straight line
\begin{equation}\label{eq:path1}
\vec s(t)=  \vec s_b + (\vec s_e-\vec s_b)\,t\,, \qquad  t\in [0,1]\,.
\end{equation}
The initial point $\vec s_b$, where we assume ${\bf I}$ is known, 
provides the boundary condition required to solve the differential
equation, and the final point $\vec s_e$
denotes the point in the space of Mandelstam variables where we wish to evaluate
the integrals.
Along the path, the differential equation \eqref{eqn:DE} 
degenerates onto a system of univariate ordinary differential equations
depending on the parameter $t$,\footnote{
	To avoid introducing new notation, we set ${\bf I}(t)\equiv {\bf I}(\vec s(t))$.
	For instance, ${\bf I}(0)= {\bf I}(\vec s_b)$ and ${\bf I}(1)= {\bf I}(\vec s_e)$.
}
\begin{equation}
\label{eq:DE along contour}
\frac{d}{d t} {\bf I}(t,\epsilon)=\epsilon\mathbf{A}(t){\bf I}(t,\epsilon)\,,\qquad 
\mathbf{A}(t)=\frac{1}{\epsilon}\frac{d {\bf M}(\vec s(t))}{dt}\,.
\end{equation}
As discussed at the start of section \ref{sec:AnalyticStructure}, such a system
admits an iterative solution in $\epsilon$,
\begin{equation}
    \label{ExactSolutionIteInt t}
    {\bf I}(t,\epsilon) = \sum_{i=0} {\bf I}^{(i)}(t)\,\epsilon^i\,,\qquad
    {\bf I}^{(i)}(t) = \int \mathbf{A}(t){\bf I}^{(i-1)}(t)\,dt + \mathbf{c}^{(i)}\,,
\end{equation}
where we assumed that the integrals are normalized such that their Laurent series in 
$\epsilon$ have no negative powers.
The $\mathbf{c}^{(i)}$ are integration constants of the differential equation,
uniquely fixed by the boundary condition at $t=0$, which in this section
we assume to be known. The starting point of the iterative solution is
the integration constant ${\bf I}^{(0)}=\mathbf{c}^{(0)}$. We recall
that there is a concept of weight associated with solutions to differential equations
of the type of eq.~\eqref{eq:DE along contour}, which is aligned
with the coefficient in the Laurent expansion in $\epsilon$ of the solution. 
We will sometimes refer to ${\bf I}^{(i)}(t)$ as the contribution of weight $i$
to ${\bf I}(t)$.

While in principle the integrals in \eqn{ExactSolutionIteInt t} are expected to be
computable in terms of multiple polylogarithms, in practice the symbol alphabet
can make this a daunting task. Firstly, to apply direct integration
procedures one must simultaneously rationalize all square roots, which may not
be possible (in appendix \ref{appendix:kinematics} we discuss some parametrizations
that rationalize a subset of the square roots). 
Secondly, it can be complicated to accurately handle both
spurious and physical branch points in any resulting expression. This problem can be
exacerbated by introducing variables that rationalize the alphabet. 
Fortunately, these issues can either be sidestepped or clarified with locally-valid
solutions written in terms of (generalized) power series. Such local
solutions are only valid in a well-defined region. The first task in solving
eq.~\eqref{eq:DE along contour} on the path of eq.~\eqref{eq:path1} with this approach
is then to split the path into segments, each with its own local solution. More explicitly
we write the solution ${\bf I}^{(i)}(t)$ as
\begin{equation}
\label{eq:sol gamma}
{\bf I}^{(i)}(t)=\sum_{k=0}^{N_e-1}  \chi_k(t)\, {\mathbf I}^{(i)}_{k}(t)\,, \quad t \in [0,1]\,,
\end{equation}
with,
\begin{equation}
	\chi_k(t)=\left\{\begin{array}{ll} 1, &\quad t\in[t_k-r_k,t_k+r_k) \\
				0, &\quad \mbox{otherwise}  
	\end{array}\right. \,.
\end{equation}
Here, $t_k$ is the expansion point of the local power series solution,
$r_k$ is the radius of the region where the local solution is used, i.e.,
the radius of segment $k$, and $N_e$ is the number of segments. 
The path segment centered at $t_k$ with radius $r_k$ is denoted
$S_k = [t_k-r_k, t_k+r_k)$.
Our goal is to compute the value of ${\bf I}^{(i)}(t)$ at $t=1$, which is given by
\begin{equation}
{\mathbf I}^{(i)}(1)= {\mathbf I}^{(i)}_{N_e - 1}(1)\,.
\end{equation}
In the following, we will first discuss the construction
of the local solutions ${\bf I}^{(i)}_k(t)$, and then discuss how to construct a
segmentation of the path. 

A local solution ${\bf I}^{(i)}_k(t)$ is one that is valid in some region
around the point $t_k$. We can easily construct such a solution through series
expansion of the integrand in \eqn{ExactSolutionIteInt t}. This series
has a finite radius of convergence and so the
solutions will only be valid locally. 
The matrix $\mathbf{A}(t)$ determines the form of the
series expansion of the integrand. 
Given the form of the alphabet discussed in the previous section,
in our case it contains both simple poles
and square-root branch cuts. The
series expansion around the point $t_k$ then takes the form
\begin{equation}
\label{eq:Aexp}
\mathbf{A}(t)=\sum_{i=-2}^\infty \mathbf{A}_{i,k} (t-t_k)^{\frac{i}{2}},
\end{equation}
where the $\mathbf{A}_{i,k}$ are constant matrices. 
Through iterated integration, the series solution then takes
the form of a half integer power series with logarithmic terms 
\begin{align}\begin{split}
    {\bf I}^{(i)}_{k}(t) &= \mathbf{c}^{(i)}_k + \sum_{j=-2}^\infty \mathbf{A}_{j,k} \int  (t-t_k)^{\frac{j}{2}}{\bf I}^{(i-1)}_{k} (t) dt \\
  &=\sum_{j_1=0}^\infty 
 \sum_{j_2=0}^{N_{i,k}}\mathbf{c}_{k}^{(i,j_1,j_2)}(t-t_k)^{\frac{j_1}{2}}\log{(t-t_k)}^{j_2},
    \label{eq:SeriesFISingular}
\end{split}\end{align}
where we have exchanged the order of integration and summation. Here,
$\mathbf{c}_k^{(i)}=\mathbf{c}_{k}^{(i,0,0)}$ are integration constants and
$\mathbf{c}_{k}^{(i,j_1,j_2)} $ are constant vectors determined iteratively from
the matrices $\mathbf{A}_{j,k}$. $N_{i,k}$ is the maximum power of the logarithm
in the local solution at iteration~$i$. When $t_k$ is a regular point of
$\mathbf{A}(t)$, the ${\mathbf I}^{(i)}_{k}(t)$ simplify to a Taylor series. 
The radius of convergence of this solution is the same as that of
the expansion of the matrix $\mathbf{A}(t)$ in \eqn{eq:Aexp}.
We note that the solution in eq.~\eqref{eq:SeriesFISingular} introduces
logarithmic and square-root branch points at $t_k$ that must be handled with care.
This will be discussed in section \ref{sec:analyticContinuation}.

Let us briefly discuss how the integration constants $\mathbf{c}_{k}^{(i,0,0)}$
associated with the local solution around $t_k$ are related to the 
boundary condition ${\bf I}^{(i)}(0)$ of the full solution.
First, the constant of integration $\mathbf{c}_0^{(i,0,0)}$ is obtained by requiring 
that the $k=0$ local solution ${\bf I}_0^{(i)}(t)$ matches the known boundary value at
$t=0$, that is
\begin{equation}
  {\bf I}_0^{(i)}(0) = {\bf I}^{(i)}(0).
\end{equation}
Note that this only requires ${\bf I}_0^{(i)}(t)$ to be valid at $t=0$, rather than
centered there. 
The remaining integration constants are then iteratively determined by exploiting 
the continuity of the full solution \eqn{eq:sol gamma} at the boundary of each segment.
Explicitly,
\begin{equation}
\label{eq:cont cond}
{\bf I}^{(i)}_{k}(t_{k}-r_{k}) =
{\bf I}^{(i)}_{k-1}(t_{k-1}+r_{k-1})\,,\quad k=1,2,\ldots, N_e-1\,,
\end{equation}
where the right-hand side should be understood as the limit of ${\bf I}^{(i)}_{k-1}(t)$
as $t\to t_{k-1}+r_{k-1}$, which exists by construction.
In this way, the integration constant in each local solution can be determined from ${\bf I}^{(i)}(0)$. 
We will discuss how to compute ${\bf I}^{(i)}(0)$ in section \ref{sec:boundary conditions}.

To make the solution in eq.~\eqref{eq:SeriesFISingular} practical, 
it will be necessary to work with truncated series expansions and control
the numerical error associated with the truncation. As is well known, the
convergence rate of the series decreases as one approaches the radius of convergence
of the series. We must thus be careful with how we construct
the segments $S_k$, in particular in balancing the size and the number of segments
used to cover the integration path: 
they should be small enough so that the truncated series solution
converges fast enough on each segment, but there should not be too many segments as the complexity of 
the algorithm scales linearly with the number of segments. The remainder of this 
section is devoted to describing the segmentation of the path. We choose to work
under the constraint that segments should never be larger than half the radius
of convergence of the associated series solution to guarantee that convergence
is fast enough on each segment. We note nevertheless that this constraint can be modified
at the price of having more segments
if we want to build local solutions that converge at a different rate.
Finally, we note that the segmentation of the path is the same for all orders
in the $\epsilon$ expansion, that is for all $i$ in eq.~\eqref{ExactSolutionIteInt t}.

A segmentation of the path is a collection of non-overlapping segments
(or intervals) $S_k = [t_k - r_k, t_k + r_k)$ such that the union of all of the
segments covers the interval $[0,1]$, that is 
\begin{equation} [0,1] \, \, \subset \, \, \bigcup_{k=0}^{N_e-1} S_k.
\end{equation}
Each segment is specified by its center $t_k$ and radius $r_k$.
The choice of the pairs $(t_k,r_k)$ is primarily dictated by the singular
points of the differential equation (\ref{eq:DE along contour}).
These singular points may occur for both real and complex values of $t$.
Let us denote the set of real singular points $R=\{\sigma_k \}_{k=1,\ldots,N_s}$ and
the set of complex singular points $C=\{\lambda_k\}_{k=1,\ldots,N_c}$. The
complex-valued singular points will also affect the convergence
properties of neighboring series solution. 
In order to avoid using complex arithmetic, we define the set of
real regular points $C_r=\{ {\rm Re}(\lambda_k) - {\rm Im}(\lambda_k), 
{\rm Re}(\lambda_k), {\rm Re}(\lambda_k) +
{\rm Im}(\lambda_k)\}_{k=1,\ldots,N_c}$.
Considering these real-valued points effectively accounts for the effect of
the complex valued singularities.
It is clear that not all points in $R\cup C_r$ affect the series solution in $[0,1]$, but it
is also clear that it is not sufficient to consider the points that are in $[0,1]$.
Given our constraint of only using a series solution in half its radius of convergence,
it is sufficient to consider the points $t_k\in R \cup C_r$ such that 
$t_k\in(-2,3)$.\footnote{
	We note that this interval is dependent on the constraint that the
	segments should never be larger than half the radius
	of convergence of the associated series solution.
}
To each $t_k$ we associate a radius $r_k$,
chosen to be half the distance between $t_k$ and the closest point in $R \cup C_r \cup \{-2,3\}$.

The above procedure may not cover the full interval $[0,1]$.
For these uncovered regions we turn to bisection, that is we add segments centered 
at regular points in the middle of the uncovered intervals of $(-2,3)$ that overlap with $[0,1]$.
The associated radii are chosen to be the minimum of the following two quantities,
\begin{itemize}
\item half the distance between $t_k$ and the closest point in $R \cup C_r \cup \{-2,3\}$,
\item the distance between $t_k$ and the closest segment already determined.
\end{itemize}
We iterate the bisection until the $[0,1]$ interval is covered.
We note that if $R \cup C_r$ does not contain any point $-2<t_k<3$, 
there is a single regular expansion point at $t_0=1/2$.
Finally, we note that the segmentation procedure we described may
have produced segments with no overlap with $[0,1]$ which we simply remove.

\subsection{Analytic continuation}
\label{sec:analyticContinuation}

As was noted below eq.~\eqref{eq:SeriesFISingular}, 
a local solution $\mathbf{I}^{(i)}_k(t)$ of the differential equation will in general
have a branch cut if the associated expansion point $t_k$ is either a singular point 
or a square-root branch point of $\mathbf{A}(t)$. At each such point $t_k$,
a subset of the letters in the symbol alphabet will either vanish or become
infinity. In this section, we will classify the different types of branch-points
we can encounter and then explain how we deal with the analytic continuation 
across different types of branch points.

Let us first introduce our naming for three different classes of branch points.
The simplest to define are the `square-root branch points', which arise from
terms with non-integer exponents in eq.~\eqref{eq:SeriesFISingular}.
The remaining two cases are logarithmic branch cuts. 
We distinguish those that are `physical thresholds'
from those that are `non physical thresholds' as follows. It is well known
that Feynman integrals with massless propagators have logarithmic branch cuts 
when either of the Mandelstam variables in $\vec s$ vanishes. These are 
the physical thresholds. From the alphabet we have determined in section 
\ref{sec:AlphabetResults}, it is nevertheless clear that there are many other 
potential branch points. To contrast these against the physical thresholds we
call them non physical thresholds.

Consider now a $t_k$ that is associated with a logarithmic branch-point
in eq.~\eqref{eq:SeriesFISingular}.
Given the distinction between physical and non physical thresholds, we would 
like to determine to which class $t_k$ belongs. 
To achieve this, it is natural to make a connection with the letters
of the symbol alphabet, since we expect that some of them should either
vanish or become infinity at $t_k$.
Na\"ively, one might say that if $t_k$ corresponds to a physical
threshold, it should be associated with one of the letters $W_1$ through $W_6$,
and if it is a non physical threshold it should be associated to any of the other
letters. This is however not exactly the case, as we now show in an example.
Consider a point $t_k$ where $\offShellScale\to0$. It is clear that at
this point $W_1=0$. Nevertheless, this is not the only letter that vanishes.
For instance, letter $W_{33}$ can be written as
\begin{equation}
	W_{33}=\frac{\offShellScale+s_{45}-s_{23}+\sqrt{\Delta_3}}
	{\offShellScale+s_{45}-s_{23} - \sqrt{\Delta_3}}	
	=\offShellScale \,\widehat{W}_{33}\,,
	\quad
	\textrm{with}
	\quad
	\widehat{W}_{33}=\frac{4s_{45}}{\left(\offShellScale+s_{45}-s_{23} - \sqrt{\Delta_3}\right)^2}\,,
\end{equation}
and will thus also vanish as $\offShellScale\to0$ if $(s_{45}-s_{23}) < 0$. This observation might cast a doubt
on whether $t_k$ should correspond to a physical threshold or not. It is nonetheless
true that $t_k$ is a physical threshold, and the fact that $W_{33}$ vanishes is,
geometrically, a consequence of the fact that, due to the square root, the zero set of an odd letter
does not correspond to an irreducible algebraic variety 
(with our choice of alphabet, each even letter defines an irreducible variety).
This situation should however be distinguished from the case where, at a given
point $t_k$, both $\offShellScale\to0$ and $\widehat{W}_{33}\to0$.
Then the point $t_k$ corresponds to an overlapping singularity, where two independent
singular surfaces intersect. To make the different singular surfaces associated with each
letter manifest, we can explicitly compute their \dlog{} using the variables
$\vec s$ as coordinates. In the case of $W_{33}$ we would find 
\begin{equation}
  d \log (W_{33}) = \left[  \frac{\sqrt{\Delta_3} (s_{13}+s_{12} + 
  \sqrt{\Delta_3})^2}{ 2 s_{45} \, \Delta_3 \, \offShellScale{}} \right]
  \left[ \frac{s_{13}+s_{12}}{2} 
  \left(  \frac{ d \offShellScale{} }{\offShellScale{}} - 
  \frac{ d s_{45}}{s_{45} } \right)- d s_{23}\right],
\end{equation}
and identify the irreducible singular surfaces $s_{45}=0$, $\offShellScale=0$ and $\Delta_3=0$.
In summary, the classification of $t_k$ into physical or non physical thresholds should be done
with care. The first step is to check if a given $t_k$ is associated with the vanishing
of one of the entries of~$\vec s$. If it is, one should check the behavior of the other
letters. If they vanish (or become infinity) only because of the same entry of $\vec s$, then
$t_k$ is a physical threshold. Otherwise, it is associated with an overlapping singularity.
Finally, if $t_k$ is not associated with the vanishing of one of the entries of $\vec s$
then it is a non physical threshold. Non physical thresholds might also appear together
in overlapping singularities, but this classification is immaterial for our
purposes.

Now that we have classified all types of branch points we can encounter, 
we discuss the analytic continuation across each one of them.
\paragraph{Physical thresholds:}
Analytic continuation is determined by Feynman's $i\varepsilon$-prescription. 
Assuming that the threshold is associated with variable $s_i$, we take
\begin{equation}
s_{i}(t)\rightarrow s_{i}(t)+i\varepsilon=s_{bi}+(s_{ei}-s_{bi}) t+i\varepsilon, \quad \varepsilon>0\,.
\end{equation}
This can then be implemented by performing a deformation of the $t$-contour in
the segment centered at $t_k$,
\begin{equation}
t\rightarrow t+i\,\text{sign}( s_{ei}-s_{bi} )\,\varepsilon, \quad \varepsilon>0.
\end{equation}
As $\varepsilon$ is taken infinitesimally small, this only has an effect in the
 logarithmic terms of~(\ref{eq:SeriesFISingular}) which are then defined as
\begin{equation}
\log(t-t_k)=
\begin{cases}
\log(t-t_k) & \text{for } t>t_k \,, \\
\log(t_k-t) +i \,\text{sign}(s_{ei}-s_{bi} ) \pi & \text{for } t<t_k \,.
\end{cases}
\end{equation}

\paragraph{Non physical thresholds:}
It is well known that logarithmic singularities associated with non physical thresholds 
are absent in Feynman integrals in the Euclidean region. 
Therefore, the associated logarithmic terms drop out
of ${\mathbf I}_k^{(i)}$ in \eqn{eq:SeriesFISingular}. Indeed, we will return to
this observation in section \ref{sec:boundary conditions} and use it to determine
the boundary condition.
In the Euclidean region there is thus no analytic continuation to perform through
these branch points.
In the physical region, a path might cross a non physical threshold.
In order to avoid having to continue through such a threshold, 
in practice we instead take another path with the same
end point. Given that our paths are always straight lines, this means that we
start from a different initial point $\vec s_b$ to reach the desired point $\vec
s_e$.

\paragraph{Square-root branch points:}
Square-root branch points are an artefact of our choice of basis of master
integrals. Indeed, they are absent from genuine Feynman integrals (in the language
of section \ref{sec:AnalyticStructure}, Feynman integrals are invariant under the
action of the Galois groups
associated with each of the square roots), and  are introduced in the pure bases
when Feynman integrals are normalized by a square root.
We can thus freely chose the analytic continuation prescription of these branch points
as the effect drops out when we relate the pure basis back to Feynman integrals
(provided we are consistent with this prescription in the normalizations).
We use the prescription
\begin{equation}
(t-t_k)^{\frac{j_1}{2}}=
\begin{cases}
(t-t_k)^{\frac{j_1}{2}} & \text{for} \quad t>t_k\,, \\
i(t_k-t)^{\frac{j_1}{2}} & \text{for} \quad t<t_k\,. 
\end{cases}
\end{equation}

\paragraph{Overlapping singularities}

As discussed above, a given $t_k$ on a given path might correspond to an overlapping
singularity. In practice we have never encountered such a situation. Nevertheless,
we have implemented a check for this eventuality and, if such a situation were
detected, we would simply veto that path and choose an alternative path to the
end point.

\subsection{Boundary conditions}
\label{sec:boundary conditions}

Up to this point, we have assumed knowledge of the numerical value
of the integrals ${\bf I}(t)$ at some point in the space of Mandelstam invariants 
and elaborated on how to use generalized series expansions and differential
equations to transport this to another point in Mandelstam space. 
More precisely, we assumed that ${\bf I}(0)$ is known, and discussed
how to obtain ${\bf I}(1)$. In this section we discuss how to determine ${\bf I}(0)$.
Our approach will be based on arguments analogous to those that were used in
section \ref{sec:strucSymb}, where the symbol of the integrals
was determined by imposing the `first-entry condition'. This condition is a consequence of 
the fact that Feynman integrals have no branch-cuts in the bulk of the Euclidean region.
We now show how, by imposing this behavior, we can determine ${\bf I}(0)$
up to an overall normalization.

Our approach to the determination of the boundary condition is most conveniently formulated
order-by-order in the $\epsilon$ expansion. Throughout this discussion, we will
thus assume that we have fully determined the function ${\bf I}^{(i-1)}(t)$,
and use it to compute the boundary value ${\bf I}^{(i)}(0)$.
This will be achieved by enforcing that ${\bf I}^{(i)}(t)$ does not introduce
spurious logarithms at order $i+1$.
To build such a constraint, we consider a choice of path for which a spurious
logarithmic singularity occurs at $t=t_k$. 
Here, by spurious we mean a point $t_k$ where
${\mathbf A}(t)$ has a pole and none of the first entry Mandelstam invariants
are zero, that is
\begin{equation} \label{eq:expSpurious}
{\mathbf A}(t) = \frac{ 1 }{t-t_k} \mathbf{A}_{-2,k} + \mathcal{O}[(t-t_k)^0]
\end{equation} 
and all entries of $\vec{s}(t_k)$ are different from zero.
In the language of local solutions, a spurious logarithmic
singularity manifests itself as an explicit logarithm in the generalized series
solution associated to the point $t_k$. If we consider the computation of the
weight $(i+1)$ solution through \eqn{eq:SeriesFISingular}, it is clear that such a
logarithm arises if the contribution of the pole term from the right-hand side of 
eq.~\eqref{eq:expSpurious} is non-zero. 
We therefore see that requiring the absence of this spurious logarithm is
equivalent to the condition
\begin{equation}\label{eq:boundk}
  {\bf A}_{-2, k} \left[  {\bf I}_{k}^{(i)}(t_k) \right] = 0\,.
\end{equation}
Given our assumption that ${\bf I}^{(i-1)}(t)$ has been fully determined, it must
also satisfy eq.~\eqref{eq:boundk}, and the primitive in eq.~\eqref{eq:SeriesFISingular}
is thus regular at $t=t_k$. Using the continuity conditions of eq.~\eqref{eq:cont cond},
we can explicitly relate ${\bf I}_{k}^{(i)}(t_k)$ to  ${\bf I}^{(i)}(0)$.
More explicitly,
\begin{equation}
	{\bf I}_{k}^{(i)}(t_k)={\bf I}^{(i)}(0)+{\bf v}_k^{(i)}\,,
\end{equation}
where ${\bf v}_k^{(i)}$ is fully known. For instance, a useful implementation
strategy is to note that it can be computed as a difference of local solutions,
\begin{equation}
	{\bf v}_k^{(i)}={\bf I}_{k}^{(i)}(t_k) - {\bf I}_{0}^{(i)}(0)\,,
\end{equation}
which is independent of the boundary condition ${\bf I}^{(i)}(0)$.
Imposing eq.~\eqref{eq:boundk} then becomes an explicit constraint on ${\bf I}^{(i)}(0)$:
\begin{equation}
  {\bf A}_{-2, k} \left[ {\bf I}^{(i)}(0) \right] = -{\bf A}_{-2, k} \left[ {\bf v}_k^{(i)} \right].
  \label{eq:BoundaryCondition}
\end{equation}
We note that the constraint (\ref{eq:BoundaryCondition}) is particularly simple for 
$i=0$ as the integrals are constants. This implies that ${\bf v}_k^{(0)}=0$,
and we reproduce the conditions determined in section \ref{sec:strucSymb}
to constrain the symbol of the integrals.

It is clear that the above discussion can be repeated for a series of spurious
singularities to build more and more constraints on the value of ${\bf I}^{(i)}(0)$.
Searching for these singularities can be implemented in many ways. In our case,
we considered a piecewise straight-line path in the Euclidean region. More
concretely, we consider the vertices
\begin{equation}
\label{eq:eu-list}
\begin{array}{ccccccc}
 \vec s_{\text{eu-}1} =&\bigg(-11, & -1, & -\dfrac{5}{2}, & -\dfrac{7}{2} ,& -3, & -\dfrac{153}{14}\bigg), \\[7pt]
 \vec s_{\text{eu-}2} =&\Big(-11 ,& -10 ,& -\dfrac{5}{2}, & -\dfrac{7}{2}, & -4, & -12\Big), \\[7pt]
 \vec s_{\text{eu-}3} =&\Big(-11 ,& -10, & -\dfrac{5}{2}, & -\dfrac{7}{2} ,& -30, & -12 \Big),\\[7pt]
 \vec s_{\text{eu-}4} =&\Big(-11 ,& -12 ,& -\dfrac{5}{2}, & -32 ,& -50 ,& -12 \Big),\\[7pt]
 \vec s_{\text{eu-}5} =&\Big(-11 ,& -12, & -80, & -32 ,& -50, & -42 \Big),\\
\end{array}
\end{equation}
and the path $\vec s_{\text{eu-}1}\to \vec s_{\text{eu-}2} \to 
\vec s_{\text{eu-}3}\to \vec s_{\text{eu-}4}\to \vec s_{\text{eu-}5}$.
For each spurious singularity we encounter, we use \eqref{eq:BoundaryCondition}
to build a further set of of linear constraints on ${\bf I}^{(i)}(\vec
s_{\text{eu-}1})$.
In practice, we find that if we  combine all the constraints
determined along this path we are able to fix the value of ${\bf I}^{(i)}(0)$ up to a
single degree of freedom. Indeed, if we
consider this analysis for $i=0$ this is no surprise. As noted below 
\eqn{eq:BoundaryCondition}, the weight $0$ solution is kinematically
independent and so simultaneously lives in the kernel of all ${\bf A}_{-2, k}$.
Therefore, the general solution ${\bf I}^{(i)}(0)$ to  \eqn{eq:BoundaryCondition}
can be written in terms of any particular solution ${\bf I}^{(i)}_p(0)$ to
\eqn{eq:BoundaryCondition} and the weight zero solution, i.e.,
\begin{equation}
{\bf I}^{(i)}(0) = {\bf I}^{(i)}_p(0) +f^{(i)} {\bf I}^{(0)}(0),
\end{equation}
where $f^{(i)}$ is a constant we are yet to determine.
We note that this is the order by order in $\epsilon$ incarnation of the fact
that the differential equation \eqref{eq:DE along contour} is invariant under overall rescaling of
${\bf I}(t)$ by a kinematically independent but $\epsilon$-dependent function.
The value of $f^{(i)}$ can then be determined by computing a simple master
integral, such as a factorized bubble-type integral, with the normalization
chosen in eq.~\eqref{eq:intNorm}.

We end by emphasizing that, for a given topology, the boundary condition
only has to be computed once with this approach. It
can then be transported to other regions of phase-space with the procedure
described in section \ref{sec:localSolution}, and the result obtained in this
way used as a boundary condition for subsequent evaluations in each region.

\subsection{Numerical precision of integrals}
\label{sec:precision}

In section \ref{sec:localSolution} we already noted that our approach
to solving the differential equation \eqref{eq:DE along contour}
relies on truncated series expansions. Here, we describe how to fix the
truncation order in order to reach a given precision in the evaluation of ${\bf I}(t)$, 
which we define as the number $p$ of correct digits after the decimal point.
Throughout this discussion, we will refer to $p$ as the precision of the integrals.
The precision will be affected by two distinct factors. One is the 
precision associated with the boundary condition we compute with
the procedure described in section \ref{sec:boundary conditions}, 
and the other is the precision
of the numerical transportation of the solution along the integration path. We
will first discuss the transportation precision, and then comment on the
precision of the boundary condition.

The coefficients of the generalized series expansions in eq.~\eqref{eq:SeriesFISingular}
are represented by finite-precision numbers. In practice, we take
these coefficients to be much more precise than $p$ digits so that there is
no error associated with them.
The precision of a numerical evaluation of an integral is then controlled by two 
factors: the precision of a local solution on the boundary of a segment and the
accumulation of these errors along a path. That is, when computing
${\bf I}(t)$ along a given path, one needs to concatenate multiple segments,
and the truncation error of the integrals at the end point of the path is
obtained by combining the error on each segment. We determine the required
truncation order on a given segment by using the connection matrix ${\mathbf
A}(t)$, which is known exactly. We introduce the expression 
${\mathbf A}_{[k]}(t)$ for the truncated expansion over the $k$-th segment,
\begin{equation}
\mathbf{A}_{[k]}(t)=\sum_{i=-2}^{n_k} \mathbf{A}_{i,k} (t-t_k)^{\frac{i}{2}}\,,
\end{equation}
where, unlike in eq.~\eqref{eq:Aexp}, the expansion is truncated at order $n_k$.
For each segment, $n_k$ is determined by requiring that 
each element of ${\mathbf A}_{[k]}(t)$ approximates the matrix ${\mathbf A}(t)$
within a certain tolerance. That is, we fix $n_k$ by requiring
\begin{equation}
\label{eq:truncation order}
 \max_{i,j} \left|A_{[k], ij}(t) - A_{ij}(t)\right| < 10^{-(p+\delta)}, 
\quad t\in [t_k-r_k,t_k+r_k)\,,
\end{equation}
where we introduced another (positive) parameter $\delta$, which must
be determined so that the precision of the integrals at the end of the
integration procedure is indeed larger than~$p$.

In order to understand how to determine $\delta$, we must first understand how the truncation 
error of the local solutions accumulate to an error at the end point of the path.
We start by estimating the truncation error on each segment.
While we can in principle perform a detailed analysis of the
error propagation, we find it more practical to obtain an estimation of 
the error from Cauchy's convergence criterion.
Specifically, for a given local solution we consider the last $m$ terms
of the series expansion and estimate the truncation error at the end point
of the associated segment as,
\begin{equation}
    \label{error k}
    \Delta_{k} =\max_{i,a} \left|\sum_{j_1=n_k-m}^{n_k}
\sum_{j_2=0}^{N_{i,k}}c_{k,a}^{(i,j_1,j_2)}(r_k)^{\frac{j_1}{2}}(\log{r_k})^{j_2}\right|\,,
\end{equation}
where $m$ is a small integer compared to $n_k$ (we use $m=\left\lceil\frac{n_k}{50} \right\rceil$)
and the index $a$ labels the different integrals in ${\bf I} (t)$.
That is, we conservatively assign the worst
estimate across all weights $i$ and all master integrals $a$ to all integrals
and weights. 
In practice we observe that, for a given choice of $p$ and $\delta$,
the error of the local solution is of order $10^{-(p+\delta)}$ for each segment,~i.e.,
\begin{equation}\label{eq:errSeg}
	\Delta_k\sim10^{-(p+\delta)}\,.
\end{equation}
Next, we consider how the errors associated with the segments accumulate 
when matching multiple local solutions along the path.
Conservatively, we estimate that the error increases along the path by the
sum of the errors $\Delta_{k}$, that is
\begin{equation}
\label{eq:truncation error}
\Delta(\vec{s}_e) - \Delta(\vec{s}_b)= \sum_{k=0}^{N_e-1} \Delta_k,
\end{equation}
where $\Delta(\vec{s}_e)$ is the error on the value at the end of the path and
$\Delta(\vec{s}_b)$ is the error on the value at the beginning of the path, i.e., 
on the boundary condition. 
Note that, according to eqs.~(\ref{error k}) and~(\ref{eq:truncation error}), the
error associated to $\mathbf{I}(t)$ is the same for all
the master integrals in $\mathbf{I}(t)$ and for all the weights $i$.

We now have all the tools required to determine $\delta$. We distinguish
two cases for which our evaluation strategy is slightly different: 
the evaluation of $\mathbf{I}(t)$ at a single phase-space point
$\vec s_e$ using a known boundary value at $\vec s_b$, and the evaluation
of $\mathbf{I}(t)$ at multiple phase-space points.
Let us first discuss the case of a single evaluation. We assume that the boundary
value at $\vec s_b$ was computed
as described in \ref{sec:boundary conditions} to a precision much higher than
$p$, that is $\Delta(\vec s_b)\sim0$. 
In this case, the error is fully determined from the accumulation of the truncation
errors along the path, and from eqs.~\eqref{eq:errSeg} and \eqref{eq:truncation error}
it is of order $n_s10^{-(p+\delta)}$. Choosing $\delta \ge \log_{10}(n_s)$
then ensures that we obtain $\mathbf{I}(1)$ with precision $p$.
Let us now discuss the case of multiple evaluations.
When evaluating many times in a give region of phase space,
we may take previous evaluations as boundary points. Then, the error on the
boundary value $\Delta(\vec{s}_b)$ is no longer negligible, 
but is given by a previously calculated $\Delta(\vec{s}_e)$ which 
can therefore be calculated by iteration of \eqn{eq:truncation error}.
To guarantee that all evaluations have a precision of at least $p$ digits, we
take $\delta \ge \log_{10}(\bar{n}_s n_{ps})$ where $\bar{n}_s$ is the average
number of segments required for each evaluation and $n_{ps}$ is the number of
phase-space points under consideration. As we will show in section
\ref{sec:physical evaluation}, the average number of segments per path 
is of order two. Therefore, setting $\delta \ge\log_{10}(2 n_{ps})$ ensures that 
all evaluations of $\mathbf{I}(1)$ have a precision of at least $p$.
Finally, we note that, to err on the side of caution, in practice we always take $\delta$ to be an 
integer greater than the estimates we have discussed in this paragraph.

We finish this section by discussing the precision of the boundary conditions
determined with the procedure described in section \ref{sec:boundary conditions}.
Aside from the truncation error associated with the different segments required
to reach each vertex of the pentagon in eq.~\eqref{eq:eu-list}, which can be 
estimated with the same analysis as above, there is a new source of error
associated with the numerical solution of the conditions of eq.~\eqref{eq:boundk}.
This error is harder to determine, but can be estimated a posteriori as follows.
We start by noting that in the Euclidean region the integrals are either purely
real or purely imaginary. This condition is however broken by the fact
that eq.~\eqref{eq:boundk} only holds up to a certain numerical accuracy.
This leads to a residual imaginary part in real integrals, and real part in
imaginary integrals. The magnitude of these residual contributions are then a 
measure of the precision of the zero in eq.~\eqref{eq:boundk}, and can thus
be used to estimate the precision of the determination of the boundary
condition. This estimation is applicable only when a Euclidean region exists,
as it does in our case. Nonetheless, we expect that  similar methods
can be applied to estimate the error for boundary values computed at non-Euclidean points. 
We leave this analysis for future work.

%% file: numerics.tex

\section{Numerical evaluation of master integrals}
\label{sec:numerics}

In this section we illustrate the power of our approach to obtain numerical
values for Feynman integrals. We demonstrate this by computing high-precision
benchmark values, and by evaluating and plotting the integrals over a sub-region
of physical phase space. Finally, we discuss the validation of our numerical
results.

\subsection{High-precision evaluations}
\label{sec:highPrecision}

We first show that our approach can be used
for high-precision evaluation of master integrals.
To this end, we have computed the full set of master integrals in the
$\mzz$, $\zmz$, $\zzz$ and 1-loop topologies at a sample phase-space point in
each of the kinematic regions listed in \tab{tab:regions}. We recall that
these are the different physical regions relevant for vector-boson production in
association with two jets and the Euclidean region. For the Euclidean region, we
used the point $\vec{s}_{\text{eu-1}}$ of eq.~(\ref{eq:eu-list}). For the
physical regions, we chose the following points:
\begin{equation}
\label{eq:p_phys}
\begin{array}{ccccccc}
\vec{s}_{\text{ph-}1} =&\bigg( \dfrac{137}{50}, & -\dfrac{22}{5}, & \dfrac{241}{25}, & -\dfrac{377}{100}, & \dfrac{13}{50}, & \dfrac{249}{50} \bigg),\\ [7pt]
\vec{s}_{\text{ph-}2} =& \bigg( \dfrac{137}{50}, & -\dfrac{22}{5}, & -\dfrac{91}{100}, & -\dfrac{377}{100} ,& -\dfrac{9}{10} ,& \dfrac{249}{50} \bigg),\\[7pt]
\vec{s}_{\text{ph-}3} =& \bigg( \dfrac{137}{50}, & -\dfrac{22}{5}, & -\dfrac{91}{100} ,& \dfrac{13}{50}, & -\dfrac{9}{10} ,& -\dfrac{9}{4}\bigg), \\[7pt]
\vec{s}_{\text{ph-}4} =& \bigg( \dfrac{137}{50}, & \dfrac{357}{50}, & -\dfrac{91}{100}, & \dfrac{241}{25} ,& -\dfrac{9}{10} ,& \dfrac{249}{50}\bigg), \\[7pt]
 \vec{s}_{\text{ph-}5} =&\bigg( \dfrac{137}{50}, & \dfrac{357}{50}, & -\dfrac{91}{100} ,& -\dfrac{161}{100} ,& -\dfrac{9}{10} ,& -\dfrac{9}{4}\bigg), \\[7pt]
 \vec{s}_{\text{ph-}6} =&\bigg( \dfrac{137}{50}, & \dfrac{357}{50}, & \dfrac{13}{50}, & -\dfrac{161}{100} ,& \dfrac{241}{25}, & -\dfrac{9}{4}\bigg). \\
\end{array}
\end{equation}
For each point, we have computed high-precision benchmark values for the integrals with at least
128 digits. 
They can be found in the ancillary files \texttt{anc/f/numIntegrals-f.m} with
\texttt{f}=$\mzz$, $\zmz$, $\zzz$ or 1loop. See also
\texttt{anc/usageExample.m} for more details. As an example, in \tab{tab:targets}
we show the values for the weight-four contribution of the (non-vanishing) 
top integrals in each topology, at the point
$\vec{s}_{\text{ph-}1}$, truncated to fit the confines of the table. 
Our motivation for presenting these results is two-fold. First, it demonstrates
that our approach is able to compute the master integrals to a very high level
of precision. Thus, if one wishes to obtain numerical values for the integrals,
our approach is competitive with an analytic solution of the master integrals in
terms of multiple polylogarithms. 
In section~\ref{sec:numChecks} we will comment on how our evaluation timings compare
to those of a fully analytic solution.
Second, these high-precision benchmark values
can be used as initial conditions in each region when solving the differential
equations.

\begin{figure}[ht] 
\centering
\begin{subfigure}{0.45\textwidth}\centering
\includegraphics[scale=0.55]{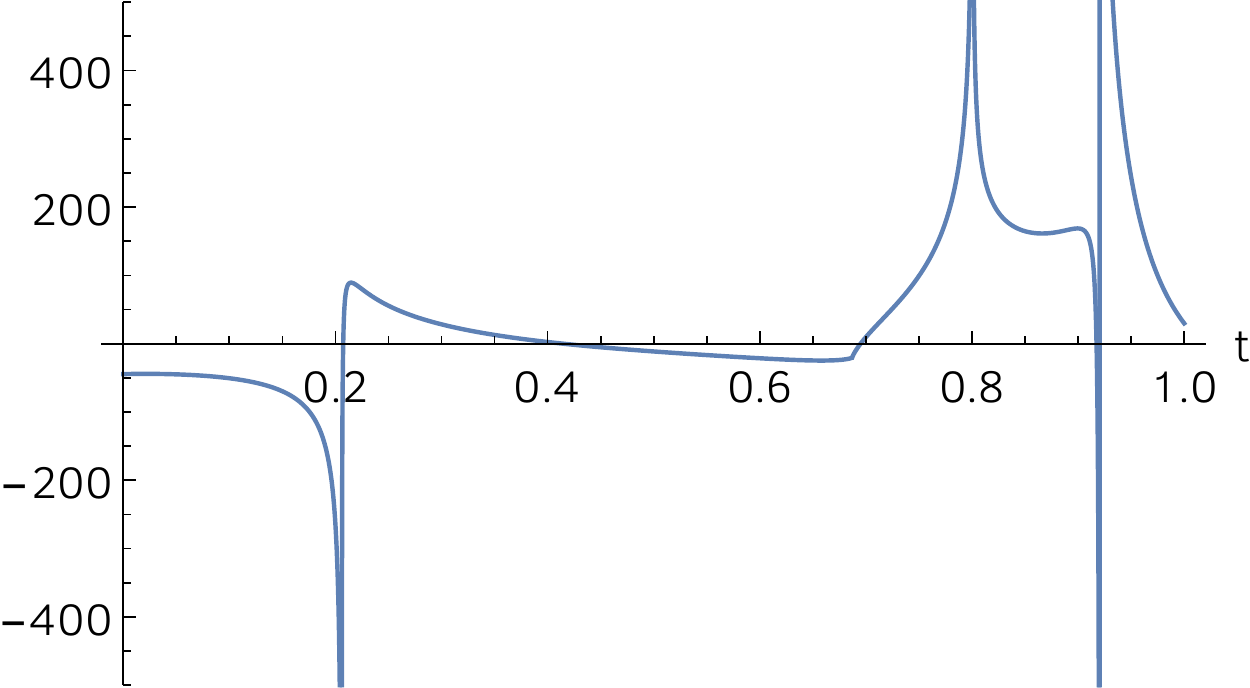}
    \caption{${\rm Re}(I^{(4)}_3)$ of family \mzz.}
\label{fig:mzzRe}
\end{subfigure}\hspace{10mm}
\begin{subfigure}{0.45\textwidth}\centering
\includegraphics[scale=0.55]{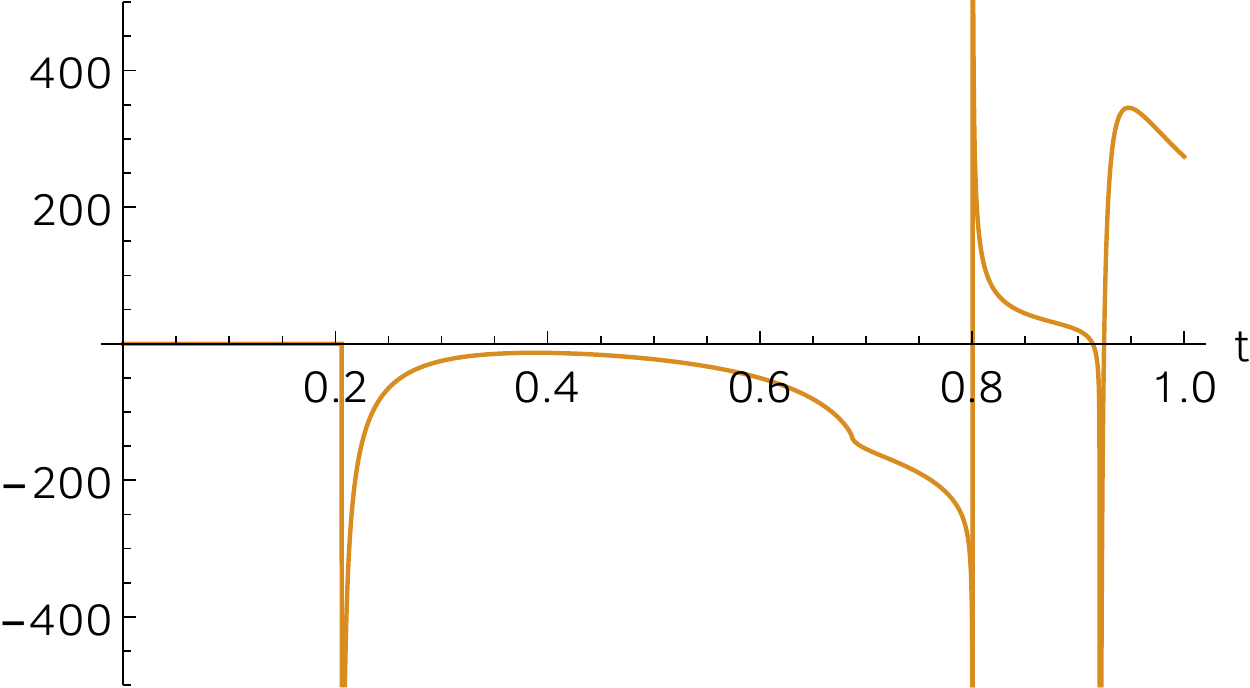}
    \caption{${\rm Im}(I^{(4)}_3)$ of family \mzz.}
\label{fig:mzzIm}
\end{subfigure} \\[1cm]
\begin{subfigure}{0.45\textwidth}\centering
\includegraphics[scale=0.55]{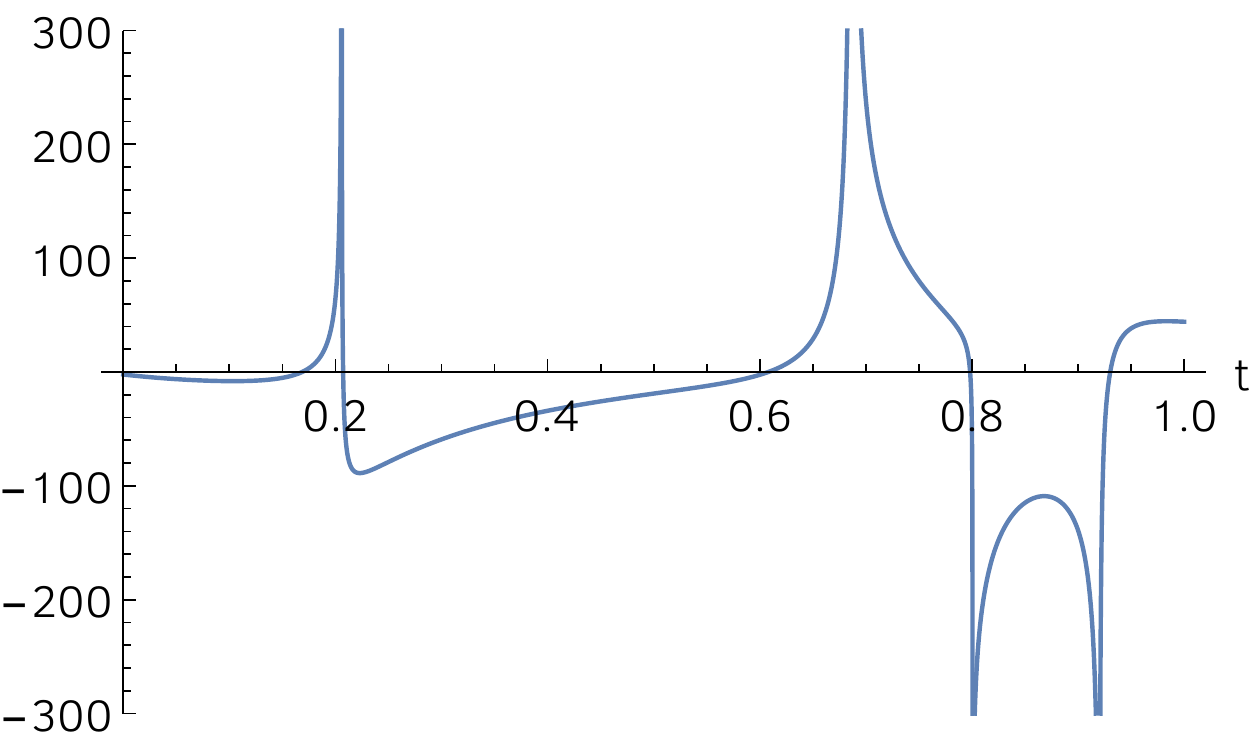}
    \caption{${\rm Re}(I^{(4)}_3)$ of family \zmz.}
\label{fig:zmzRe}
\end{subfigure}\hspace{10mm}
\begin{subfigure}{0.45\textwidth}\centering
\includegraphics[scale=0.55]{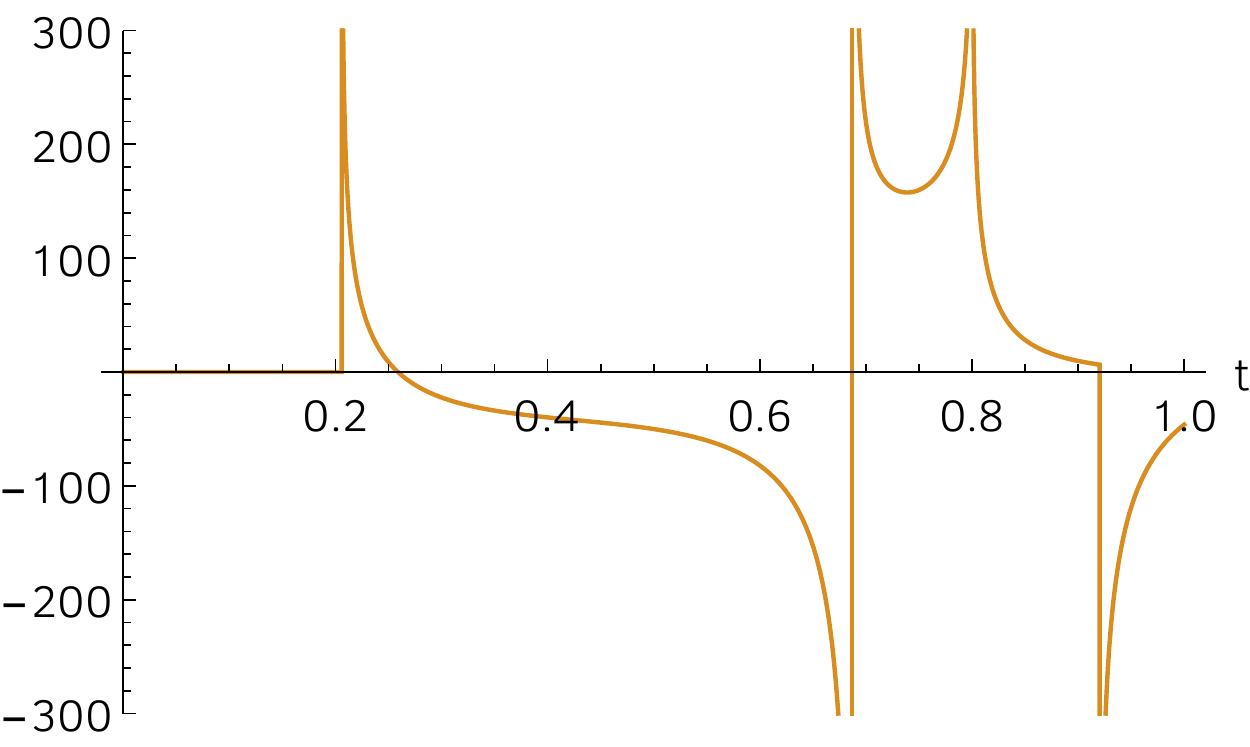}
    \caption{${\rm Im}(I^{(4)}_3)$ of family \zmz.}
\label{fig:zmzIm}
\end{subfigure} \\[1cm]
\begin{subfigure}{0.45\textwidth}\centering
\includegraphics[scale=0.55]{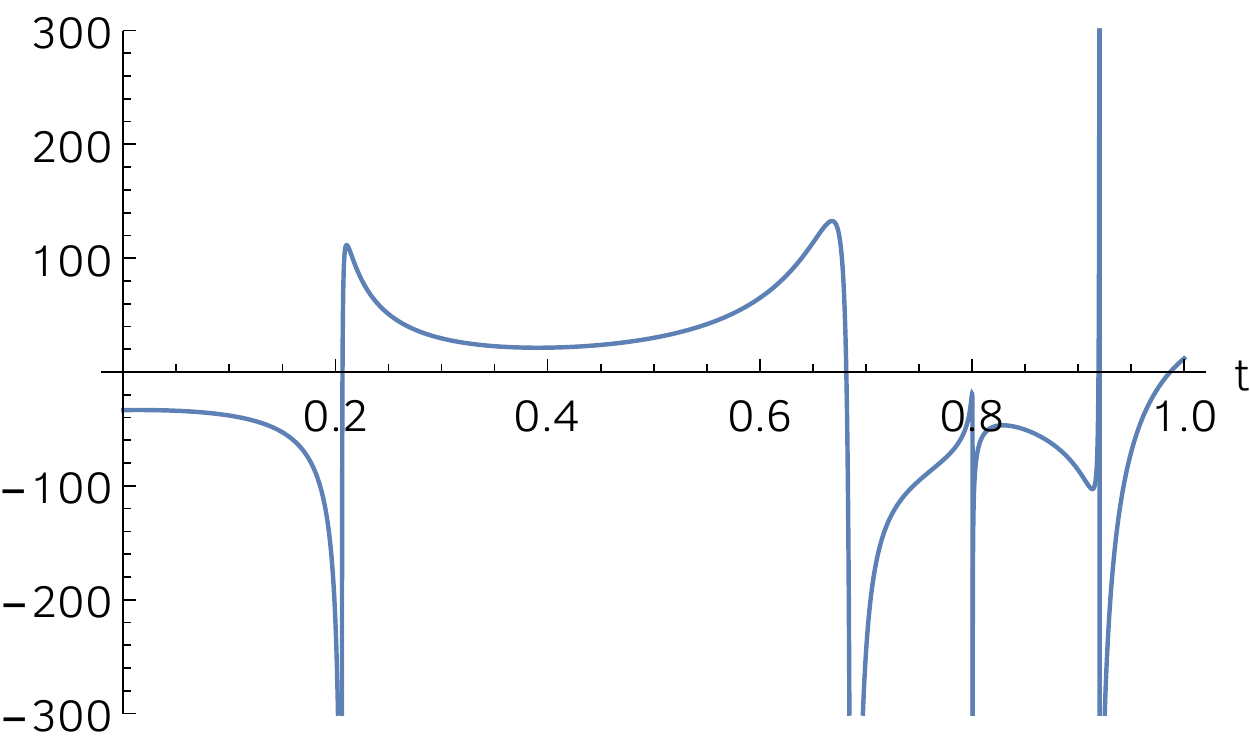}
    \caption{${\rm Re}(I^{(4)}_3)$ of family \zzz.}
\label{fig:zzzRe}
\end{subfigure}\hspace{10mm}
\begin{subfigure}{0.45\textwidth}\centering
\includegraphics[scale=0.55]{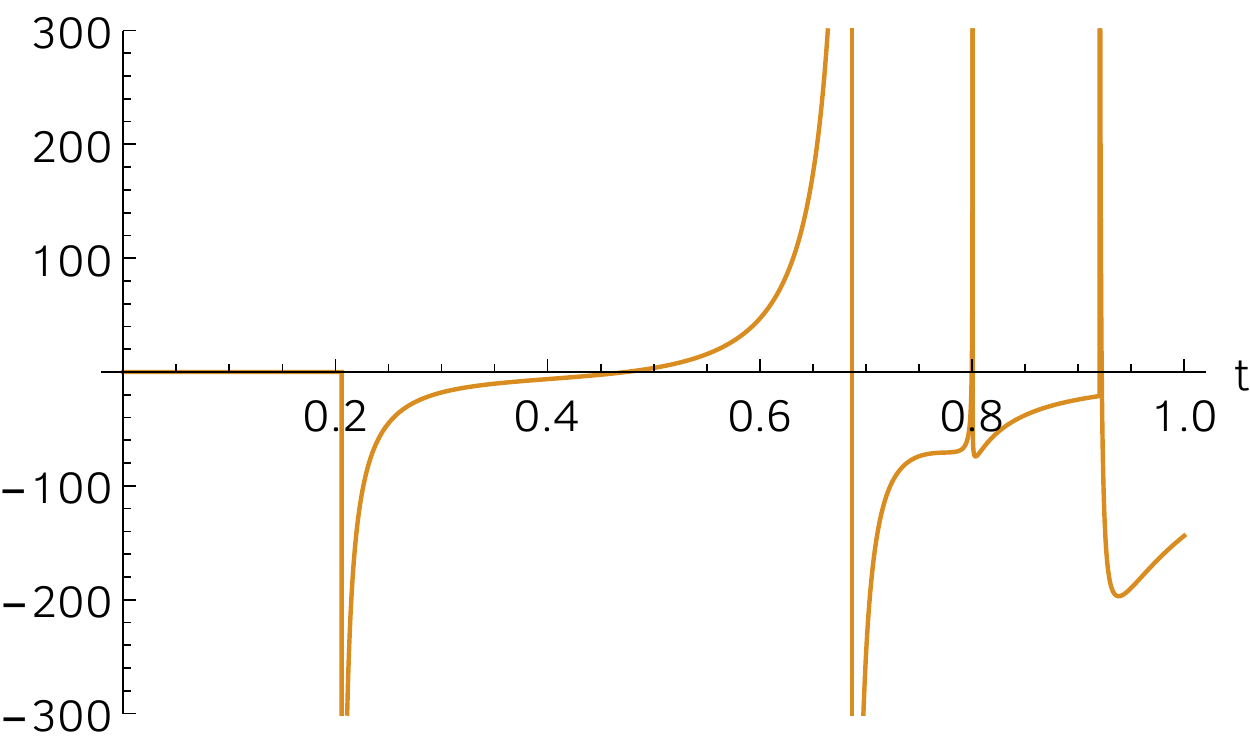}
    \caption{${\rm Im}(I^{(4)}_3)$ of family \zzz.}
\label{fig:zzzIm}
\end{subfigure}

\caption{Integrals evaluated over a path from $\vec{s}_{\text{eu-1}}$ to $\vec{s}_{\text{ph-1}}$.  
Real and imaginary parts of the integrals are
displayed separately. The visible singularities and discontinuities are associated to the physical
thresholds for which analytic continuation is required. 
The kink in plots (\ref{fig:mzzRe}) and (\ref{fig:mzzIm}) near $t=0.7$ 
corresponds to a threshold singularity which locally behaves as 
$(t-t_k) \left[\log (t-t_k)\right]^3$ .}
\label{fig:Line}
\end{figure}

\begin{figure}[ht] 
\begin{subfigure}{0.45\textwidth}\centering
\includegraphics[scale=0.55]{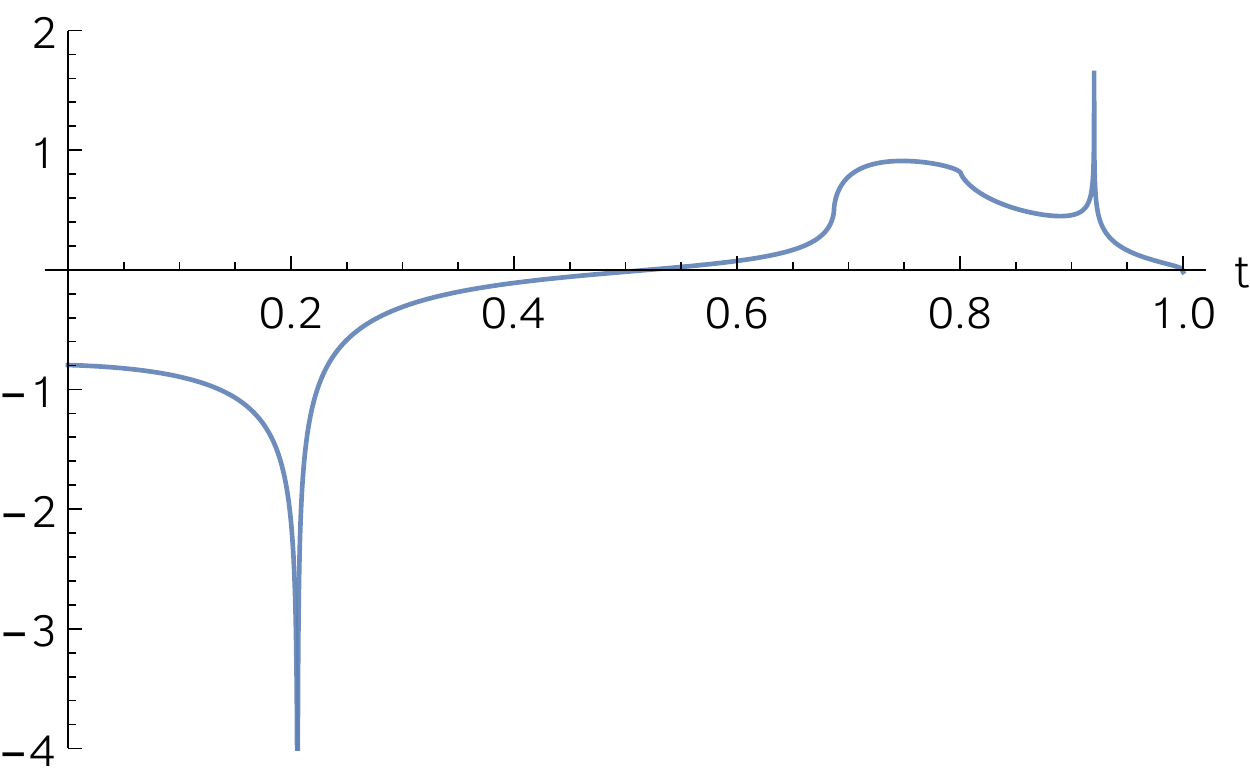}
    \caption{${\rm Re}(I^{(4)}_1)$ of 1-loop family.}
\label{fig:1loopRe}
\end{subfigure}\hspace{10mm}
\begin{subfigure}{0.45\textwidth}\centering
\includegraphics[scale=0.55]{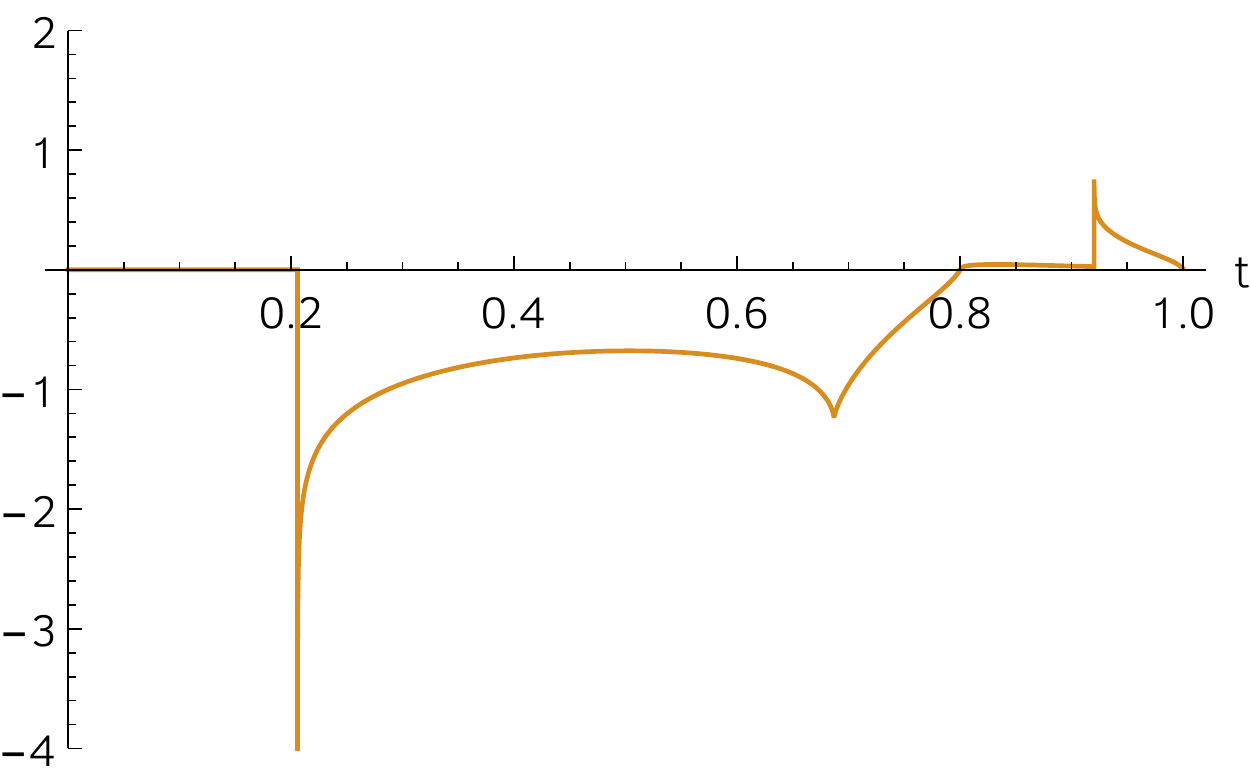}
    \caption{${\rm Im}(I^{(4)}_1)$ of 1-loop family.}
\label{fig:1loopIm}
\end{subfigure} 
\caption{1-loop pentagon at weight 4, evaluated over a path from 
$\vec{s}_{\text{eu-1}}$ to $\vec{s}_{\text{ph-1}}$.
Real and imaginary parts of the integrals are
displayed individually. The visible kinks, 
singularities and discontinuities are associated with the physical
thresholds for which analytic continuation is required.}
\label{fig:Line 1-loop}
\end{figure}

We end by briefly describing how the high-precision results were obtained.
First, the results at the Euclidean point $\vec{s}_{\text{eu-1}}$
were obtained with the procedure described in section~\ref{sec:boundary conditions}, 
that is by requiring that the integrals have no spurious branch cuts in the 
Euclidean region. The high-precision evaluations in the physical regions
were then performed by transporting the solution from $\vec{s}_{\text{eu-1}}$
to the different physical points, using the approach described 
in section~\ref{sec:localSolution}. To illustrate this procedure, 
in figs.~\ref{fig:Line} and \ref{fig:Line 1-loop} we plot the
same functions that we tabulated in Tab.~\ref{tab:targets} along
the path connecting $\vec{s}_{\text{eu-1}}$ and $\vec{s}_{\text{ph-1}}$.
Since $\vec{s}_{\text{ph-1}}$ has four positive Mandelstam variables, we
expect to see the effect of four logarithmic physical thresholds. 
We indeed observe non-trivial behaviors at four points 
(at $t\sim 0.2$, $t\sim 0.7$, $t\sim 0.8$ and $t\sim 0.92$), in the form
of divergences or kinks in the curves of both the real and imaginary parts.

\begin{table}[t]
\centering
 \begin{tabular}{|c|c|c|l|} 
\hline
\multirow{2}{*}{$I_3^{(4)}$}&\multirow{2}{*}{zzz}&Re&$+11.908529680841593329567378444341231494621544817813763$\\
&&Im&$-143.83838235097336513553728991658286648264414416047763$\\\hline
\multirow{2}{*}{$I_3^{(4)}$}&\multirow{2}{*}{zmz}&Re&$+44.162165744735300867233118554182853322209473851043647$\\
&&Im&$-46.218746133850339969944403077556678434364686840750803$\\\hline
\multirow{2}{*}{$I_3^{(4)}$}&\multirow{2}{*}{mzz}&Re&$+29.802763651793108812023893217593351307350121722845006$\\
&&Im&$+273.86627846266515113913295225572416419016316389639992$\\\hline
\multirow{2}{*}{$I_1^{(4)}$}&\multirow{2}{*}{1-loop}&Re&$-12.997557921493867410660219778141561158754063252253784$\\
&&Im&$-34.691238289230523215562386582080833547255858602481034$\\\hline
  \end{tabular}
\caption{Sample numerical values for weight-four integrals evaluated at the physical 
    point $\vec{s}_{\text{ph-}1}$ defined in eq~(\ref{eq:p_phys}). 
    For \texttt{f}=$\mzz$, $\zmz$ and $\zzz$, $I_3$ denotes the penta-box integral
    with the insertions $\mathcal N^{(3)}_{\textrm{pb},\texttt{f}}$ given in eqs.~\eqref{eq:pbmzz}, 
    \eqref{eq:pbzmz} and \eqref{eq:pbzzz}
    respectively (the other two insertions for each penta-box are only non-zero starting at weight 5).
    For the one-loop topology, we quote the result for the one-loop pentagon with the insertion
    given in the ancillary files. The results are truncated to fit the confines
    of the table. Results with at least 128 digits of precision can be found in
    the ancillary files \texttt{anc/f/numIntegrals-f.m}.}
 \label{tab:targets}
\end{table}

\subsection{Integral evaluation over physical phase space}
\label{sec:numChecks}

Having in mind future phenomenological applications, it is not sufficient to
have precise evaluations at a single phase-space point. One also needs to have
efficient and stable evaluations across phase-space that can be used for Monte
Carlo phase-space integration. In the following, we describe how this can be
achieved with our approach.

Let us begin by elaborating on the strategy we follow when evaluating master
integrals over a large set of phase-space points. In this context it is possible
to improve the average evaluation time of the integrals by exploiting previous
evaluations. In our approach, when a vector of master integrals has already been
computed over a set of phase-space points, any of these previous evaluations can
used as a boundary point for the next integral evaluation. As such, it is
fruitful to pick the element in the set of available points that minimizes the
evaluation time which, considering the analysis of the previous sections,
depends linearly on the number of segments of the contour that connects the
boundary point to the target point.
It is therefore wise to search for a boundary point which decreases the number
of segments.
In order to find the optimal boundary point, in principle it would be necessary to consider 
the full set of available points. However, in the context of a phase-space Monte
Carlo integration this set may be prohibitively large and 
this analysis would then undermine the aim
of decreasing the average evaluation time. It is however natural to expect
that the optimal boundary point will be close to the target point in the space of Mandelstam variables.
We thus consider only the $k$ nearest points and choose the one that minimizes the
number of segments. In practice we find that $k=10$ gives an average speed-up of
40\% in
comparison to the $k=1$ case. This can be justified by noting that the
optimal boundary point is not in general the nearest one, as the number of
segments also depends on the configuration of the singular points.
An important feature of this approach is that, with each  new evaluation, the pool of
available points increases and so the average number of segments, and therefore the
evaluation time, required for each new evaluation decreases. 

\begin{figure}
\begin{center}
\includegraphics[scale=0.75]{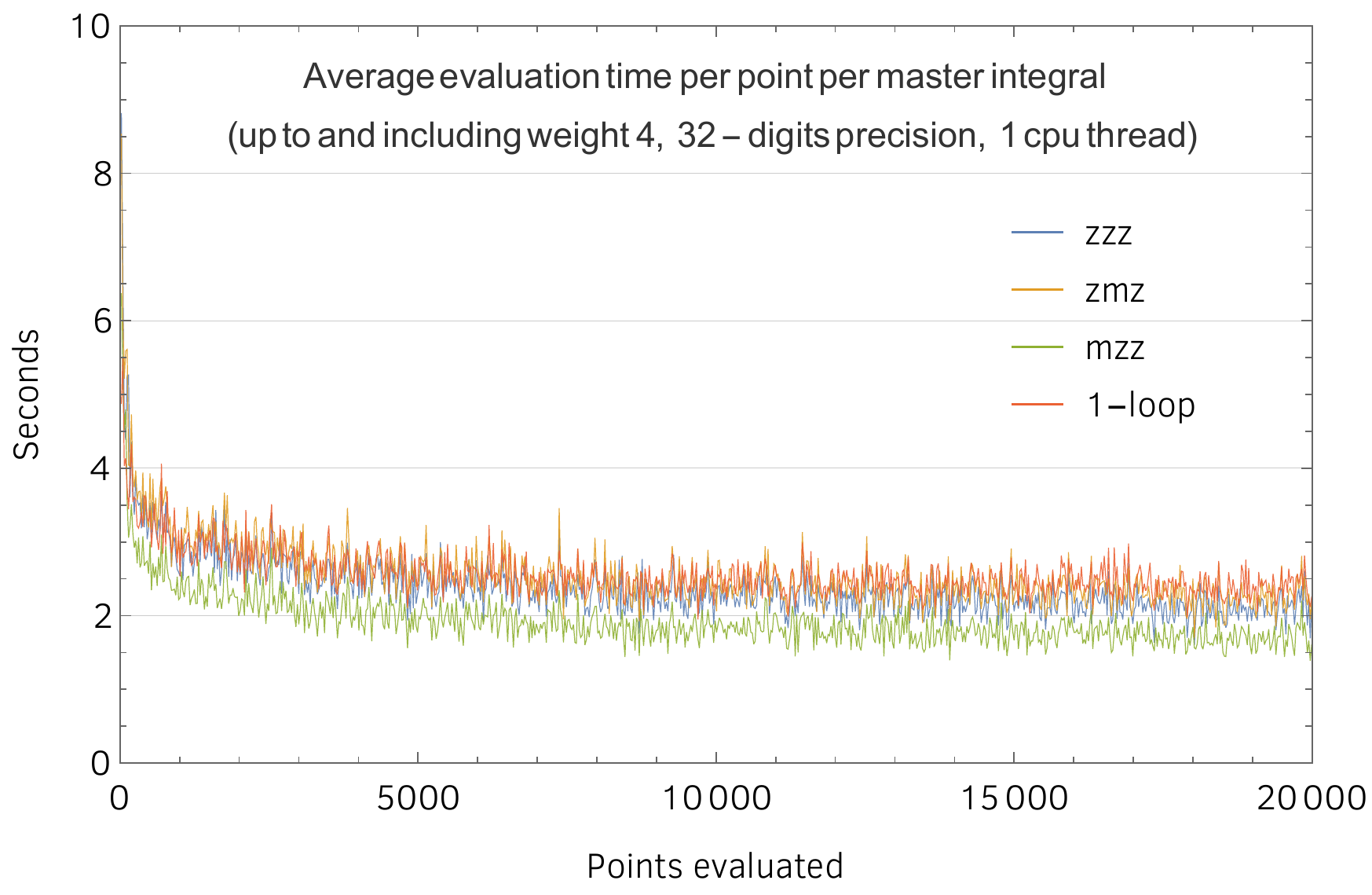}
\end{center}
\caption{Timing study over a set of 20k Monte-Carlo phase-space points in a physical scattering region.
The average evaluation time per master integral is given by the total evaluation time divided by the 
number of master integrals of each topology (see also table~\ref{tab:timings parameters}). 
Each point of the plot is obtained by averaging the timing of 25 phase-space points.}
\label{fig:timing}
\end{figure}

In order to demonstrate these features, we generated 20k sample phase-space points 
corresponding to vector-boson production at the LHC with phase-space cuts of ref.~\cite{Bern:2013zja}.
(We used the \Sherpa\ Monte-Carlo program~\cite{Gleisberg:2008ta} to generate the phase-space points.) 
The particles with momenta $p_2$ and $p_3$ are taken in the initial state.
As a seed evaluation in this physical region, we took the high-precision value
discussed in section \ref{sec:highPrecision}.
We evaluated the complete set of master integrals of the three two-loop topologies and of the 
one-loop topology 
over the 20k phase-space points and, for each evaluation, recorded both the number 
of segments per contour and the evaluation time. 
Figure~\ref{fig:timing} shows the average evaluation time per master integral as a 
function of the number of points evaluated. 
The figure corresponds to evaluations with 32-digit precision on a single CPU thread. 
As expected, the evaluation time decreases as the number of points increases, and we observe
that the evaluation time stabilizes after about 10k points. 
The asymptotic timings, along with other evaluation parameters for 32- and 16-digit evaluations, 
are presented in \tab{tab:timings parameters}. 
The number of segments per contour stabilizes around two for all the families. 
In the language of section \ref{sec:precision}, we performed the evaluations
with an offset of $\delta=6$. We thus obtained a numerical precision of at least 36
for the 32-digit run, and 20 for the 16-digit run.

Finally, we compare our timings with a fully analytic
solution of the integrals. For this, we focus
on the $\mzz$ topology and compare our evaluation timings with those of
the analytic solution of ref.~\cite{Papadopoulos:2015jft}. We point out
that it is hard to make this comparison meaningful since
the two strategies are very different. In the spirit of this section, we
thus choose to compare our asymptotic timing with the timing of a single
evaluation of the expressions of ref.~\cite{Papadopoulos:2015jft}, since
these are the relevant numbers when using the two implementations for
e.g.~Monte-Carlo phase-space integration. We find that in our approach the timings are
stable in each of the physical regions of Tab.~\ref{tab:regions}.
That is, in each region, to evaluate all master integrals of the $\mzz$ topology at a phase-space
point takes $\sim75s$ with 16 digits, $\sim150s$ with 32 digits and $\sim1150s$ with 128 digits.
We then evaluated the polylogarithms in the expressions of 
ref.~\cite{Papadopoulos:2015jft} at the six phase-space
points of eq.~\eqref{eq:p_phys} using the \texttt{GiNaC} implementation of \cite{Vollinga:2004sn}
on a single CPU core. 
We observed a very large fluctuation of the evaluation
times across the six phase-space points, which ranges from $42s$ to $2800s$ for 16 digits,
from $\sim50s$ to $\sim 4800$ for 32 digits, and from $\sim120s$ to $\sim22200s$ 
for 128 digits (we note that obtaining
integrals with 16, 32 or 128 digit precision requires running \texttt{GiNaC} 
with a slightly higher number of digits). Whilst we stress
that given the differences in the approaches the timing comparisons are not
straightforward, we conclude that our approach is competitive with a fully analytic
solution of the results, with a more stable behaviour across phase space.

\begin{table}
\centering
 \begin{tabular}{|c| c | c | c | c | c | } 
\hline
          & Family &  MI's  & time per MI (s) & total time (s)&truncation order\\\hline
\multirow{4}{*}{32 digits}&zzz      & 86 & 2.08 & 179&\multirow{4}{*}{$70<n_k<140$}\\  
            &zmz     & 75 &  2.24 & 168& \\   
            & mzz   & 74  & 1.69 & 125& \\
            &1-loop & 13& 2.38 & 31& \\\hline
\multirow{4}{*}{16 digits}&zzz     & 86 & 1.10 & 94&\multirow{4}{*}{$40<n_k<80$}\\  
            &zmz     & 75 &  1.15 & 86 &\\   
            & mzz   & 74  & 0.88 & 65 &\\
            &1-loop & 13& 1.69 & 22 &\\\hline
\end{tabular}
    \caption{Characteristics of master-integral (MI) evaluation over 
    20k phase-space points on a single CPU thread. 
    The timing (in seconds) is given for 32-digit precision and 16-digit precision 
    evaluations. 
    The evaluation times in the fourth column correspond to the asymptotic timings,
    computed by averaging over the last 2k phase-space points. 
    We also give the truncation order of the series expansions (see section \ref{sec:precision}).}
 \label{tab:timings parameters}
\end{table}

\subsection{Plots over physical phase space}
\label{sec:physical evaluation}

A further way to demonstrate the efficiency and the numerical stability of our approach 
is to produce plots of the integrals over a sub-region of physical phase space. 
Specifically, we present plots of the highest non-vanishing integrals for each family
(i.e., the same integrals for which we gave high-precision values in \tab{tab:targets}), 
over a two-dimensional sub-region of the physical region relevant for $W+2$-jet production at the LHC,
where the particles of momenta $p_2$ and $p_3$ are in the initial state. 
The sign of the independent Mandelstam variables in this phase-space region are 
given in \tab{tab:regions}. We fix the following four variables 
as,\footnote{These values correspond to a rationalization of one of the physical
points obtained from \Sherpa{} that were used in the previous section. Mandelstam
variables are normalized such that $\offShellScale$, the vector-boson mass, is set to 1.}
\begin{align}
\begin{split}
\label{eq:fixed invariants}
&\offShellScale=1\,,\qquad s_{12}=-\frac{154120668029}{42334495831}=-3.64055\ldots\,,\\
&s_{15}=\frac{1619721713191}{211672479155}=7.65202\,,\qquad
s_{45}=\frac{761855318631}{42334495831}=17.9961\ldots.
\end{split}
\end{align} 
The remaining Mandelstam variables are $s_{23}>0$ and $s_{34}<0$, but they do
not take arbitrary values if they are to correspond to a physical phase-space
point in the region under consideration. We shall now characterize this region
explicitly. We base our analysis on the observation that, in this region, the
Gram matrix $G(p_1,p_2,p_3,p_4)$ has three negative
eigenvalues~\cite{Byers:1964ryc}. We note that this is a stronger condition than simply
requiring the determinant to be negative.

\begin{figure}
\begin{center}
\includegraphics[scale=0.7]{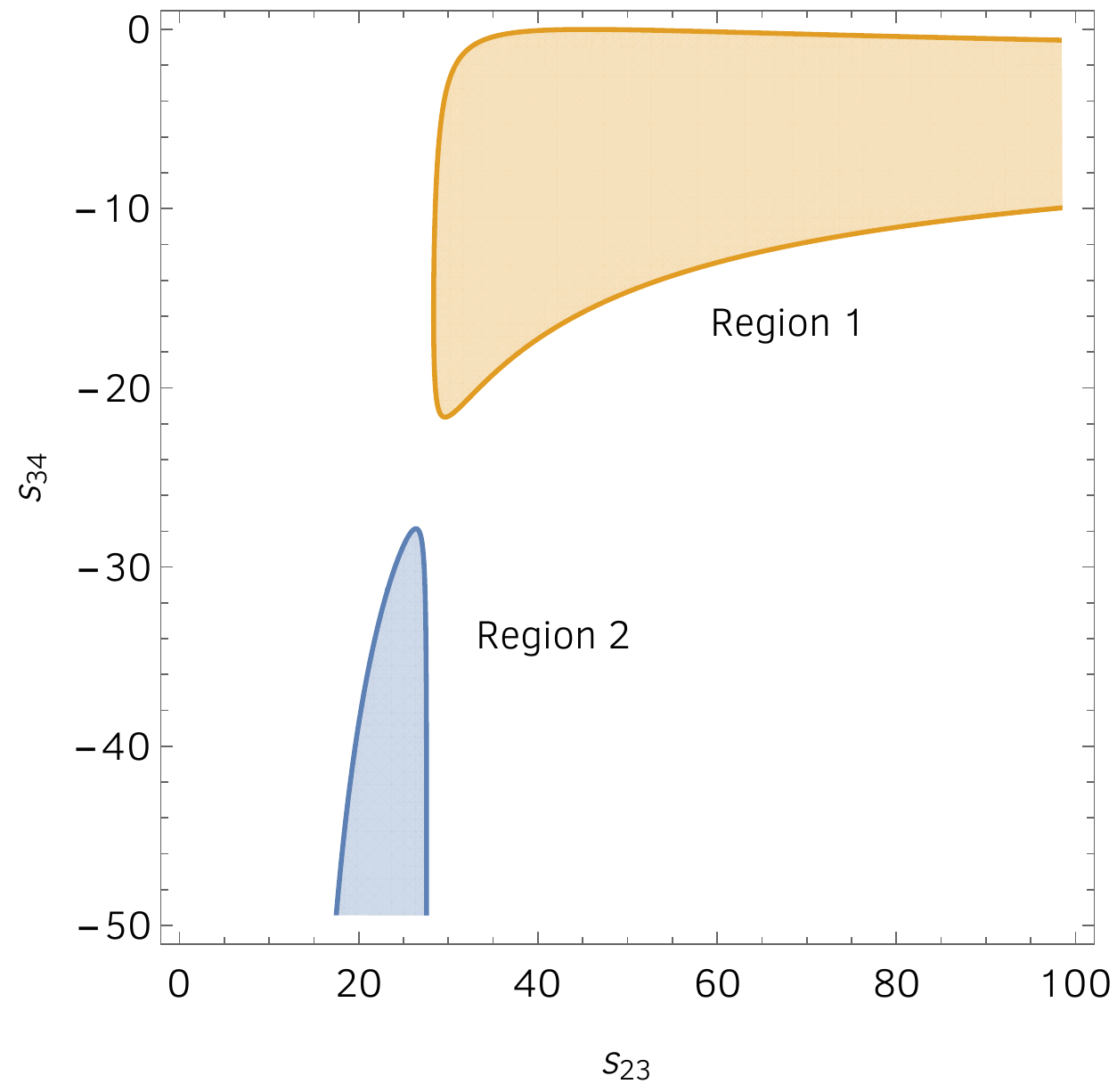}
\end{center}
\caption{Regions where $\det \,G(p_1,p_2,p_3,p_4)<0$ in the $s_{23}>0$ and $s_{34}<0$ quadrant
and under the conditions of eq.~\eqref{eq:fixed invariants}. Region 1 is unbounded
as $s_{23}\to\infty$, and Region 2 as $s_{34}\to-\infty$.}
\label{fig:regions}
\end{figure}

In \fig{fig:regions} we depict the two disconnected 
regions in the $s_{23}>0$ and $s_{34}<0$ quadrant where
$\det \,G(p_1,p_2,p_3,p_4)<0$. To determine the boundary of these regions,
we first solve $\text{det} \,G(p_1,p_2,p_3,p_4)=0$ with respect to $s_{34}$,
finding
\begin{equation}
\label{eq:sol s34}
s_{34}^{(\pm)}=\frac{\mathcal{N}\pm \sqrt{\Delta}}{\mathcal{D}}\,.
\end{equation}
These $s_{34}^{(\pm)}$ are functions of $s_{23}$.
The discriminant is,
\begin{equation}
\Delta=s_{23} s_{45} \left(s_{15} s_{45}+s_{12} \left(s_{12}+s_{23}-s_{45}\right)-\offShellScale{} \left(s_{12}-s_{45}\right)\right) \left(\left(s_{15}-s_{23}\right) \left(s_{15}-\offShellScale{}\right)\right),
\end{equation}
while the polynomials $\mathcal{N}$ and $\mathcal{D}$ are,
\begin{align}
\mathcal{N}&=2 \left(s_{15} s_{45}^2-s_{45} \left(s_{15} \left(\offShellScale{}+s_{12}\right)+s_{23} 
\left(s_{12}+s_{15}-2 \offShellScale{}\right)\right)+s_{12} \left(s_{15}-s_{23}\right) 
\left(\offShellScale{}-s_{23}\right)\right),\nonumber\\
\mathcal{D}&=2 \left(\offShellScaleSquared{}+\left(s_{23}-s_{45}\right){}^2-2 \offShellScale{}\left(s_{23}+s_{45}\right)\right).
\end{align}
The requirement that the $s_{34}^{(\pm)}$ be real means that $\Delta$ must be positive. This
gives a condition on the values of $s_{23}$,
\begin{equation}
  0< s_{23}<s_{23}^{(1)}\,,\quad \text{   or  } \quad s_{23}^{(2)}< s_{23}<\infty\,,
\end{equation}
where $s_{23}^{(1)}$ and $s_{23}^{(2)}$ are the non-trivial solutions to $\Delta=0$,
\begin{equation}
\;s_{23}^{(1)}=-\frac{\left(s_{12}-s_{45}\right) \left(s_{12}-\offShellScale{}\right)}{s_{12}}\,,\qquad
s_{23}^{(2)}=\frac{s_{15} \left(s_{15}+s_{45}-\offShellScale{}\right)}{s_{15}-\offShellScale{}}\,.
\end{equation}
These two intervals for $s_{23}$, together with eq.~\eqref{eq:sol s34}, 
correspond to the  two regions in \fig{fig:regions}. 
However, at this stage we have not yet imposed
the condition that three eigenvalues of the Gram matrix should be negative. This excludes
one of the regions, Region 2 in \fig{fig:regions}, 
leaving us with the relevant phase-space region, Region 1 in \fig{fig:regions}, 
defined as,
\begin{equation}
\label{eq:2dslice}
R=\left\{ \left(s_{23},s_{34}\right)
 \;|\; s_{23}^{(2)}< s_{23}<\infty,\; 
s_{34}^{(-)}<s_{34}<s_{34}^{(+)}\right\}.
\end{equation}
Let us note once more that the $s_{34}^{(\pm)}$ are functions of $s_{23}$, while $s_{23}^{(2)}$ 
is a constant determined by the values given in eq.~\eqref{eq:fixed invariants}.\

We expect the master integrals to have interesting behaviors
near the branch points that we can approach in this region.  
These are $s_{34}\to0$ and $s_{23}\to\infty$. Given the constraint of remaining
in the region $R$ these correspond not to dimension 1 surfaces but to the points
\begin{align}\label{eq:singularPoints}
  P_1=\left\{s_{23}=\frac{s_{15}(s_{12}-s_{45})}{s_{12}}\,, s_{34}=0\right\}
  \quad\textrm{and}\quad
  P_2=\left\{s_{23}=\infty\,, s_{34}=s_{12}\right\}\,.
\end{align}
In practical applications, however, we are not interested in approaching the point
$s_{23}\to\infty$. Instead, we introduce a cut-off at
$s_{23}=(13\;\text{TeV}/80\;\text{GeV})^2$, i.e., at
the LHC center-of-mass energy divided by a scale similar to the $W$-boson mass,
which we use as the regularization scale.

\begin{figure}[h!] 
\begin{subfigure}{0.45\textwidth}\centering
\includegraphics[scale=0.50]{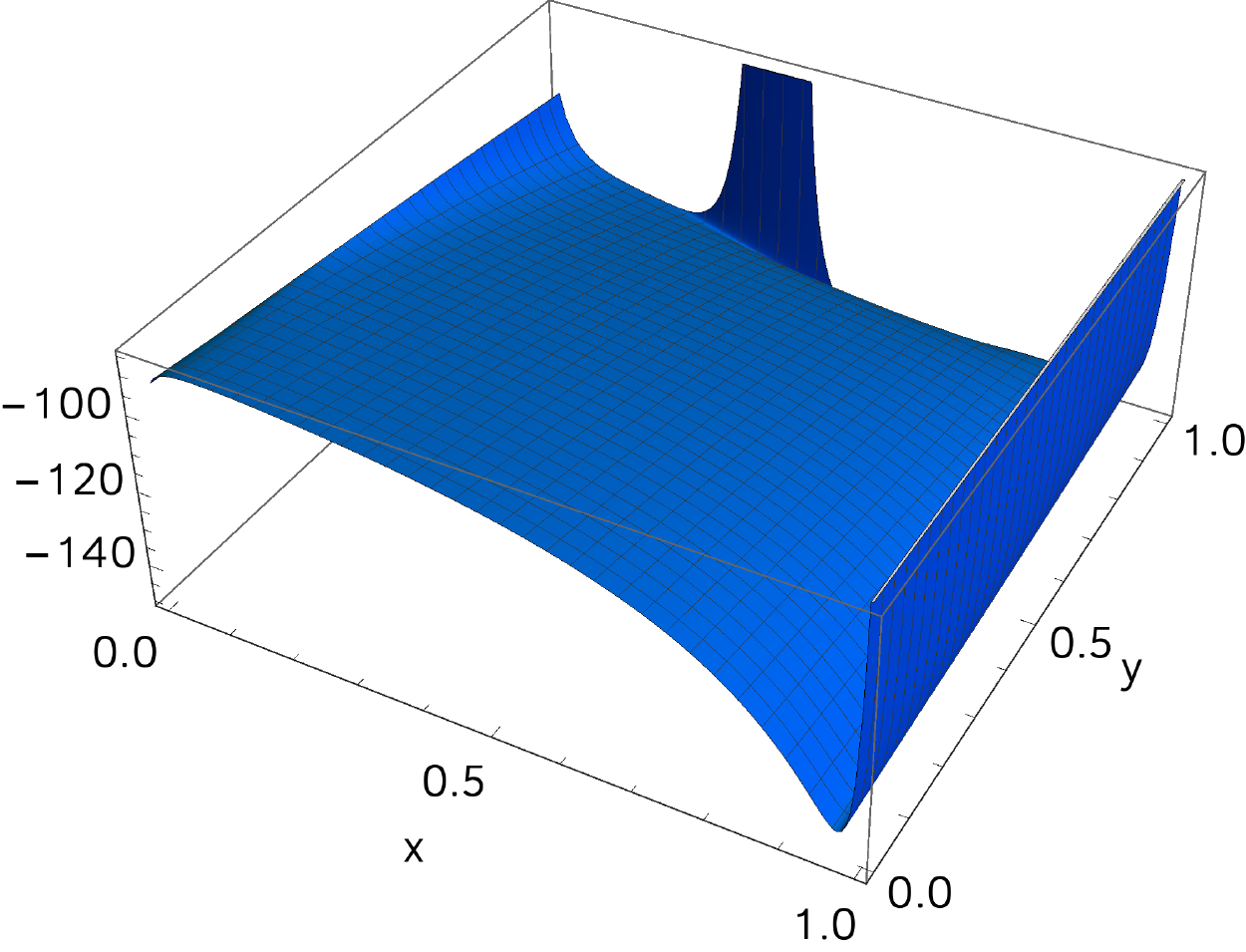}
    \caption{${\rm Re}(I^{(4)}_3)$ of \mzz{} topology.}
\label{fig:mzzRe2d}
\end{subfigure}\hspace{10mm}
\begin{subfigure}{0.45\textwidth}\centering
\includegraphics[scale=0.50]{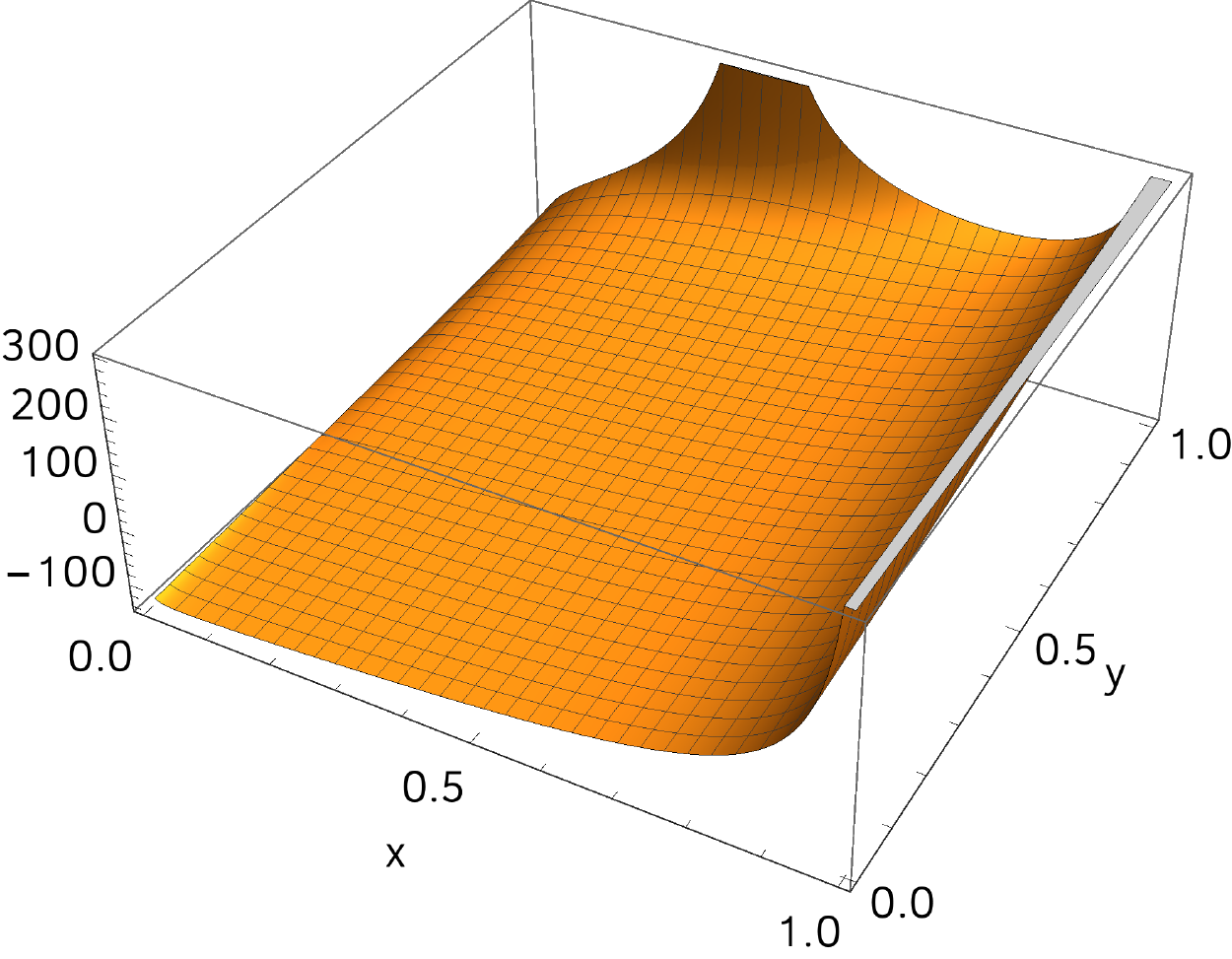}
    \caption{${\rm Im}(I^{(4)}_3)$ of \mzz{} topology.}
\label{fig:mzzIm2d}
\end{subfigure}\\[7mm]
\begin{subfigure}{0.45\textwidth}\centering
\includegraphics[scale=0.50]{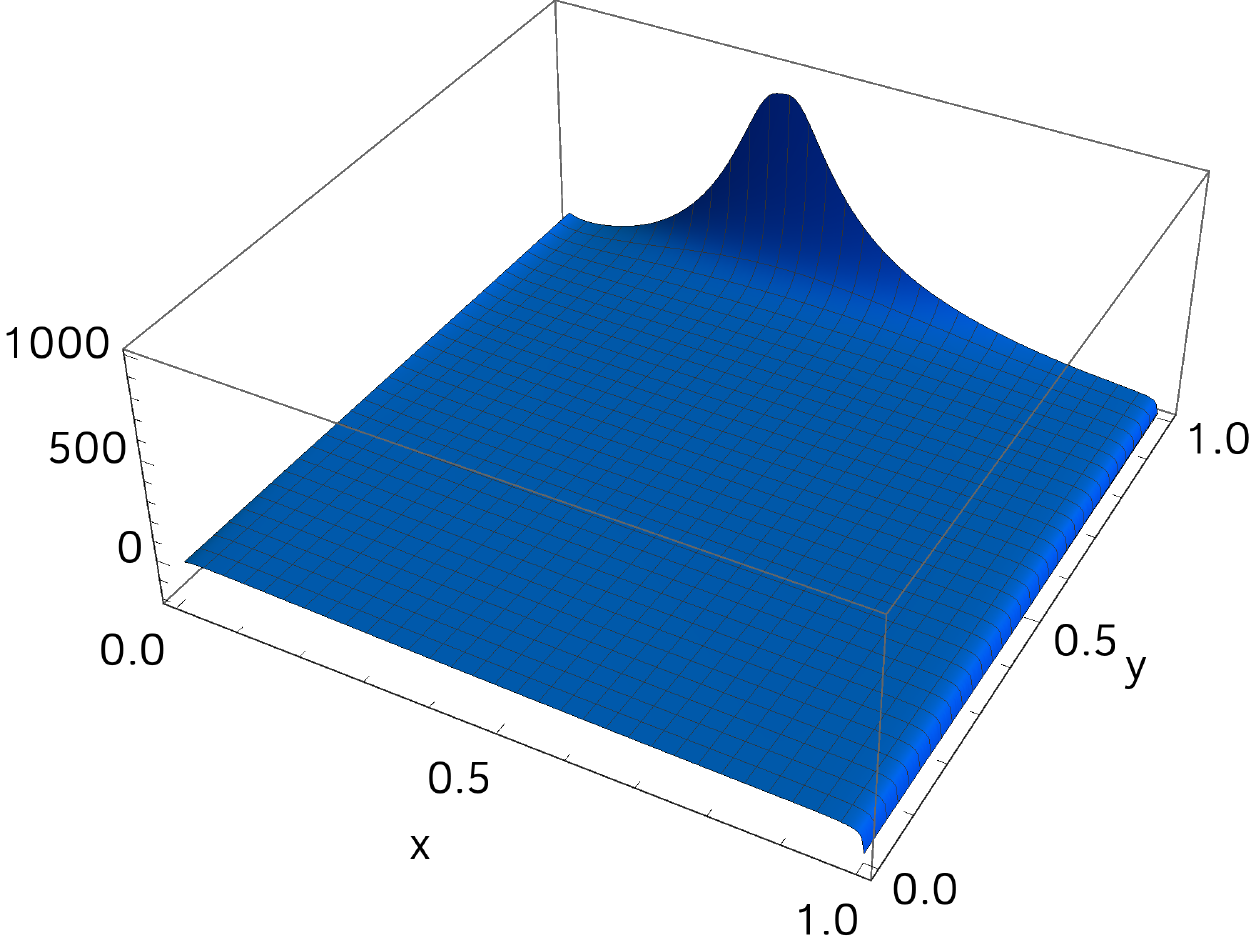}
    \caption{${\rm Re}(I^{(4)}_3)$ of \zmz{} topology.}
\label{fig:zmzRe2d}
\end{subfigure}\hspace{10mm}
\begin{subfigure}{0.45\textwidth}\centering
\includegraphics[scale=0.50]{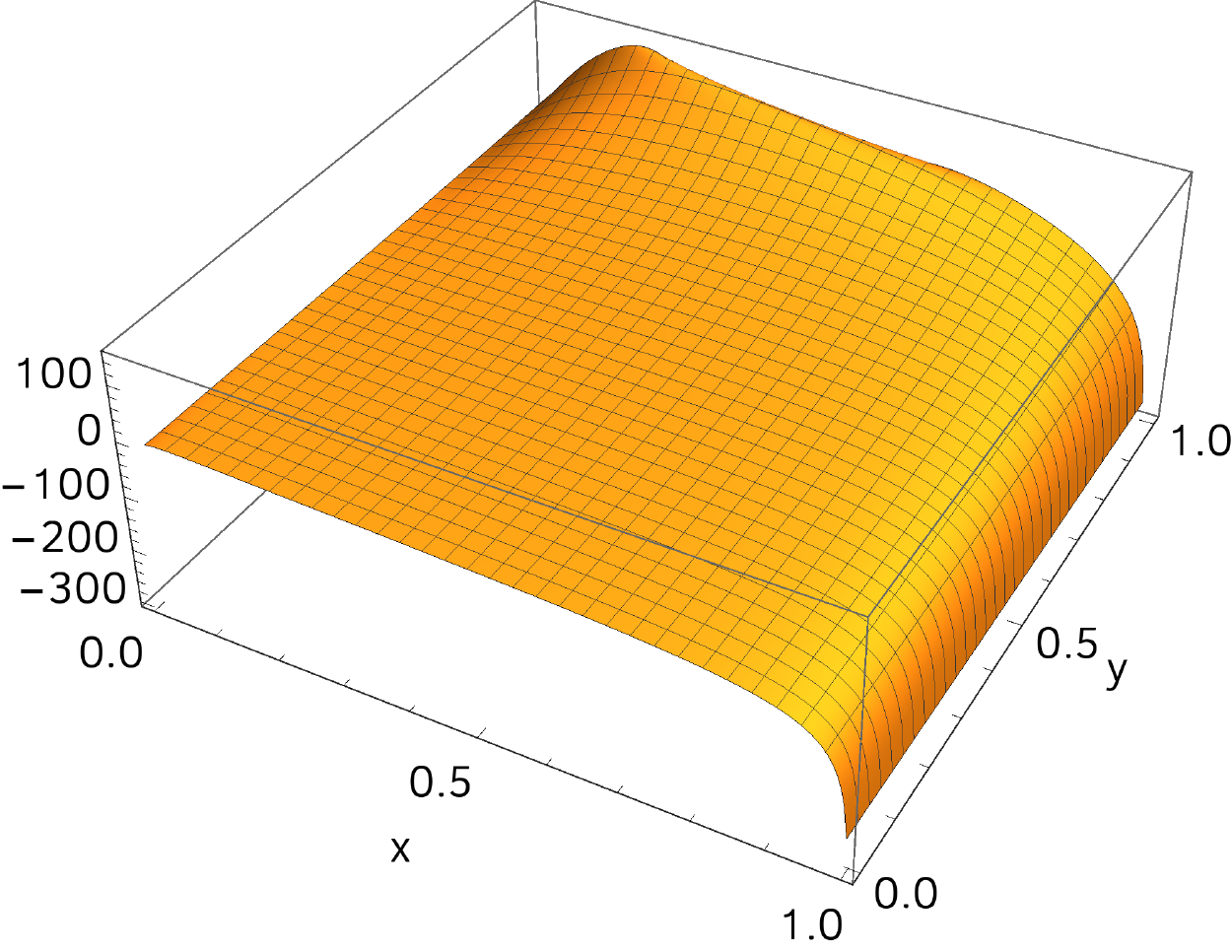}
    \caption{${\rm Im}(I^{(4)}_3)$ of \zmz{} topology.}
\label{fig:zmzIm2d}
\end{subfigure}\\[7mm]
\begin{subfigure}{0.45\textwidth}\centering
\includegraphics[scale=0.50]{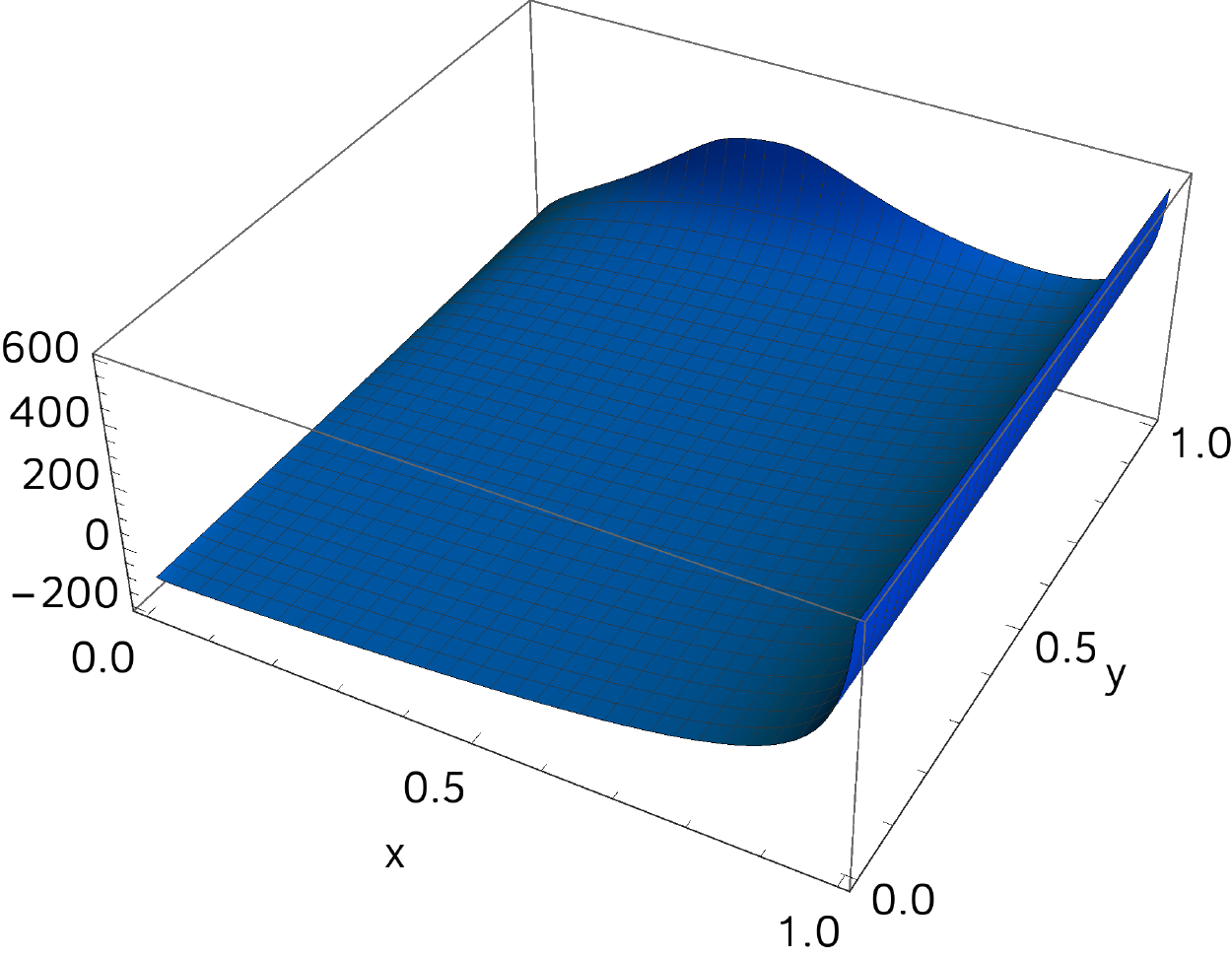}
    \caption{${\rm Re}(I^{(4)}_3)$ of \zzz{} topology.}
\label{fig:zzzRe2d}
\end{subfigure}\hspace{10mm}
\begin{subfigure}{0.45\textwidth}\centering
\includegraphics[scale=0.50]{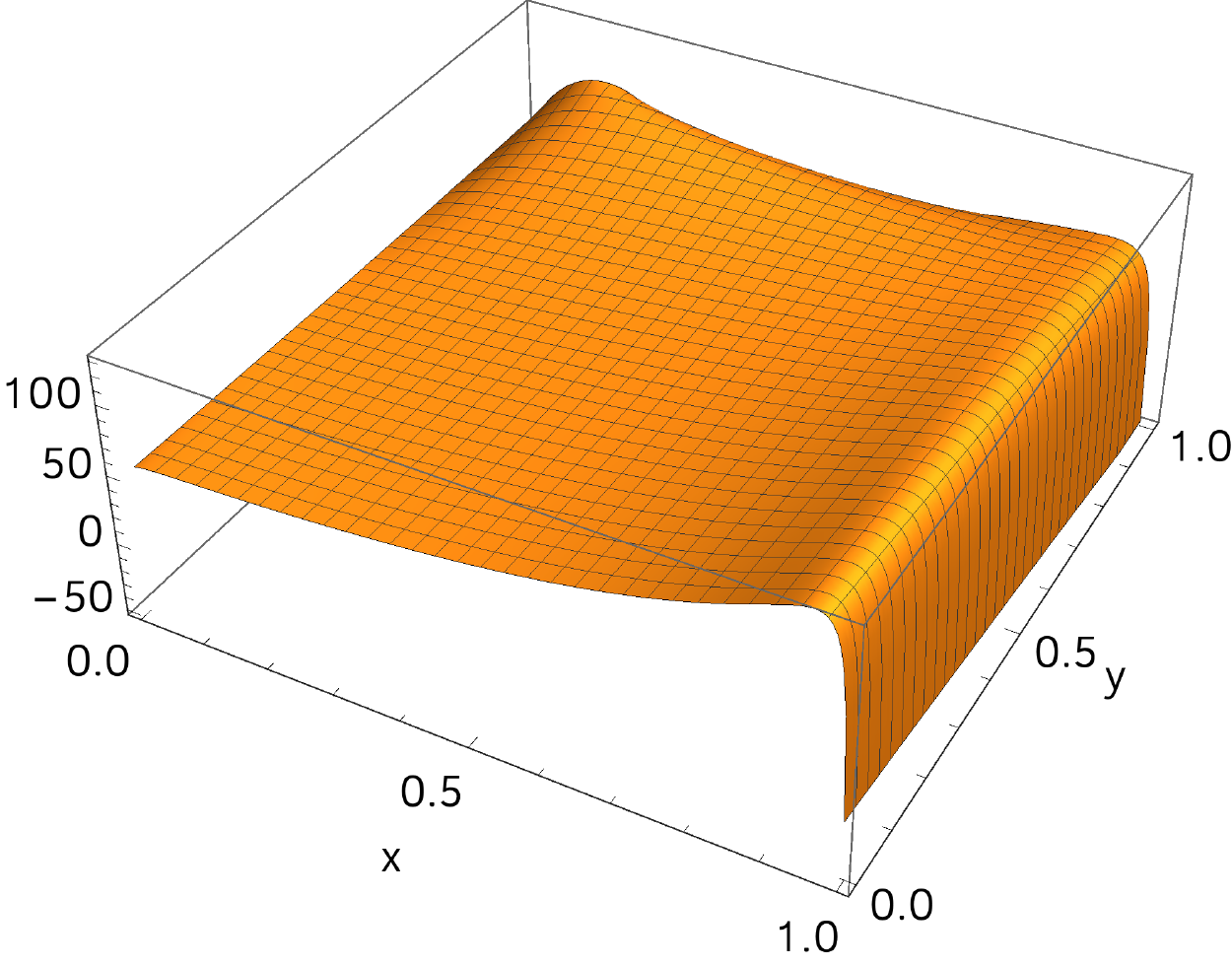}
    \caption{${\rm Im}(I^{(4)}_3)$ of \zzz{} topology.}
\label{fig:zzzIm2d}
\end{subfigure}
    \caption{Integrals plotted over $\bar{R}$, a two-dimensional sub-region of the physical region
    defined in eq.~(\ref{cut R}). The integrals are singular at the point 
    $P_1$ of eq.~\eqref{eq:singXY} on the 
    top edge ($y=1$) of the unit square, 
    and on the right edge ($x\rightarrow 1$) of the unit square, 
    corresponding to the threshold at the point $P_2$ of eq.~\eqref{eq:singXY}.  }
\label{fig:2dSlice}
\end{figure}

\begin{figure}[h!] 
\centering
\begin{subfigure}{0.45\textwidth}\centering
\includegraphics[scale=0.50]{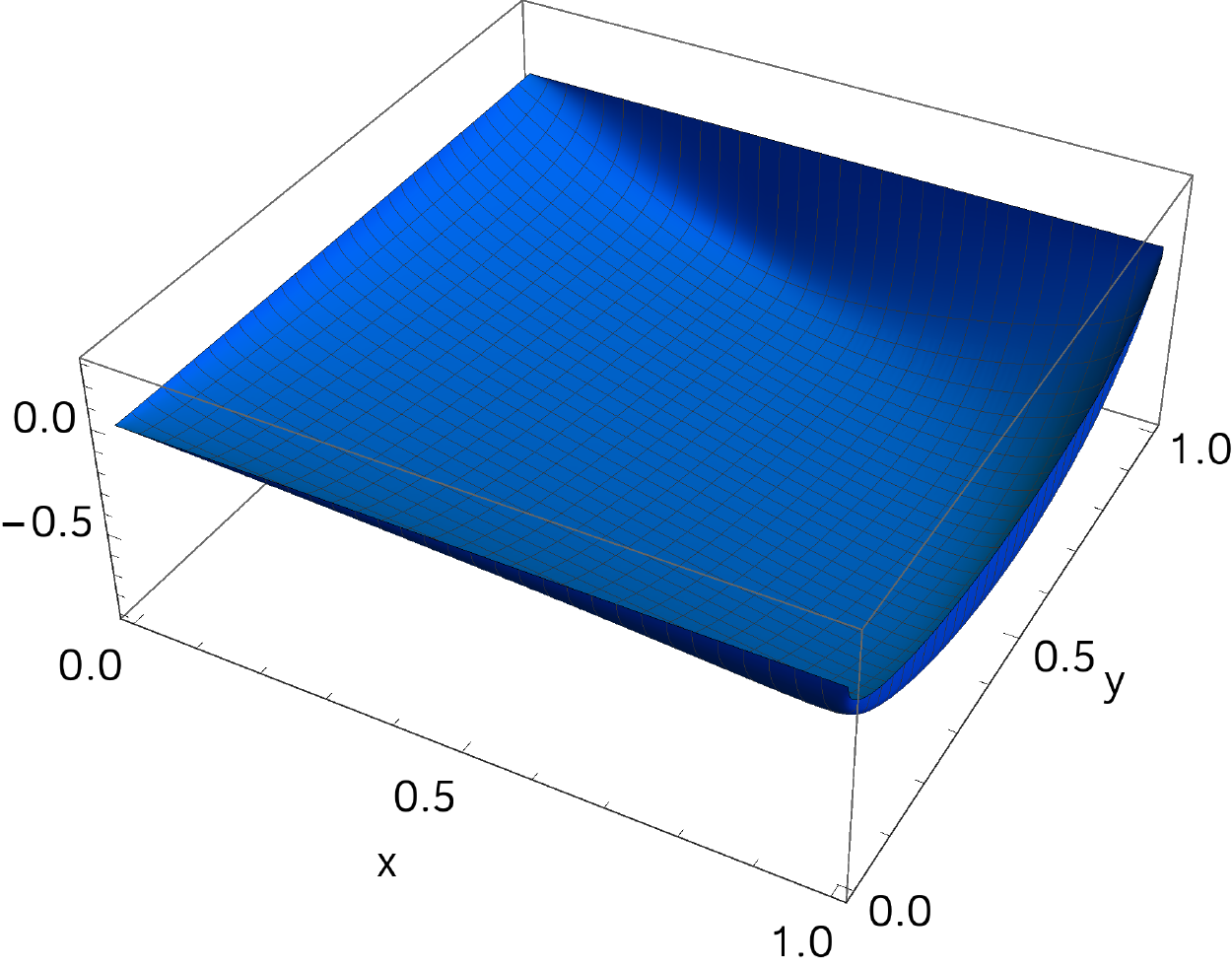}
    \caption{${\rm Re}(I^{(4)}_1)$ of 1-loop topology.}
\label{fig:1loopRe2d}
\end{subfigure}\hspace{10mm}
\begin{subfigure}{0.45\textwidth}\centering
\includegraphics[scale=0.50]{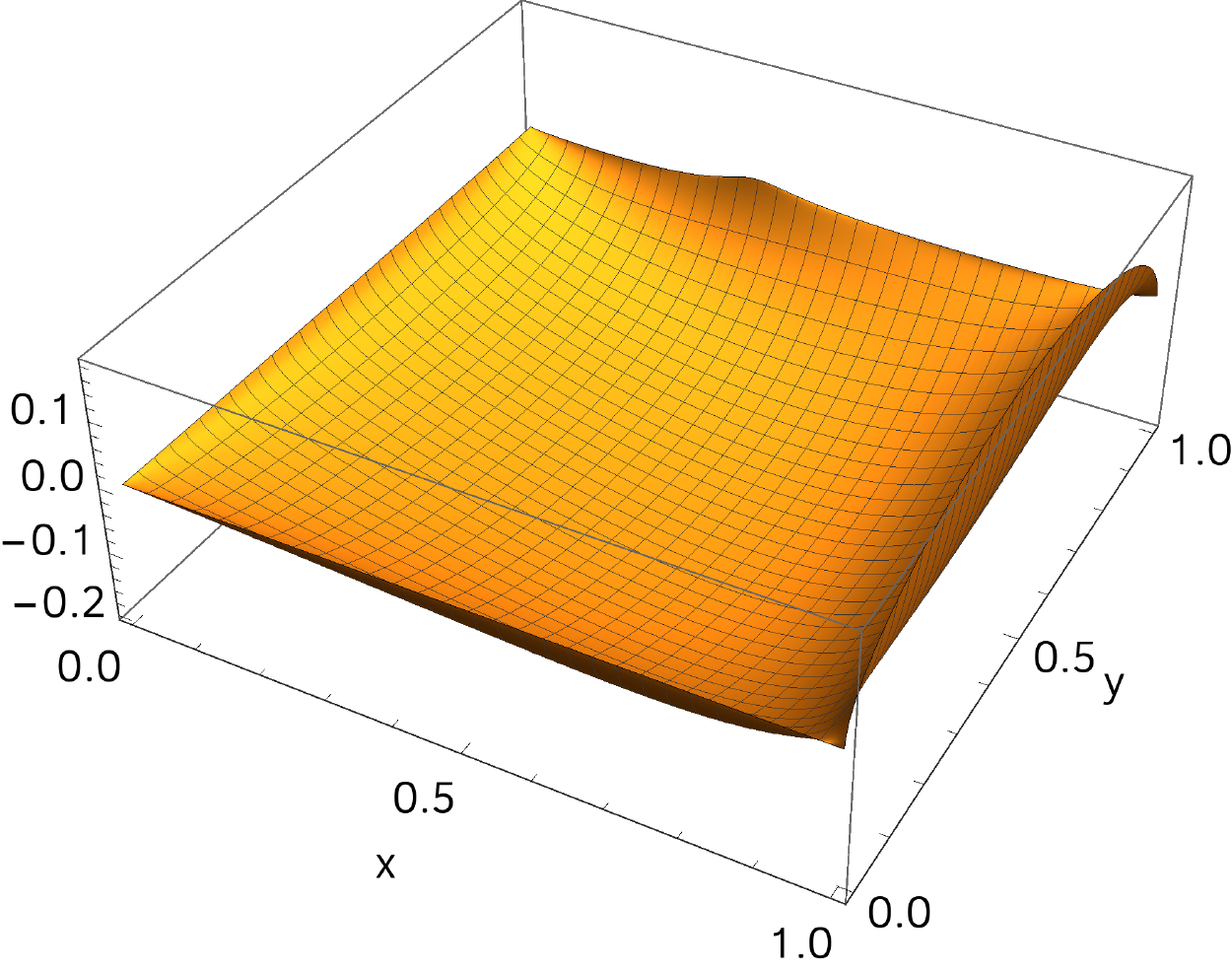}
    \caption{${\rm Im}(I^{(4)}_1)$ of 1-loop topology.}
\label{fig:1loopIm2d}
\end{subfigure}\hspace{10mm}
\caption{Weight-four contribution to the 
    pure 1-loop pentagon plotted over the region $\bar{R}$ defined in eq.~(\ref{cut R}).
    Due to its normalization, the function should vanish at $\text{det}\;G(p_1,p_2,p_3,p_4)=0$,
    corresponding to the edges of the unit square.
    Because of the cut-off at $x=0.998926\ldots$, 
    the function does not vanish on the $x=1$ edge.}
\label{fig:2dSlice 1-loop}
\end{figure}

In order to plot the functions, it is convenient to map $R$ to a finite region. 
We thus map $R$ to a unit square with the following change of variables
\begin{equation}
\label{eq: x y }
s_{23}(x)=\frac{s_{23}^{(2)}}{\left(1-x\right)}\,, \quad
s_{34}(x,y)=y\,s_{34}^{(+)}(x)+(1-y)\,s_{34}^{(-)}(x)\,,\quad 0<x<1,\; 0<y<1\,,
\end{equation}
where we highlight that $s_{34}^{(\pm)}$ are functions of $x$, following from their dependence
on $s_{23}$.
Under this change of variables, the points in eq.~\eqref{eq:singularPoints} are
mapped to
\begin{align}\label{eq:singXY}
  P_1=\{x=0.376542\ldots,\; y=1\}
  \quad\textrm{and}\quad
  P_2=\{x=1,\; 0\leq y\leq1\}\,.
\end{align}
The cut-off at $s_{23}=(13\;\text{TeV}/80\;\text{GeV})^2$ translates
to a cut-off at $x=0.998926\ldots$. We expect the plots to have interesting features
around $P_1$ and $P_2$ of eq.~\eqref{eq:singXY}. Despite the cut-off not allowing us
to reach $x=1$, the cut-off is sufficiently close to 1 that we expect to see the effect of the
threshold.
In summary, the phase-space region shown in figs.~\ref{fig:2dSlice} and
\ref{fig:2dSlice 1-loop} is
\begin{equation}
\label{cut R}
\bar{R}= \left\{(x,y)\;| 
\;0<x<0.998926\dots,\; 0<y<1\right\}.
\end{equation}

\begin{figure}[h!] 
\begin{subfigure}{0.45\textwidth}\centering
\includegraphics[scale=0.55]{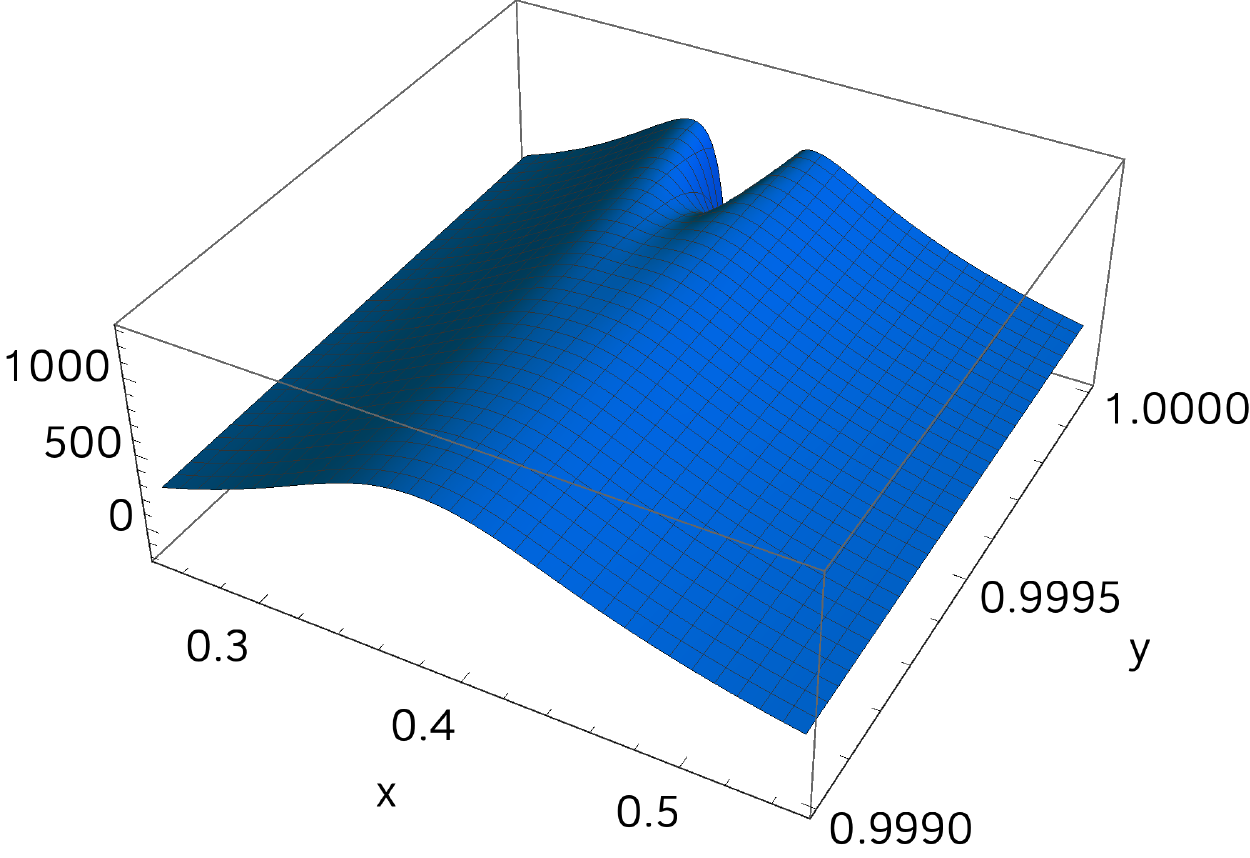}
    \caption{${\rm Re}(I^{(4)}_3)$ of \zmz{} topology.}
\label{fig:zmzRe2dzoom}
\end{subfigure}\hspace{10mm}
\begin{subfigure}{0.45\textwidth}\centering
\includegraphics[scale=0.55]{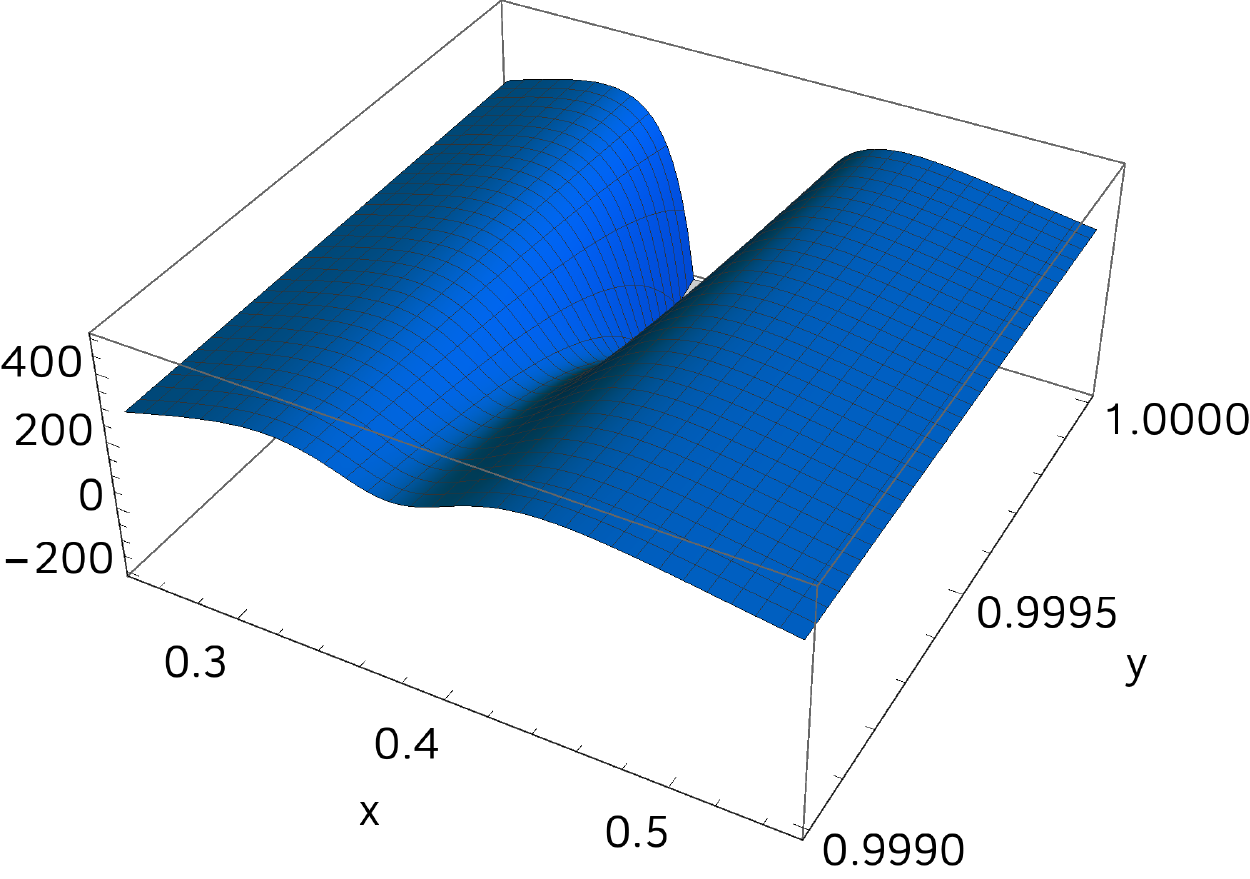}
    \caption{${\rm Re}(I^{(4)}_3)$ of \zzz{} topology.}
\label{fig:zzzRe2dzoom}
\end{subfigure}
\caption{Enlarged view of the integrals in figs.~\ref{fig:zmzRe2d}
and \ref{fig:zzzRe2d} near the threshold $s_{34}=0$.}
\label{fig:zoom 2dSlice}
\end{figure}

\begin{figure}[h!] 
\centering
\begin{subfigure}{0.45\textwidth}\centering
\includegraphics[scale=0.55]{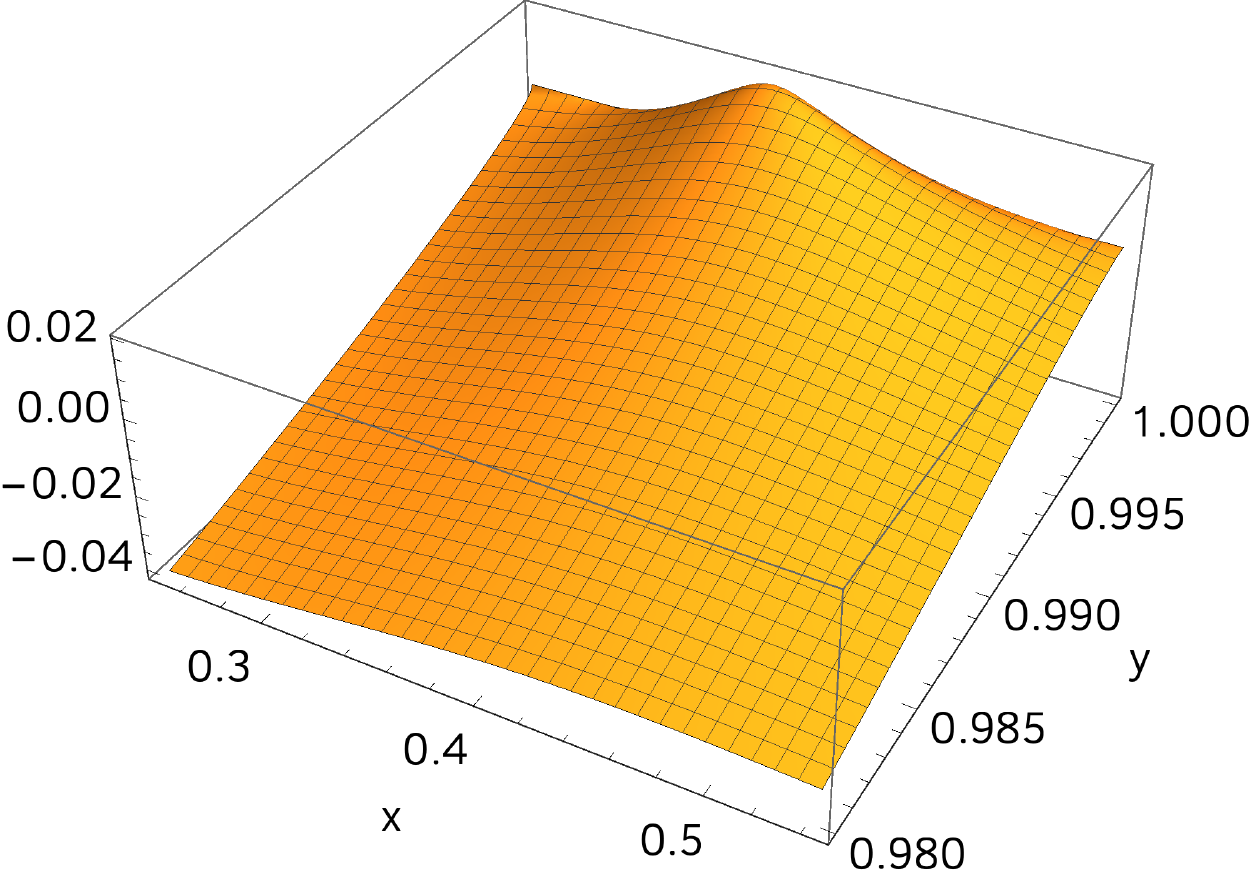}
\caption*{${\rm Im}(I^{(4)}_1)$ of 1-loop topology.}
\label{fig:1loopIm2dzoom}
\end{subfigure}\hspace{10mm}
\caption{Enlarged view of the integral in \fig{fig:1loopIm2d} near the threshold $s_{34}=0$.  
}
\label{fig:zoom 2dSlice 1-loop}
\end{figure}

In the remainder of this section we discuss the plots of master integrals in
this region shown in figs.~\ref{fig:2dSlice} through \ref{fig:zoom 2dSlice 1-loop}.
They were generated by computing selected integrals over a set of 200k points in the 
interval $0<y<0.9$, and over a set of 200k points in the interval $0.9\leq y <1$, 
where the integrals exhibit fine structures and large variance. 
In each interval, the points are evenly distributed over 200 equally-spaced parallel 
lines in the direction of the $y$ axis. This gives us enough granularity to 
observe the smoothness of the functions in the bulk of the phase-space region
we are exploring, as well as the behavior around the singular points $P_1$ and $P_2$.

The plots of the highest non-vanishing two-loop integrals at weight four over $\bar{R}$ 
are presented in \fig{fig:2dSlice}. 
For each topology \texttt{f}=$\mzz$, $\zmz$ and $\zzz$, we show the
real and imaginary parts of the 
penta-box integrals with the insertions $\mathcal N^{(3)}_{\textrm{pb},\texttt{f}}$ 
given in eqs.~\eqref{eq:pbmzz}, \eqref{eq:pbzmz} and \eqref{eq:pbzzz}
(the other two insertions for each penta-box are only non-zero starting at weight 5).
As expected, the integrals have interesting features near the singular points,
some of which are not always apparent in the plots due to the perspective. 
The plots in figs.~\ref{fig:mzzRe2d} and \ref{fig:mzzIm2d}
have clear logarithmic divergences at each threshold point, where
the integral tends to $+\infty$. 
The plots in figs.~\ref{fig:zmzRe2d} and \ref{fig:zmzIm2d} are consistent with a
divergence to $-\infty$ at $P_2$. Regarding the behavior at $P_1$, the start of a logarithmic
dip towards $-\infty$ can be seen in the imaginary part (see fig.~\ref{fig:zmzIm2d}), 
but the behavior of the real part in fig.~\ref{fig:zmzRe2d} is more intricate. 
We thus take a closer look at this behavior in the region around $P_1$ 
in fig.~\ref{fig:zmzRe2dzoom}, to show that the real part also has a logarithmic divergence
towards $-\infty$ at $P_1$. Similar conclusions hold for the integral of
the $\zzz$ topology: the real part diverges to $+\infty$ at $P_2$ (see \fig{fig:zzzRe2d})
and $-\infty$ at $P_1$ (see \fig{fig:zzzRe2dzoom}). The imaginary part diverges to $-\infty$
at both $P_1$ and $P_2$, see \fig{fig:zzzIm2d}.

The same analysis was performed for the one-loop pentagon integral at weight four.
In \fig{fig:2dSlice 1-loop}, we plot the real and imaginary part of the pure integral,
which we recall is normalized by a factor of $\trFive$ (see the pure basis
in \texttt{anc/1loop/pureBasis-1loop.m}). This implies that the function is
odd under $\trFive\to-\trFive$, see e.g.~the discussion in section \ref{sec:pureBasis}.
It must thus vanish at the edge of the region $R$, where $\trFive=0$. Given the
cut-off at $x = 0.998926\ldots$, the integral must vanish on all but this
edge of $\bar{R}$. This is indeed what we observe (the vanishing is not apparent
in the $y=1$ edge of \fig{fig:1loopIm2d}, but this is because of the perspective
we chose). Figure~\ref{fig:1loopIm2d} has a peak around the singular 
point $P_1$, consistent with the start of a logarithmic divergence. In 
\fig{fig:zoom 2dSlice 1-loop} we close in on that region and see that 
the condition that the function should vanish at $y=1$ eventually wins and,
as expected, pushes the integral back to zero on the edge.



\subsection{Validation}

We have performed several checks on the results obtained with our approach 
for the numerical evaluation of the integrals.
Aside from verifying that we obtain the correct values for 
integrals that are trivial to evaluate
at one and two loops,
we have validated our program with the following checks:
\begin{itemize}
\item Two independent implementations of the approach were made to check for
      internal consistency.
\item We compared the high-precision evaluations to values obtained from
the~\pySecDec{} program \cite{Borowka:2017idc}. All one-loop integrals were
validated up to weight four on the physical and Euclidean phase-space points.
All two-loop integrals were checked to match the \pySecDec{} results within
error estimates on the Euclidean point.
\item The integrals of the $\mzz$ topology were validated against the results of
ref.~\cite{Papadopoulos:2015jft}.  We tested at least one integral per sector on all 6
physical phase-space points. We found agreement, including for a
high-precision comparison with 128 digits.
\end{itemize}

\FloatBarrier

%% file: conclusions.tex

\section{Conclusions}
\label{sec:Conclusions}

In this paper we described the computation of the full set of planar two-loop
master integrals with one
massive and four massless legs. These integrals are the complete set
required to compute the amplitudes necessary for NNLO predictions of $W$-boson production in
association with two jets in the leading-color approximation at the LHC.
Furthermore, they are also a crucial ingredient for these amplitudes beyond leading color,
and for $Z$- or Higgs-boson production in association with two jets at the LHC.

We computed the master integrals by obtaining canonical
differential equations which we solved using generalized series-expansion techniques.
In order to construct the differential equations we found
a basis of pure master integrals, all with surprisingly compact integrand
representations. It would be interesting to further explore the mathematical
properties which make them pure. The analytic differential equations were then
constructed using numerical techniques, following the approach introduced
in \cite{Abreu:2018rcw}.
Importantly, we showed how finite fields, despite not being
algebraically closed, can be used throughout the calculation even when
intermediate stages require taking square roots, as 50\% of the elements of the field
are perfect squares.
Beyond a pure basis, this method of constructing the differential equation requires a
priori knowledge of the symbol alphabet. Remarkably, we find that the alphabet
can be constructed by considering differential equations for only maximal and
next-to-maximal cut integrals, which can easily be constructed analytically. 

The alphabet displays a number of notable features. Firstly, despite the
complex nature of the five-point one-mass kinematics, the full set of letters
can be written in a remarkably compact form. We expect the alphabet itself to be
of great use in the future. Indeed, it forms the minimal necessary information
required for the construction of pentagon functions, extending the construction
of ref.~\cite{Gehrmann:2018yef} to five-point one-mass kinematics. These are a
valuable tool for compactly presenting scattering amplitudes, which has been shown
to be of great use in the reconstruction of analytic results from numerical data
\cite{Badger:2018enw,Badger:2019djh,Abreu:2018zmy,Abreu:2019odu}.

We also considered the analytic structure of the master integrals at symbol level
and made a number of interesting observations. First, we find that certain
letters which arise in the master integrals at all orders in $\epsilon$ are in
fact not present in the symbol at weight four---that is, they decouple from the
four dimensional physics. Second, we confirm that the master integrals satisfy
the extended Steinmann relations to all orders in $\epsilon$, and also observe that
there are other as-yet unexplained similar relations.

In order to solve the differential
equations, we employed the generalized series-expansion method of 
ref.~\cite{Francesco:2019yqt} both to compute the integrals in all kinematic regions
relevant for vector-boson production in association with two jets, and to obtain
Euclidean boundary conditions from consistency conditions of the differential
equation. We demonstrated the viability of the method for applications to
LHC physics through a number of numerical studies, both at high precision for
individual phase-space points and more generically over physical regions. 

A natural next step is to consider the non-planar extension of the integrals
considered here, especially given their relevance for precise predictions for
the production of a Higgs boson in association with two jets at hadron
colliders. In the case of massless scattering, it was observed that the
non-planar symbol alphabet could be obtained through permutations of the planar
alphabet, and it would be interesting to see if this also holds here. As the
generalized-series approach is powerful and applicable to any first order linear
differential equation, it would also be interesting to develop an automated
public implementation for general Feynman integrals.

%% file: app_kin.tex

\section{Kinematic parametrizations}
\label{appendix:kinematics}
In handling expressions with five-point one-mass kinematics, it is often useful
to have different parametrizations for the kinematics. Firstly, it is useful to
be able to express all possible Mandelstam invariants in terms of the ordered
variables $\vec{s}$, 
\begin{align}\begin{split}
s_{13}&= -s_{12}-s_{23}+s_{45}+\offShellScale{}, \, s_{14}= -s_{15}+s_{23}-s_{45} + \offShellScale{},\, s_{24}= s_{15}-s_{23}-s_{34}, \\
s_{25}&= -s_{12}-s_{15}+s_{34}+\offShellScale{},\, s_{35}= s_{12}-s_{34}-s_{45}.
\end{split}\end{align}

Furthermore, we can write all the necessary Gram determinants in terms of
ordered invariants, 
\begin{align}\begin{split}
  &\Delta_5 = (-s_{12} s_{15} + s_{12} s_{23} + \offShellScale{} s_{34} + s_{15} s_{45} - s_{34} s_{45} - s_{23} s_{34})^2 \\
  &\quad \quad \quad - 
  4 s_{23}  s_{34} s_{45} (\offShellScale{} - s_{12} - s_{15} + s_{34}), \\
   &\Delta_3=s_{23}^2+s_{45}^2+\offShellScaleSquared-2 s_{23} s_{45}-2 \offShellScale{} s_{23}-2 \offShellScale{} s_{45}, \\
   &\Delta_3^{\ncg}=(s_{12}+s_{15})^2-4 \offShellScale{} s_{34}.
\end{split}\end{align}

Beyond this parametrization in terms of Mandelstam invariants, it is often
useful when handling symbols to work with a set of variables that rationalizes
(a subset of) the alphabet. One useful parametrization is that in which the
Gram determinant $\Delta_3$ is a perfect square. The variables $
\offShellScale{},\ s_{12},\ s_{15}$ and $s_{34}$ remain unchanged and we introduce new
variables $z$ and $\bar{z}$ defined via
\begin{flalign}
 s_{23}=z \bar{z} \offShellScale{},\ \ \ s_{45}=(1-z)(1-\bar{z}) \offShellScale{}.
\end{flalign}
In these variables $\sqrt{\Delta _3}$ takes the form 
\begin{flalign}
 \sqrt{\Delta _3} = \offShellScale{} (z-\bar{z}).
\end{flalign}
Another useful parametrization is that which rationalizes both
$\text{tr}_5$ and $\sqrt{\Delta _3}$ simultaneously.\footnote{
	We thank Marco Besier for building this parametrisation, based on
	the work presented in \cite{Besier:2018jen}.
}
The corresponding change of variables is given via
\begin{flalign}\begin{split}
\offShellScale{} &= \frac{\left(u_2-1\right) \left(s_{23}-s_{45} u_2\right)}{u_2},\ s_{12}=s_{12}, s_{23}= s_{23},s_{34}= s_{34},s_{45}= s_{45}\,,  \\
\ s_{15}&= \frac{s_{23} \left(s_{12}^2 u_4 \left(s_{45} u_2-u_4\right)+s_{34} \left(s_{45}-u_4\right) \left(s_{45}^2 u_2-u_4
   \left(s_{45} u_2+s_{12}\right)\right)\right)}{s_{12} s_{45} u_2 u_4
   \left(s_{12}-s_{45}+u_4\right)} \\
   &+\frac{s_{12} u_2 u_4 \left(s_{34} \left(s_{12} u_4+s_{45} u_2 \left(u_4-s_{45}\right)\right)-s_{12} u_4 \left(s_{45} \left(u_2-1\right)+s_{12}\right)\right)}{s_{12} s_{45} u_2 u_4
   \left(s_{12}-s_{45}+u_4\right)}\,.
\end{split}\end{flalign}
The explicit form of $\sqrt{\Delta _3}$ and $\trFive{}$ in the new variables is
\begin{flalign}\begin{split}
 \sqrt{\Delta_3} =& \frac{s_{23}-s_{45} u_2^2}{u_2},\\
 \trFive=&\frac{1}{s_{12} s_{45} u_2 u_4 \left(s_{12}-s_{45}+u_4\right)}
 \bigg(s_{12}^4 u_2 u_4^2+s_{12}^3 u_4^2 \left(u_2 \left(s_{45}
   \left(u_2-2\right)-s_{34}\right)+s_{23}\right)\\
   &-s_{12}^2 u_4^2 \left(s_{23}\left(s_{34}-s_{45} \left(u_2-1\right)\right)
   +s_{45} \left(s_{34}+s_{45}\right)\left(u_2-1\right) u_2\right)\\
   &+s_{23} s_{34} s_{45} s_{12} u_2
   \left(s_{45}^2-u_4^2\right)-s_{23} s_{34} s_{45}^2 u_2 \left(s_{45}-u_4\right){}^2\bigg)\,.
\end{split}\end{flalign}

%% file: app_pure.tex

\section{Pure planar five-point one-mass integrals}
\label{pure_pentagons}

In this appendix we list our choice of master integrals
for the five-point topologies in \fig{fig_master_int}.

\subsection*{Penta-boxes}

\hspace{\parindent}
\begin{minipage}{0.4\textwidth}
\begin{figure}[H]
\centering
\includegraphics[width=4.5cm]{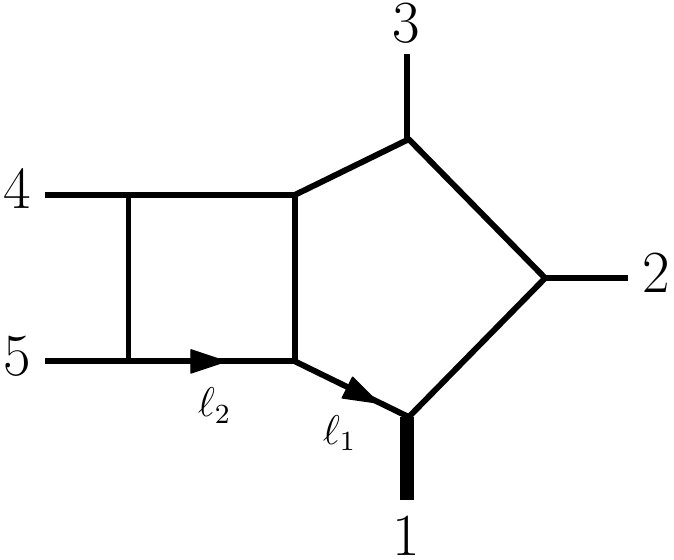}
\end{figure}
 \end{minipage}
\begin{minipage}{0.54\textwidth}
\begin{align}\begin{split}\label{eq:pbmzz}
\mathcal N^{(1)}_{\textrm{pb},\mzz} &= \epsilon^4 s_{45} \text{tr}_5 \mu_{12},\\
\mathcal N^{(2)}_{\textrm{pb},\mzz} &= \epsilon^4 \frac{1-2 \epsilon}{1+2 \epsilon} \text{tr}_5   (\mu_{11}\mu_{22}-\mu_{12}^2) \\
\mathcal N^{(3)}_{\textrm{pb},\mzz} &= \epsilon^4 s_{45}  s_{12} s_{23} (\ell_1-p_5)^2\,.
\end{split}\end{align}
\end{minipage}%

\begin{minipage}{0.4\textwidth}
\begin{figure}[H]
\centering
\includegraphics[width=4.5cm]{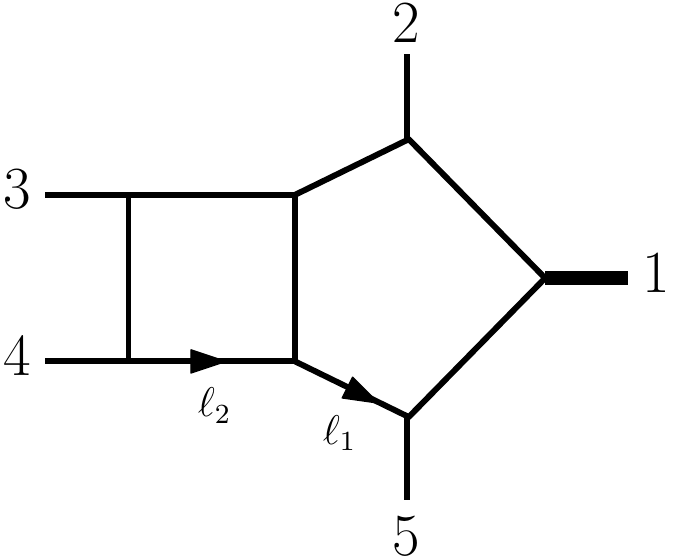}  
\end{figure}
\end{minipage}
\begin{minipage}{0.54\textwidth}
\begin{align}\label{eq:pbzmz}
\mathcal N^{(1)}_{\textrm{pb},\zmz} &= \epsilon^4 s_{34} \text{tr}_5 \mu_{12},\nonumber\\
\mathcal N^{(2)}_{\textrm{pb},\zmz} &= \epsilon^4 \frac{1-2 \epsilon}{1+2 \epsilon} \text{tr}_5    (\mu_{11}\mu_{22}-\mu_{12}^2),\\
\mathcal N^{(3)}_{\textrm{pb},\zmz} &=\epsilon^4 s_{34}  (s_{15} s_{12}-\offShellScale{} s_{34}) (\ell_1-p_4)^2\,.\nonumber
\end{align}
\end{minipage}

\begin{minipage}{0.4\textwidth}
\begin{figure}[H]
\centering
\includegraphics[width=4.5cm]{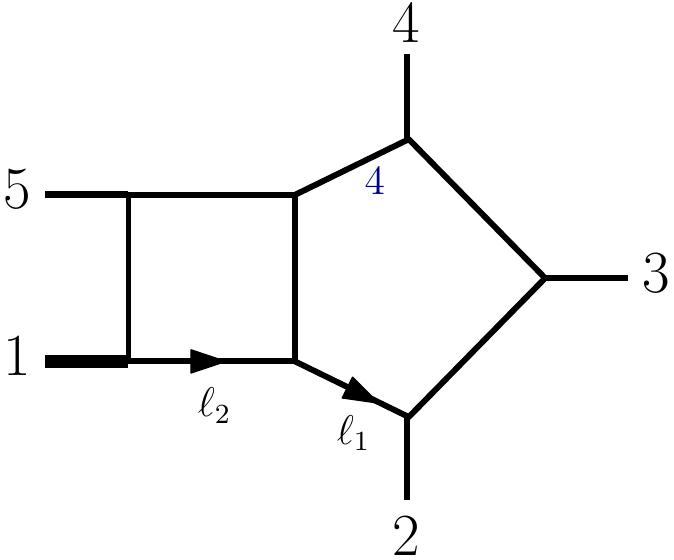}  
\end{figure}
\end{minipage}
\begin{minipage}{0.54\textwidth}
\begin{align}\begin{split}\label{eq:pbzzz}
\mathcal N^{(1)}_{\textrm{pb},\zzz} &= \epsilon^4 s_{15} \text{tr}_5  \mu_{12},\\
\mathcal N^{(2)}_{\textrm{pb},\zzz} &= \epsilon^4 \frac{1-2 \epsilon}{1+2 \epsilon}\,\text{tr}_5\,(\mu_{11}\mu_{22}-\mu_{12}^2),\\
\mathcal N^{(3)}_{\textrm{pb},\zzz} &= \epsilon^4     s_{23} s_{34}     (s_{15}(\ell_1-p_1)^2-\offShellScale{}\rho_4).
\end{split}\end{align}
\end{minipage}

\subsection*{Penta-triangle}

\hspace{\parindent}
\begin{minipage}{0.4\textwidth}
 \begin{figure}[H]
 \centering
 \includegraphics[width=4.5cm]{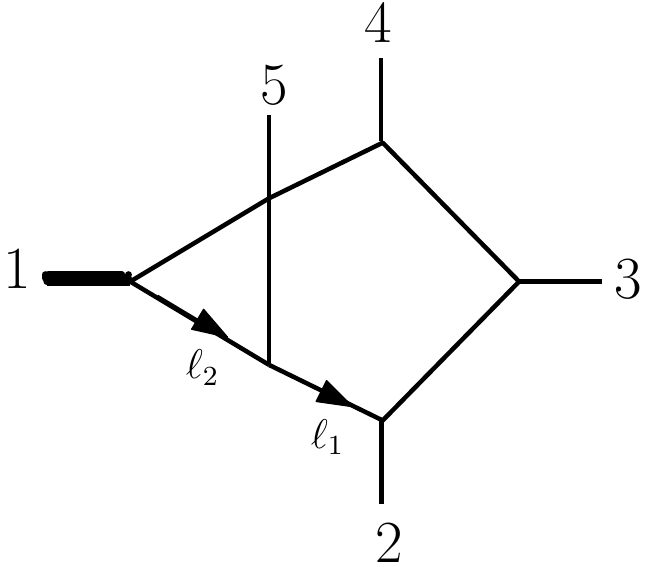}  
\end{figure}
\end{minipage}
\begin{minipage}{0.54\textwidth}
\begin{align}\begin{split}
\mathcal N_{\textrm{pt}} &=  \epsilon^4 \,\text{tr}_5\, \mu_{11} \, .
\end{split}\end{align}
\end{minipage}
\vspace{.5cm}

\subsection*{Double-boxes}

\hspace{\parindent}
\begin{minipage}{0.4\textwidth}
\begin{figure}[H]
\centering
\includegraphics[width=4.5cm]{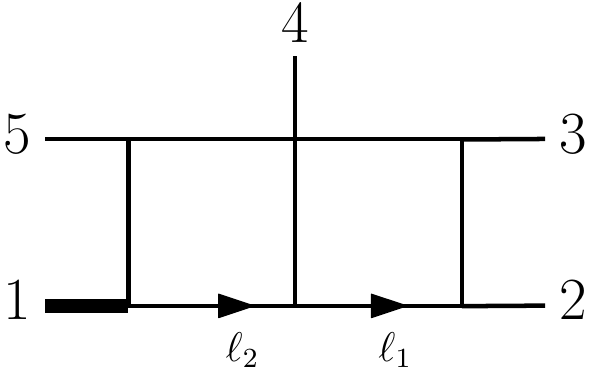}  
\end{figure}
\end{minipage}
\begin{minipage}{0.54\textwidth}
\begin{align}\begin{split}
\mathcal N^{(1)}_{\textrm{db},1} &=  \epsilon^4 s_{23} (s_{12} s_{15}-s_{34} \offShellScale{}),\\
\mathcal N^{(2)}_{\textrm{db},1} &=  \epsilon^4 s_{23}(s_{15}-\offShellScale{})(\ell_2+p_2)^2,\\
\mathcal N^{(3)}_{\textrm{db},1} &= \epsilon^4\,\trFive{}\, \mu_{12}\, .
\end{split}\end{align}
\end{minipage}

\begin{minipage}{0.4\textwidth}
\begin{figure}[H]
\centering
\includegraphics[width=4.5cm]{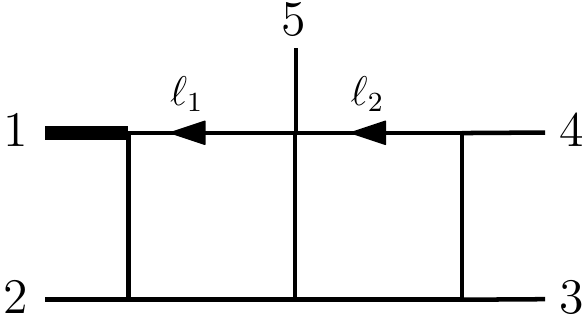} 
\end{figure}
\end{minipage}
\begin{minipage}{0.54\textwidth}
\vspace{.6 cm}
\begin{align}\begin{split}
\mathcal N^{(1)}_{\textrm{db},2} &=  \epsilon^4s_{34} s_{23} s_{12},\\
\mathcal N^{(2)}_{\textrm{db},2} &=  \epsilon^4s_{34}(s_{12}-\offShellScale{})(\ell_1-p_4)^2,\\
\mathcal N^{(3)}_{\textrm{db},2} &= \epsilon^4\,\text{tr}_5\, \mu_{12}\, .
\end{split}\end{align}
\end{minipage}

\vspace{.5cm}
\begin{minipage}{0.4\textwidth}
\begin{figure}[H]
\hspace{-0.75cm}
 \includegraphics[width=5.8cm]{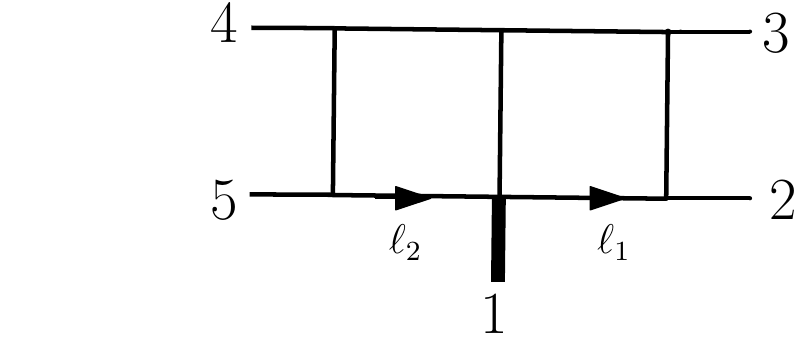}  
\end{figure}
\end{minipage}
\hspace{-1.2 cm}
\begin{minipage}{0.54\textwidth}
\vspace{-0.5 cm}
\begin{align}\begin{split}
\mathcal N^{(1)}_{\textrm{db},3} &=  \epsilon^4s_{23} s_{45} s_{34},\\
\mathcal N^{(2)}_{\textrm{db},3} &=  \epsilon^4s_{23}s_{45}(\ell_2+p_2)^2,\\
\mathcal N^{(3)}_{\textrm{db},3} &= \epsilon^4\,\text{tr}_5\, \mu_{12} \, .
\end{split}\end{align}
\end{minipage}

\subsection*{Triangle-boxes}

\hspace{\parindent}
\begin{minipage}{0.4\textwidth}
\begin{figure}[H]
\centering
\includegraphics[width=4.5cm]{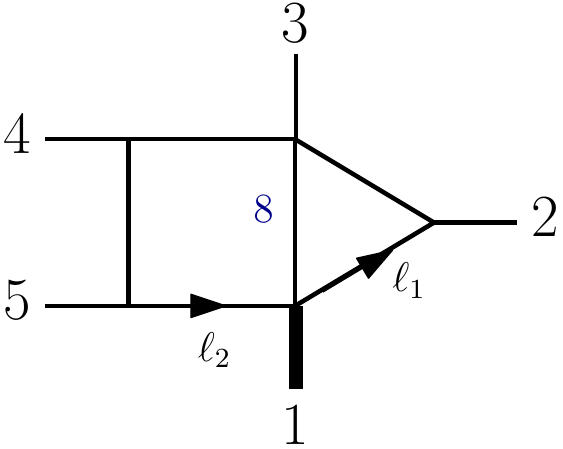}  
\end{figure}
\end{minipage}
\begin{minipage}{0.54\textwidth}
\begin{align}\begin{split}
\mathcal N^{(1)}_{\textrm{tb},1} &=  \epsilon^4 s_{45} (s_{34}-s_{15}),\\
\mathcal N^{(2)}_{\textrm{tb},1} &= \epsilon^3\,\text{tr}_5\, \mu_{22}\frac{1}{\rho_8} \, .
\end{split}\end{align}
\end{minipage}

\begin{minipage}{0.4\textwidth}
\begin{figure}[H]
\centering
 \includegraphics[width=4.5cm]{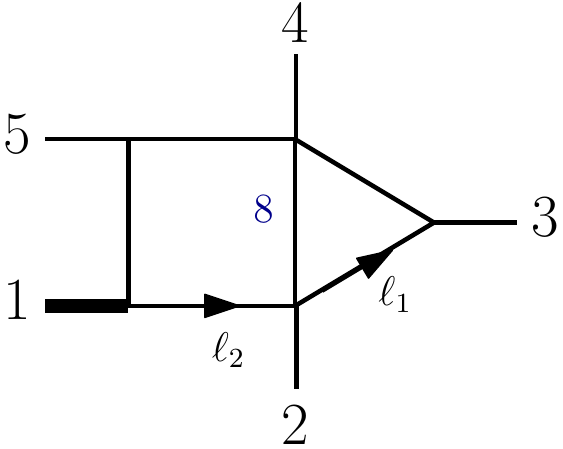}  
\end{figure}
\end{minipage}
\begin{minipage}{0.54\textwidth}
\begin{align}\begin{split}
\mathcal N^{(1)}_{\textrm{tb},2}&= \epsilon^4s_{15}(s_{12}-s_{45})-\offShellScale{}s_{34},\\
\mathcal N^{(2)}_{\textrm{tb},2}&= \epsilon^3\,\text{tr}_5\, \mu_{22}\frac{1}{\rho_8} \, .
\end{split}\end{align}
\end{minipage}

\begin{minipage}{0.4\textwidth}
\begin{figure}[H]
\centering
 \includegraphics[width=4.5cm]{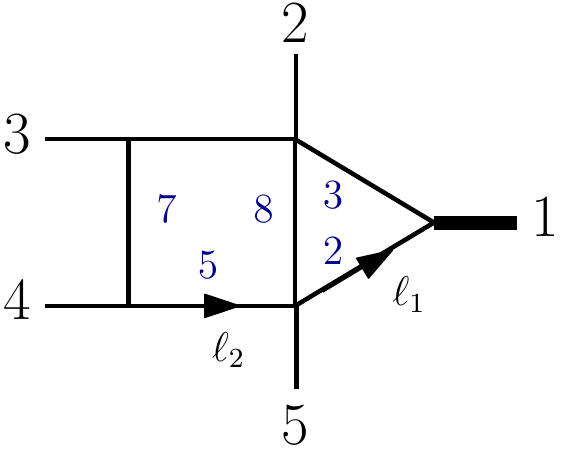}  
\end{figure}
\end{minipage}
\begin{minipage}{0.54\textwidth}
\vspace{1cm}
\begin{align}
\mathcal N^{(1)}_{\textrm{tb},3}&=  \epsilon^4 s_{34} \sqrt{\Delta_3},\nonumber\\
\mathcal N^{(2)}_{\textrm{tb},3}&= \epsilon^3\, \text{tr}_5\, \mu_{22}\frac{1}{\rho_8},\nonumber\\
\mathcal N^{(3)}_{\textrm{tb},3}&= \epsilon^4 \left(s_{34} (\offShellScale{} - s_{23} + s_{45}) - \frac{1}{\epsilon}\frac{\offShellScale{} s_{34} s_{45}}{ \rho_2}\right),\nonumber\\
\mathcal N^{(4)}_{\textrm{tb},3}&=\epsilon^4 \left(s_{34} (\offShellScale{} + s_{23} - s_{45}) -  \frac{1}{\epsilon}\frac{\offShellScale{} s_{23} s_{34}}{\rho_3}\right),\nonumber\\
\mathcal N^{(5)}_{\textrm{tb},3}&=  \epsilon^4 \left( s_{34} (\offShellScale{} + s_{23} - s_{45}) +  \frac{1}{\epsilon}s_{15} s_{34} \frac{\rho_7}{\rho_5}\right. \nonumber\\
 & \left.+\frac{1}{\epsilon}  \offShellScale{} s_{34} \frac{(\ell_1-p_4)^2}{\rho_2}\right),\nonumber\\
\mathcal N^{(6)}_{\textrm{tb},3}&=\epsilon^3\, \text{tr}_5\, \mu_{12}\frac{1}{\rho_8}\, .
\end{align}
\end{minipage}

\subsection*{Bubble-pentagons}

\hspace{\parindent}
\begin{minipage}{0.4\textwidth}
\begin{figure}[H]
\centering
\includegraphics[width=3.5cm]{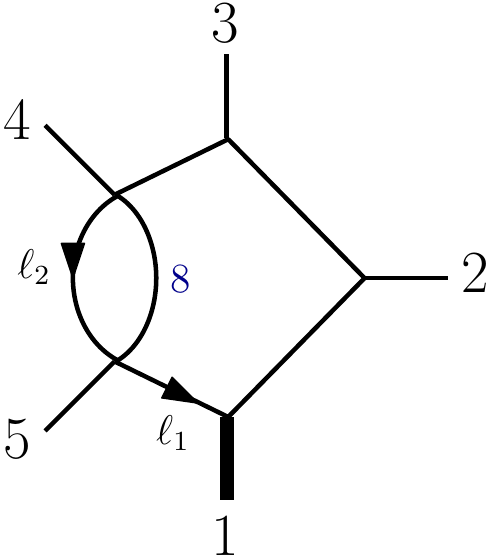}  
\end{figure}
\end{minipage}
\begin{minipage}{0.54\textwidth}
\begin{align}\begin{split}
\mathcal N^{(1)}_{\textrm{bp},1} &= \epsilon^3(1-2\eps) s_{12} s_{23},\\
\mathcal N^{(2)}_{\textrm{bp},1} &= \epsilon^3\, \text{tr}_5\, \mu_{11}\frac{1}{\rho_8}.
\end{split}\end{align}
\end{minipage}

\begin{minipage}{0.4\textwidth}
\begin{figure}[H]
\centering
 \includegraphics[width=3.5cm]{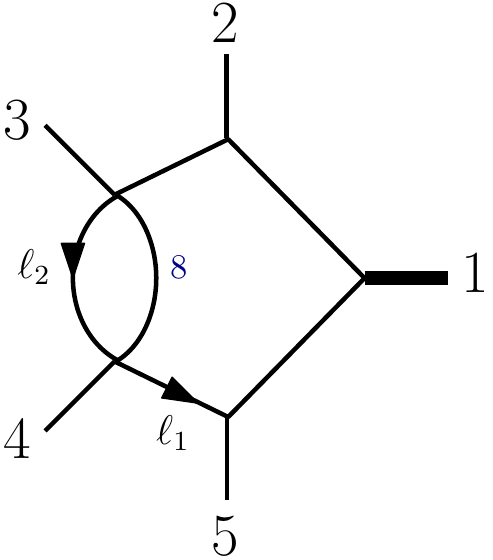}  
\end{figure}
\end{minipage}
\begin{minipage}{0.54\textwidth}
\begin{align}\begin{split}
\mathcal N^{(1)}_{\textrm{bp},2} &= \epsilon^3(1-2\epsilon) (s_{12}s_{15}-s_{34} \offShellScale{}),\\
\mathcal N^{(2)}_{\textrm{bp},2} &= \epsilon^3\, \text{tr}_5\, \mu_{11}\frac{1}{\rho_8}\, .
\end{split}\end{align}
\end{minipage}

\begin{minipage}{0.4\textwidth}
\begin{figure}[H]
\centering
  \includegraphics[width=3.5cm]{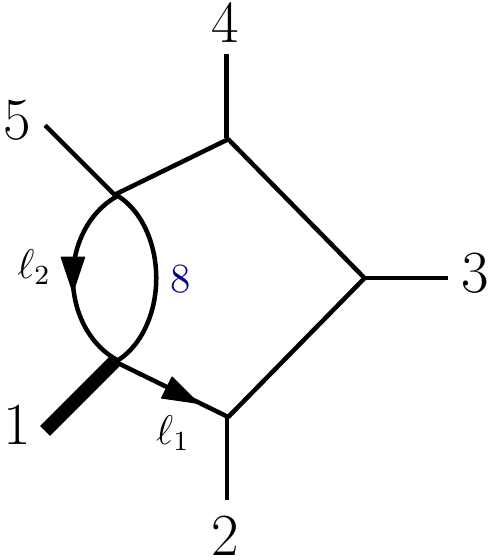}  
\end{figure}
\end{minipage}
\begin{minipage}{0.54\textwidth}
 \begin{align}\begin{split}
\mathcal N^{(1)}_{\textrm{bp},3} &= \epsilon^3(1-2\epsilon) s_{23} s_{34},\\
\mathcal N^{(2)}_{\textrm{bp},3} &= \epsilon^3\,\text{tr}_5\, \mu_{11}\frac{1}{\rho_8}\, .
\end{split}\end{align}
\end{minipage}